\documentclass[conference]{IEEEtran}
\IEEEoverridecommandlockouts
\usepackage{cite}
\usepackage{amsmath,amssymb,amsfonts}
\usepackage{graphicx}
\usepackage{textcomp}
\usepackage{xcolor}
\usepackage{empheq}
\usepackage{algorithmicx}
\usepackage{algorithm,algpseudocode}
\usepackage{algorithmicx}
\usepackage{setspace}
\usepackage{subcaption}
\usepackage{footnote}
\makesavenoteenv{tabular}
\makesavenoteenv{table}

\usepackage{float}
\usepackage{color}
\usepackage{url}
\usepackage{dblfloatfix} 
\usepackage{graphicx}

\usepackage{multirow}
\usepackage[short]{optidef}
\usepackage{balance}
\usepackage{hyperref}
\hypersetup{
    colorlinks=true,
    linkcolor=blue,
    filecolor=magenta,      
    urlcolor=cyan,
}

\def\BibTeX{{\rm B\kern-.05em{\sc i\kern-.025em b}\kern-.08em
    T\kern-.1667em\lower.7ex\hbox{E}\kern-.125emX}}
\begin{document}

\title{Spectrum Monitoring and Analysis in Urban and Rural Environments at Different Altitudes
\thanks{This research is supported in part by the NSF award CNS-1939334 and its
supplement for studying NRDZs.}
}

\author{\IEEEauthorblockN{Amir Hossein Fahim Raouf\IEEEauthorrefmark{1},
Sung Joon Maeng\IEEEauthorrefmark{1}, Ismail Guvenc\IEEEauthorrefmark{1},
{\"O}zg{\"u}r {\"O}zdemir\IEEEauthorrefmark{1}, and Mihail Sichitiu\IEEEauthorrefmark{1}
}
\IEEEauthorblockA{\IEEEauthorrefmark{1}Department of Electrical and Computer Engineering,
North Carolina State University, Raleigh, NC}
\IEEEauthorblockA{amirh.fraouf@ieee.org, \{smaeng,iguvenc,oozdemi,mlsichit\}@ncsu.edu}
}

\maketitle

\begin{abstract}
Due to the scarcity of spectrum resources, the emergence of new technologies and ever-increasing number of wireless devices operating in the radio frequency spectrum lead to data congestion and interference. 
In this work, we study the effect of altitude on sub-6 GHz spectrum measurement results obtained at a Helikite flying over two distinct scenarios; i.e., urban and rural environments. Specifically, we aim at investigating the spectrum occupancy of various long-term evolution (LTE), $5^{\text{th}}$ generation (5G) and citizens broadband radio service (CBRS) bands utilized in the United States for both uplink and downlink at altitudes up to 180 meters.
Our results reveal that generally the mean value of the measured power increases as the altitude increases where the line-of-sight links with nearby base stations is more available. SigMF-compliant spectrum measurement datasets used in this paper covering all the bands between 100~MHz to 6~GHz are also provided.  
\end{abstract}

\begin{IEEEkeywords}
5G, C-Band, CBRS, helikite, LTE, spectrum monitoring, unmanned aerial vehicles (UAV).
\end{IEEEkeywords}

\section{Introduction}
Wireless communication services and the emergence of new technologies have created a huge demand for radio frequency spectrum~\cite{islam2008spectrum}. One prominent problem is the availability of the spectrum and the increase in interference in the current wireless networks~\cite{adelantado2017understanding}. 
In addition, more aggressive frequency reuse is gaining interest recently for achieving higher link capacity in networks without introducing additional spectrum \cite{ericsson}. It is necessary to conduct occupancy studies using spectrum sensing techniques to understand and characterize  interference problems and identify spectrum sharing opportunities.  

There are various recent examples that highlight the importance of understanding spectrum occupancy characteristics, including non-terrestrial scenarios, for developing effective spectrum sharing mechanisms. The launch of $5^{\text{th}}$ generation (5G) cellular service in the United States was a concern for the commercial airline and private aircraft communities who used the radar altimeters of the aircraft industry. 
Although the assigned spectrum band for the altimeters is between 4.2-4.4~GHz, due to their poor design the current versions suffer from out-of-band leakage problem; i.e., they ignore their assigned spectrum boundaries \cite{1}. 
More specifically, Verizon and AT\&T have recently begun operating in the 3.7 GHz to 3.8~GHz spectrum range which is 400~MHz away from the altimeter band. However, this gap may not be sufficient for some aircraft to land safely. Moreover, while both Verizon and AT\&T have been delaying switching on portions of their respective 5G C-band wireless networks until July 2023, it is expected after that day that the whole 3.7-3.98~GHz C-band may be used for 5G transmissions~\cite{CBand_Ref}, introducing additional concerns. There is  a similar coexistence concern for spectrum sharing between the 5G networks to be deployed in the 3.1-3.55~GHz band in the future and the existing airborne radars using the same spectrum. In another recent debate, there is a  concern in using terrestrial nationwide network in the L-Band (i.e., 1-2~GHz) and its potential interference with GPS~\cite{GPS}.

Some existing academic studies on spectrum occupancy are summarized in~\cite{chen2014survey}.
In more recent works, \cite{al2020machine} presents a framework that captures and models the short-time spectrum occupancy to determine the existing interference for Internet-of-things (IoT) applications.
In another study~\cite{homssi2022artificial}, current state-of-the-art artificial intelligence techniques are reviewed for channel forecasting, spectrum sensing, signal detection, network optimization, and security in mega-satellite networks. In~\cite{maeng2022analysis}, authors investigate and characterize the performance of coexisting aerial radar and communication networks for spectrum overlay and time-division multiple access by utilizing stochastic geometry.
In~\cite{azari2018uplink}, the effect of
interference coming from coexisting ground networks on the
aerial link is studied, which could be the uplink (UL) of an aerial cell served by
a drone base station. By considering a Poisson field of ground interferers, they characterize aggregate interference experienced by the drone.

In this paper, by post-processing the measurements from the experiments conducted by the NSF AERPAW platform in Raleigh, NC~\cite{marojevic2020advanced} at  urban and rural environments, we analyze the spectrum occupancy in different U.S. cellular network bands as well as the citizens broadband radio service (CBRS) band. In addition, we study the effect of Helikite altitude on the signal strength pattern. In Section~\ref{data}, we describe the data structure and the overall information of the measurement campaign. Section~\ref{results} and Section~\ref{results_rural} present the spectrum monitoring results for various sub-6~Ghz bands in the urban and rural environments, respectively. Section~\ref{sec:spec_time} studies the time dependency of the spectrum occupancy for the frequency bands under consideration. Finally,  Section~\ref{conclusion} highlights the conclusions of this work.

\section{Data Structure}\label{data}
\begin{figure}
     \centering
     \begin{subfigure}[tb]{1\linewidth}
     \vspace{-1mm}
         \centering
         \captionsetup{justification=centering}
         \includegraphics[width=0.5\textwidth]{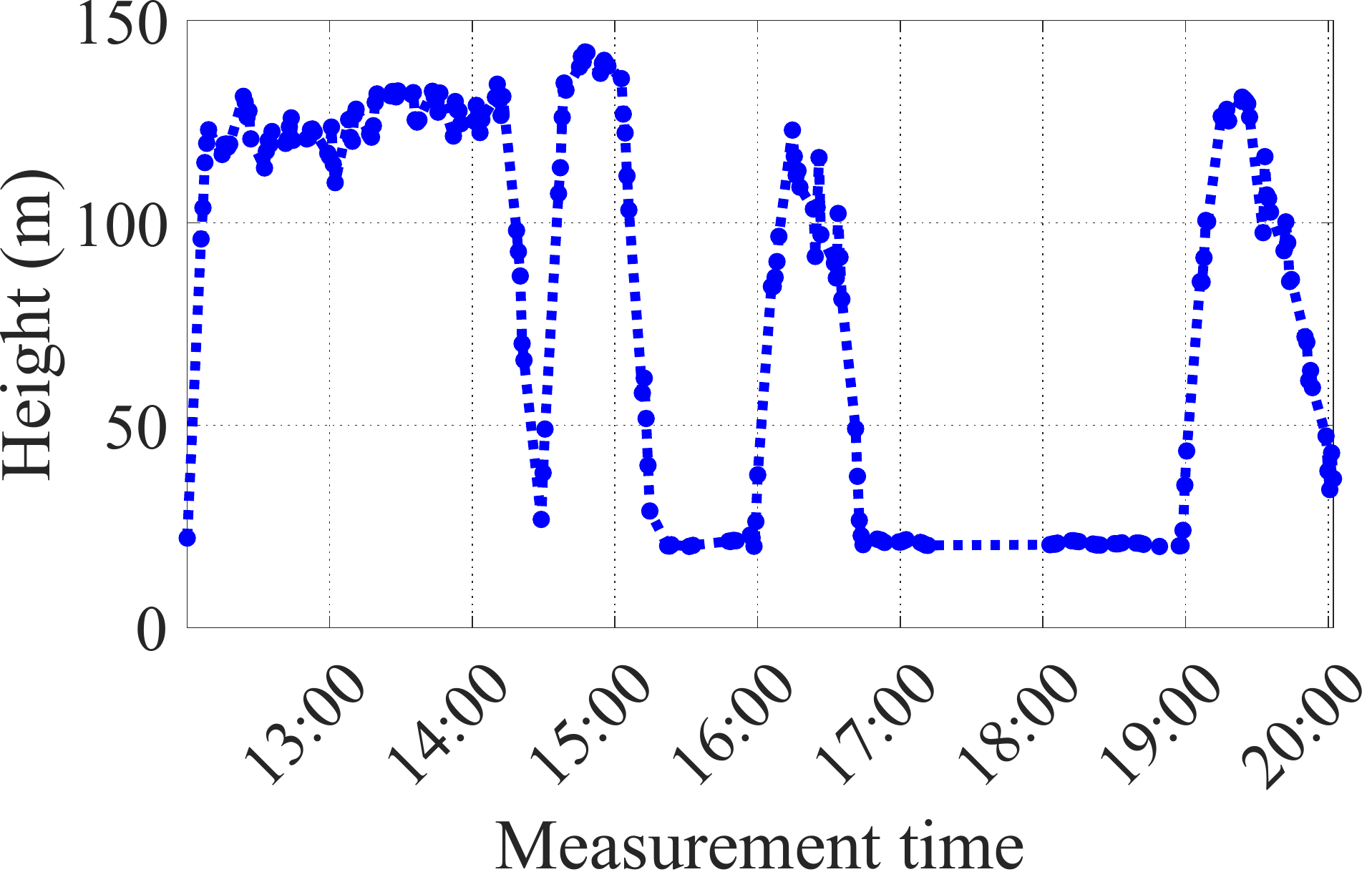}
         \includegraphics[width=0.42\textwidth]{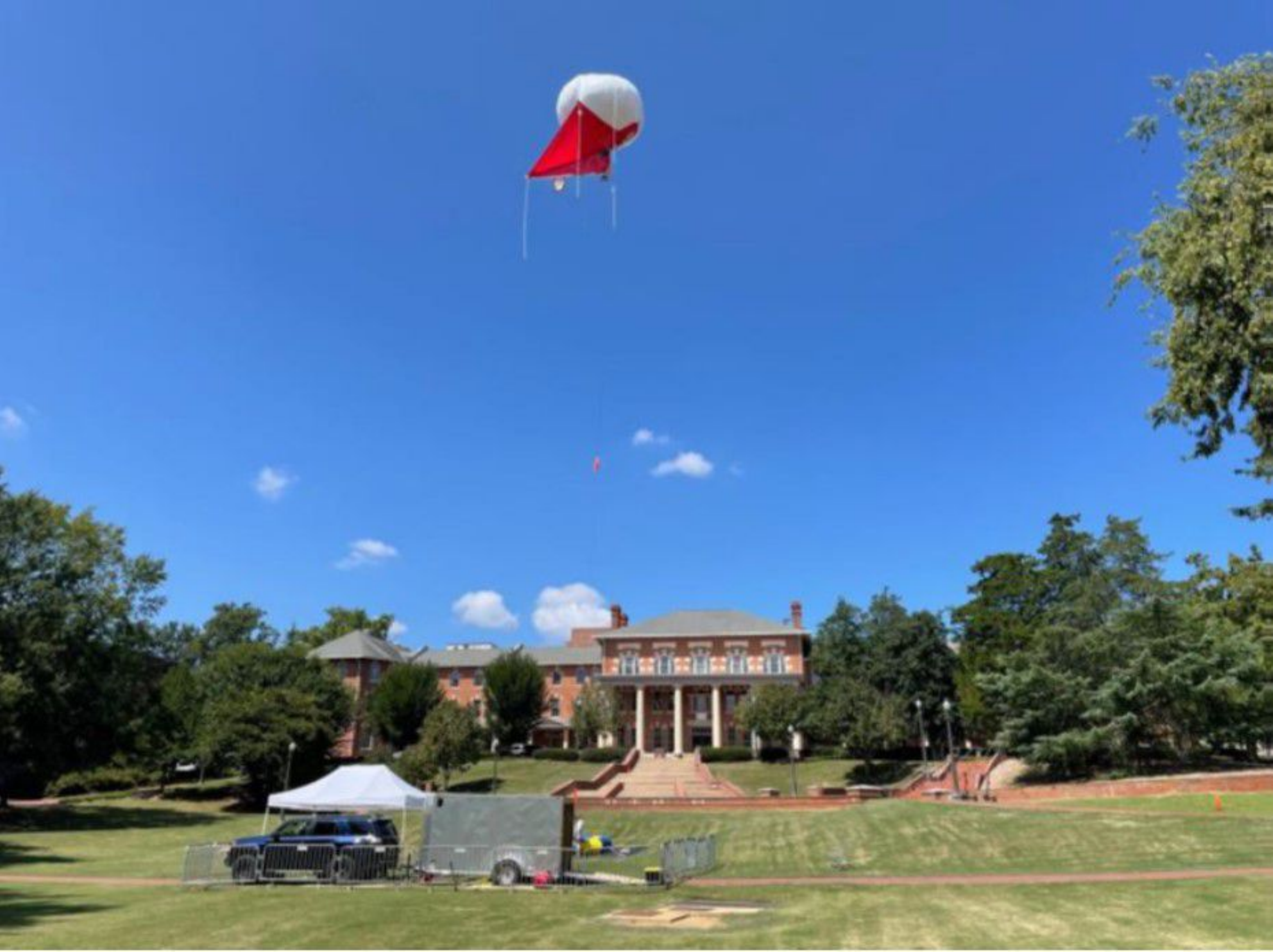}
         \caption{{Experiment scenario in NC State Main Campus (urban).}}
         \label{fig:pack}
     \end{subfigure}
     \begin{subfigure}[tb]{1\linewidth}
         \centering
         \captionsetup{justification=centering}
         \includegraphics[width=0.5\textwidth]{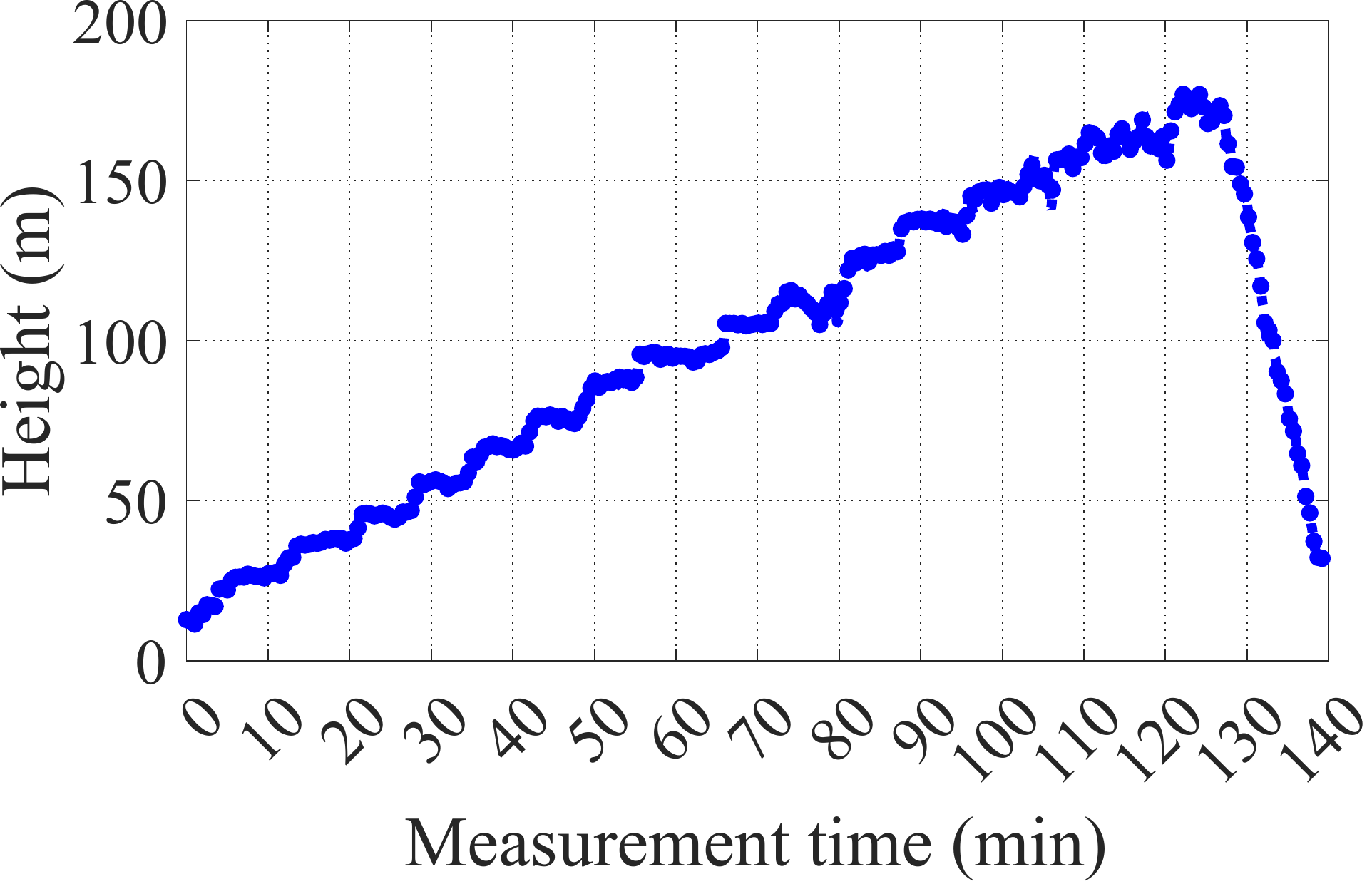}
         \includegraphics[width=0.42\textwidth]{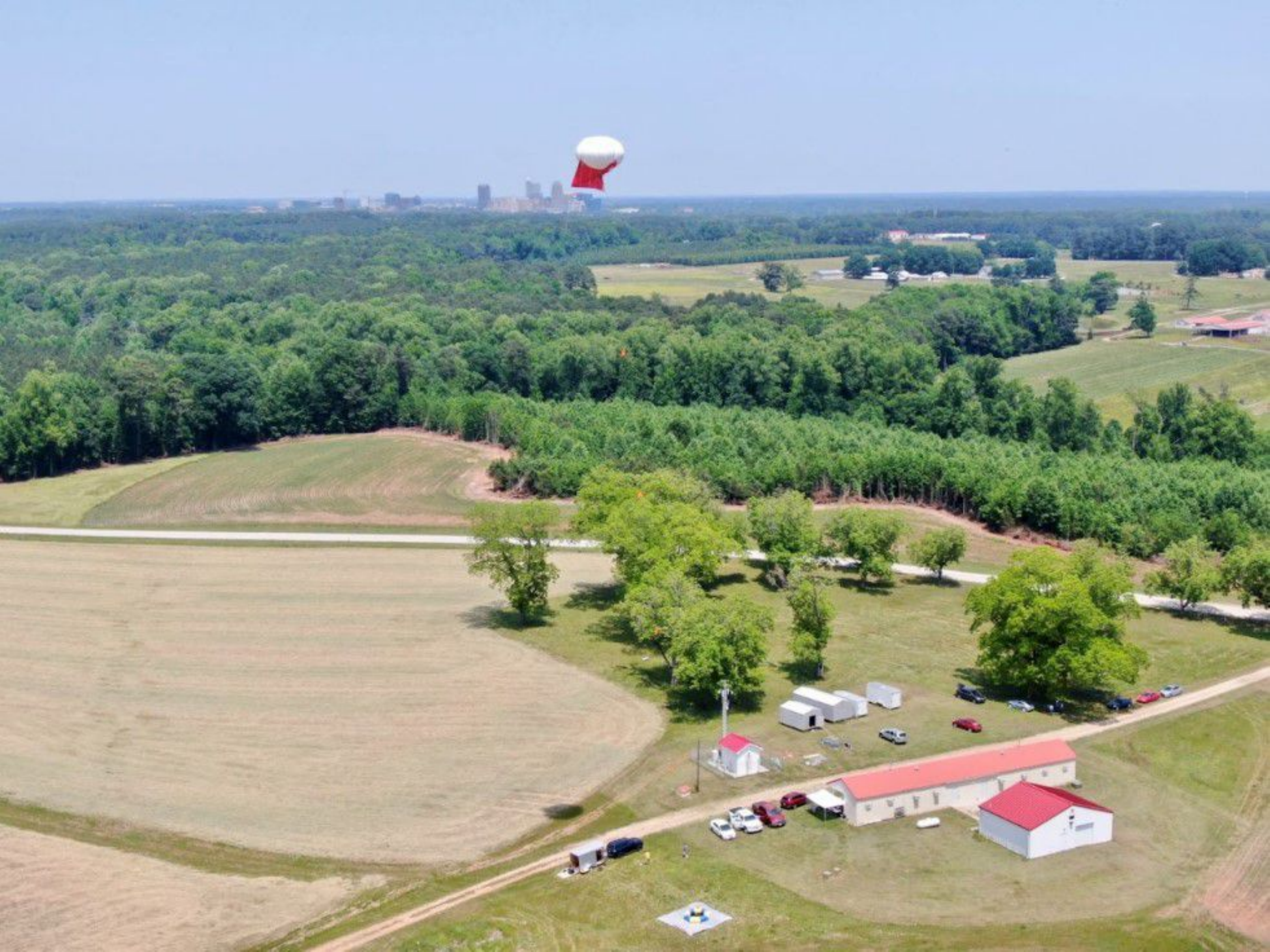}
         \caption{{Experiment scenario in NC State Lake Wheeler Field  (rural).}}
         \label{fig:wheeler}
     \end{subfigure}
        \caption{Helikite altitude and experiment scenario for: \textbf{(a)} urban environment, and \textbf{(b)} rural environment.}
        \label{fig:alt_time}
\end{figure}

The experiment for the urban environment was conducted by a Helikite flying up to 140~m on August 27, 2022. For the rural environment, the Helikite flew up to 180~m altitude on May 5, 2022.
An NI USRP B205mini SDR was mounted on the Helikite which enables executing a Python script to collect samples at the desired center frequency with the desired sampling rate.
The  datasets are SigMF compliant and include information on spectrum usage in frequency bands ranging from 89~MHz up to 6~GHz for different altitudes~\cite{Dataset1,Dataset2}. 
The data consist of time, altitude, power and Helikite location. A detailed description of the measurement setups can be found in ~\cite{maeng2023COMSNET}.
Fig.~\ref{fig:alt_time} illustrates the height of the Helikite during the operation time. 

\section{Urban Spectrum Occupancy Results}\label{results}   
In this section, we present the spectrum occupancy results for several LTE, 5G and CBRS bands. 
Table~\ref{table1} summarizes the spectrum allocations for some major cellular providers based on the technology exploited in the United States  \cite{maeng2022national}.
 In this work, we investigate the aggregate in-band power for UL and downlink (DL) spectrum of various bands.

\begin{table}[t]
\caption{Summary of LTE and 5G bands in United States.}
\label{table1}
\begin{center}
\scalebox{0.67}{
\begin{tabular}{|l|l|l|l|l|l|}
\hline
\textbf{Technology}           & \textbf{\begin{tabular}[c]{@{}l@{}}Band\\ No\end{tabular}} & \textbf{\begin{tabular}[c]{@{}l@{}}Duplex\\ Mode\end{tabular}} & \textbf{\begin{tabular}[c]{@{}l@{}}Uplink Band\\ (MHz)\end{tabular}} & \textbf{\begin{tabular}[c]{@{}l@{}}DL Band\\ (MHz)\end{tabular}} & \textbf{Operators}                                                 \\ \hline
\multirow{4}{*}{\textbf{LTE}} & 12                                                         & FDD                                                            & 698 - 716                                                            & 728 - 746                                                              & \begin{tabular}[c]{@{}l@{}}AT\&T, Verizon,\\ T-Mobile\end{tabular} \\ \cline{2-6} 
                              & 13                                                         & FDD                                                            & 777 - 787                                                            & 746 - 756                                                              & Verizon                                                            \\ \cline{2-6} 
                              & 14                                                         & FDD                                                            & 788 - 798                                                            & 758 - 768                                                              & AT\&T, FirstNet                                                    \\ \cline{2-6} 
                              & 41\footnote{It is worth mentioning that T-Mobile 5G n41   also uses the same spectrum.}                                                         & TDD                                                            & 2496 - 2690                                                          & 2496 - 2690                                                            & T-Mobile                                                           \\ \hline
\multirow{3}{*}{\textbf{5G}}  & n5                                                         & FDD                                                            & 824 - 849                                                            & 869 - 894                                                              & AT\&T, Verizon                                                     \\ \cline{2-6} 
                              & n71                                                        & FDD                                                            & 663 - 698                                                            & 617 - 652                                                              & T-Mobile                                                           \\ \cline{2-6} 
                              & n77                                                        & TDD                                                            & 3700 - 3980                                                          & 3700 - 3980                                                            & \begin{tabular}[c]{@{}l@{}}AT\&T, Verizon,\\ T-Mobile\end{tabular} \\ \hline
\textbf{CBRS}                 & n48                                                        & TDD                                                            & 3550 - 3700                                                          & 3550 - 3700                                                            & North America                                                      \\ \hline
\end{tabular}
}
\end{center}
\end{table}

\subsection{LTE Bands - Uplink}

\begin{figure}[!t]
\centering
\begin{subfigure}{0.49\columnwidth} 
\centering
\includegraphics[width=\textwidth]{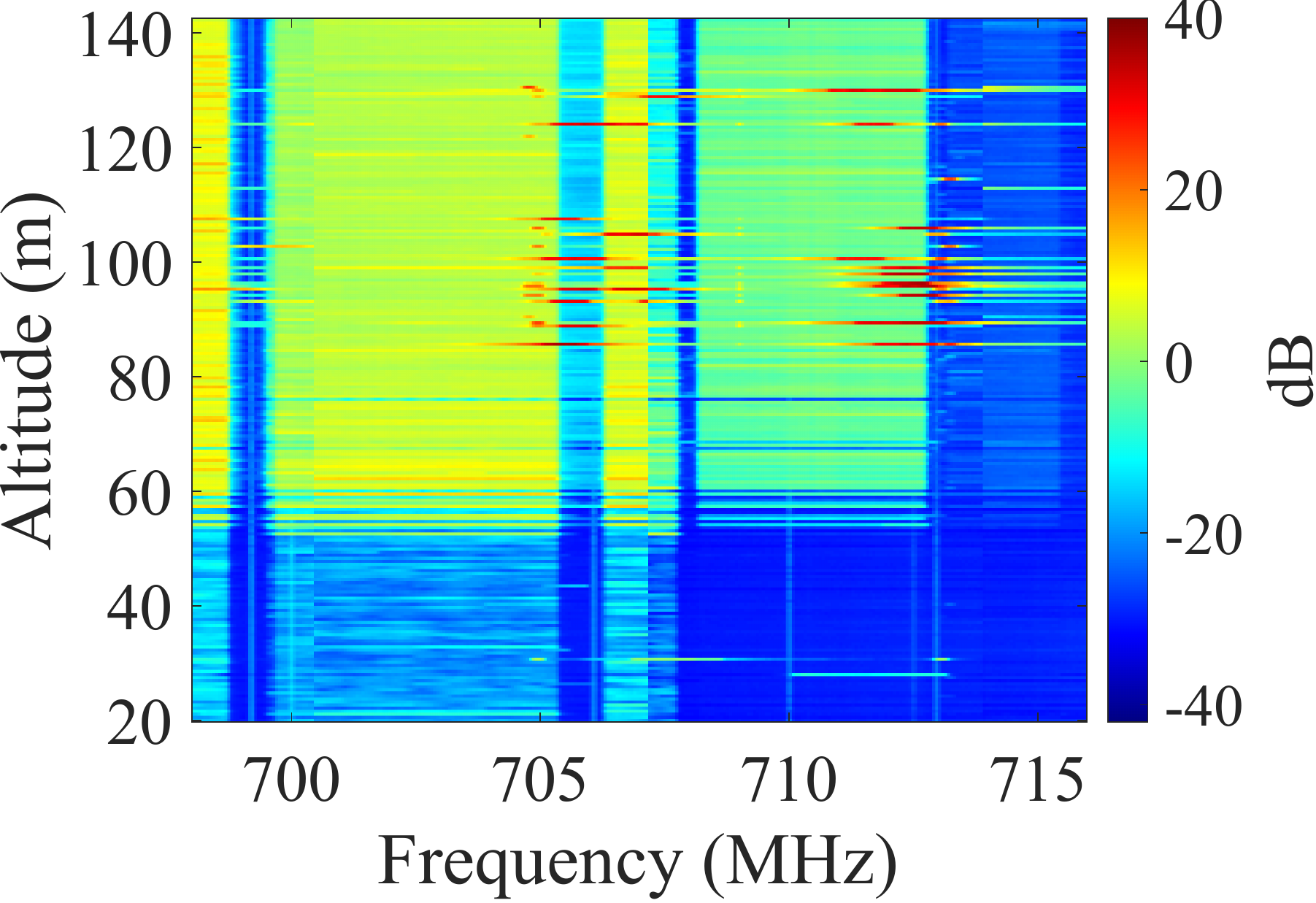} 
\caption{LTE band 12 (UL).}\label{fig:up12} 
\end{subfigure}
\begin{subfigure}{0.49\columnwidth} 
\centering
\includegraphics[width=\textwidth]{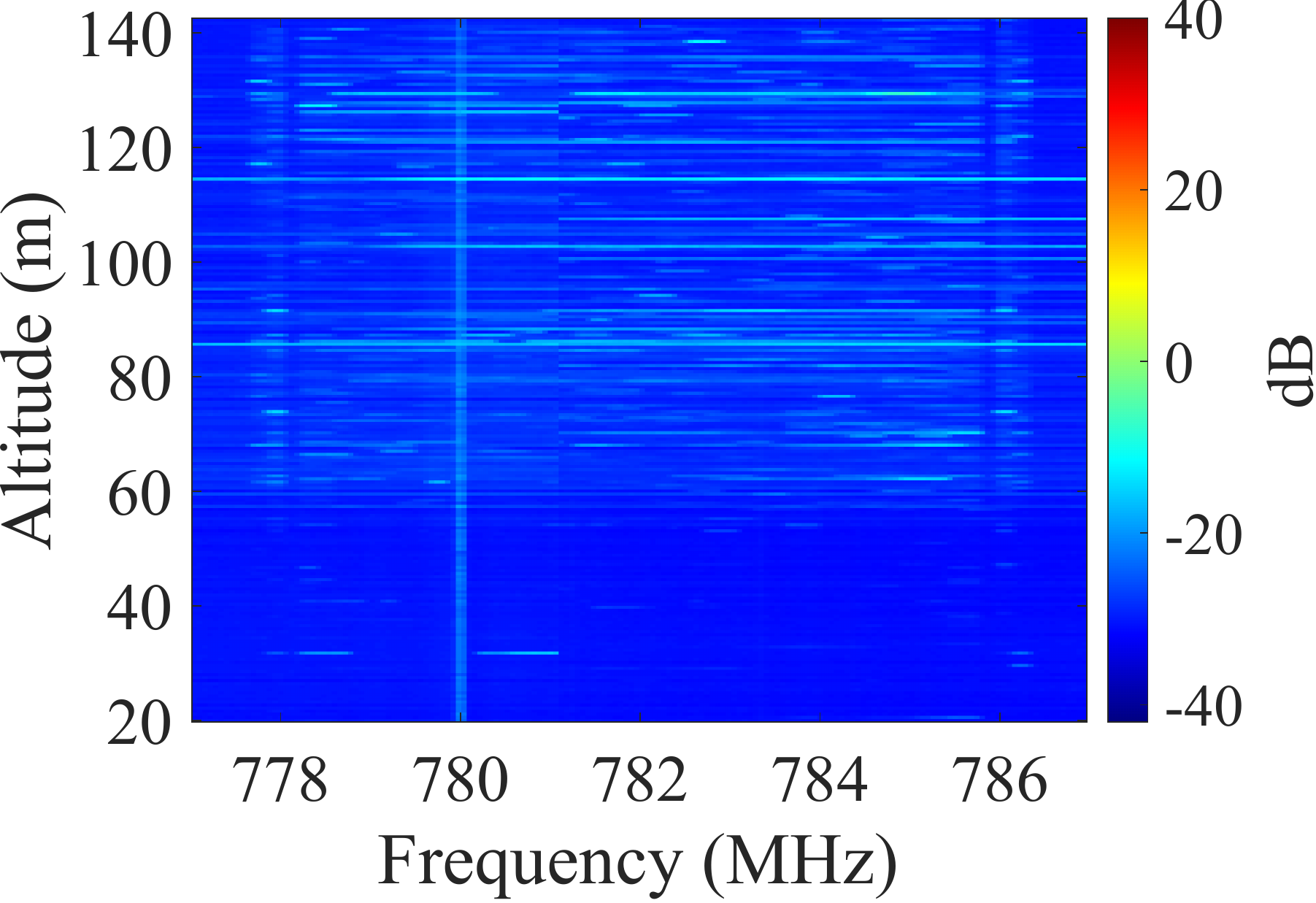}
\caption{LTE band 13 (UL).}
\label{fig:up13}
\end{subfigure}
\begin{subfigure}{0.49\columnwidth} 
\centering
\includegraphics[width=\textwidth]{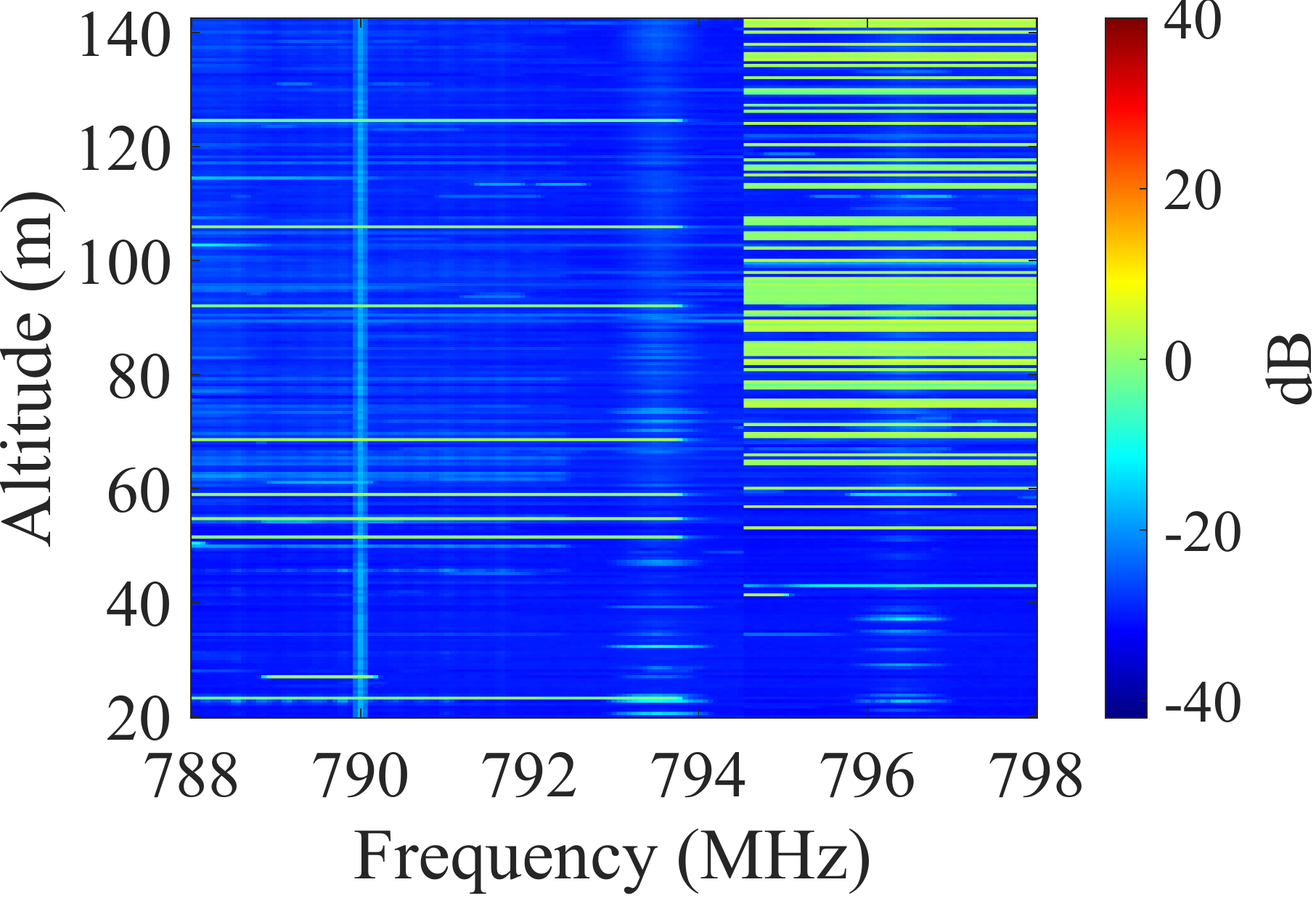} 
\caption{LTE band 14 (UL).}\label{fig:up14}
\end{subfigure} 
\begin{subfigure}{0.49\columnwidth} 
\centering
\includegraphics[width=\textwidth]{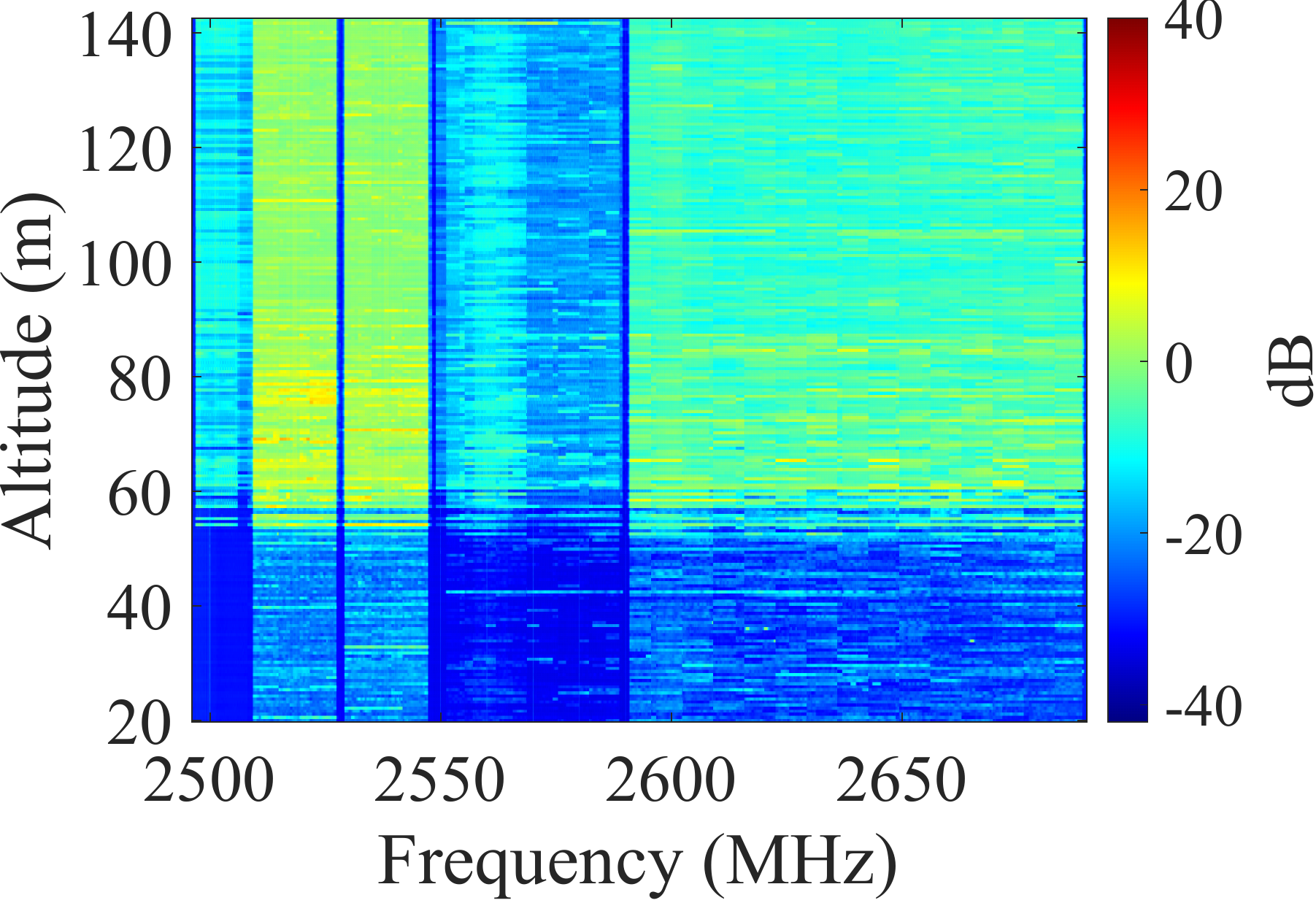} 
\caption{LTE band 41 (TDD UL/DL).}\label{fig:up41}
\end{subfigure} 
\caption{Measured LTE UL power for urban environment.}
\label{fig:LTE_power_freq_alt_up}
\end{figure}

\begin{figure}[!t]
\centering
\begin{subfigure}{0.49\columnwidth} 
\centering
\includegraphics[width=\textwidth]{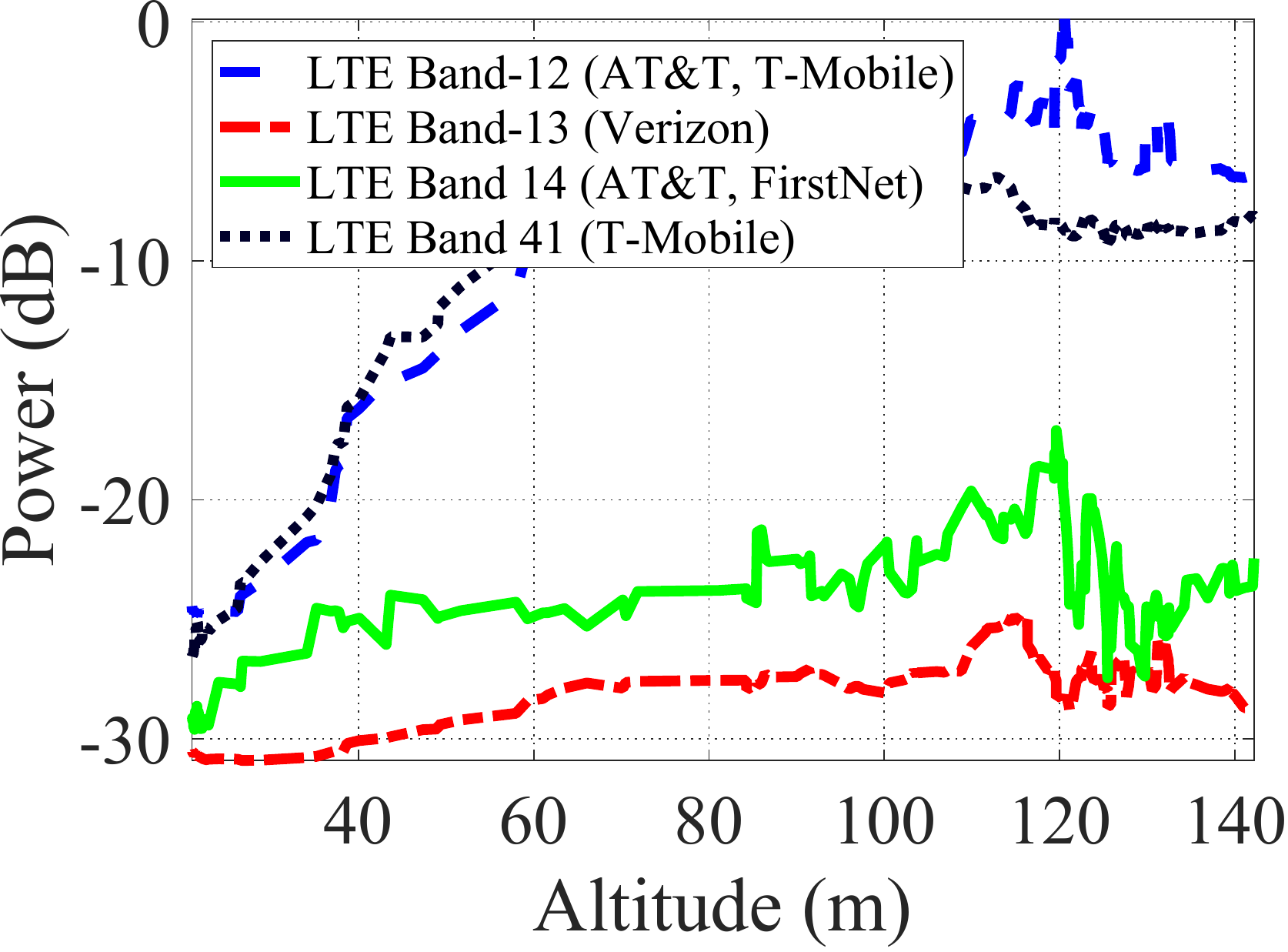} 
\caption{Mean.}\label{fig:meanLTEup} 
\end{subfigure}
\begin{subfigure}{0.49\columnwidth} 
\centering
\includegraphics[width=\textwidth]{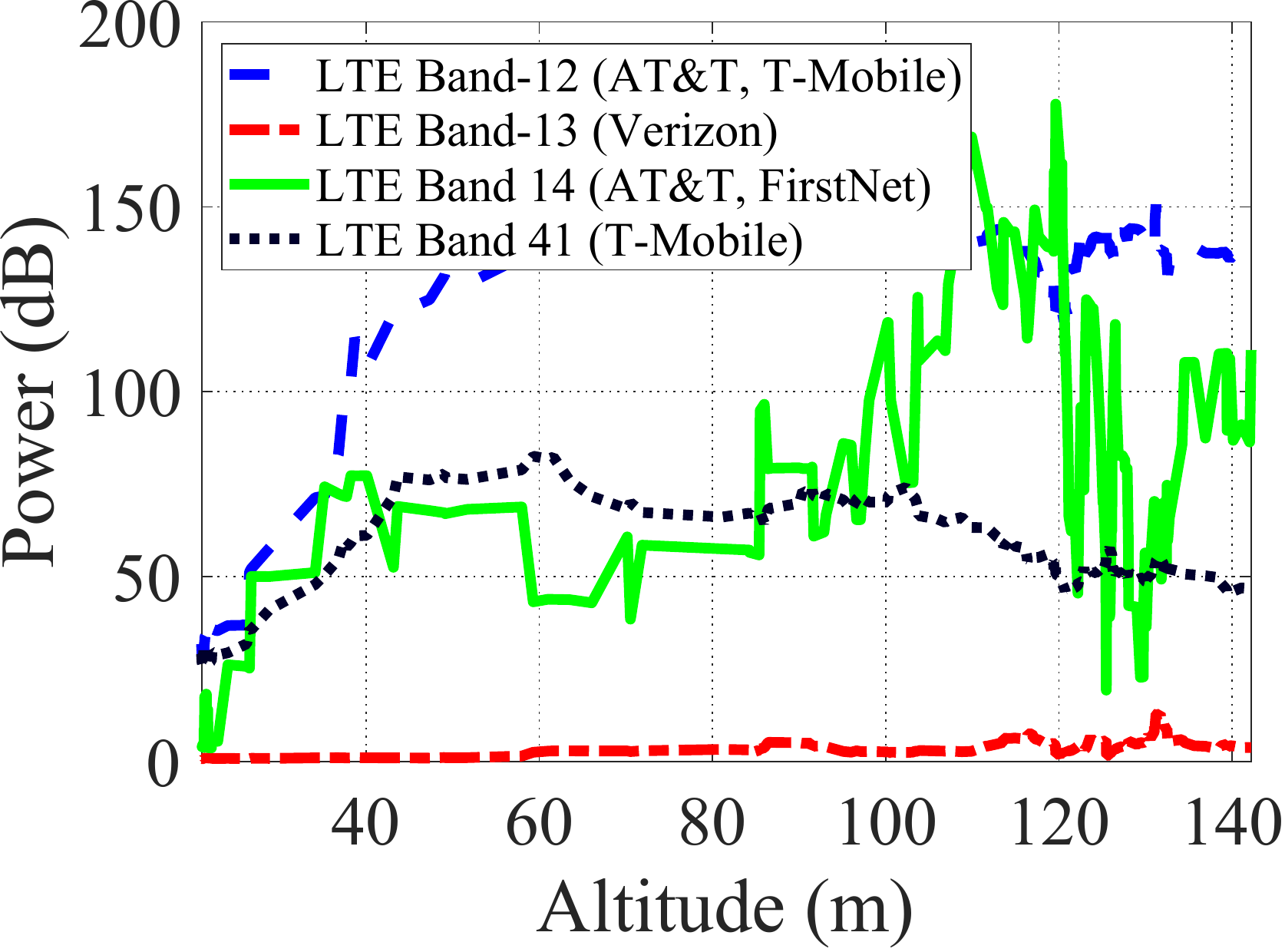}
\caption{Variance.}
\label{fig:varLTEup}
\end{subfigure}
        \caption{Spectrum occupancy versus altitude in LTE bands 12, 13, 14 and 41 (UL) for urban environment.}
        \label{fig:mean_var_LTEup}
\end{figure}

Fig.~\ref{fig:LTE_power_freq_alt_up} presents the measured power for LTE bands 13, 14, 15 and 41 considering the UL frequency spectrum ranges. As it can be seen, the spectrum of LTE 12 and LTE 41 bands are more crowded compared with LTE 13 and LTE 14 bands. It is worth mentioning that, unlike other LTE bands under consideration,  LTE 41 works in  time-division duplexing (TDD) mode and includes both UL and DL transmissions. 
The mean and variance of the measured power for various LTE bands are presented in Fig.~\ref{fig:mean_var_LTEup}. As it can be observed from Fig.~\ref{fig:meanLTEup}, generally the mean value of the measured power increases as the altitude increases. 
The mean power value for LTE bands 12 and 41 are almost identical and much higher than the other two bands under consideration. Note that band 41 has significantly larger bandwidth than band 12 and it includes both UL and DL transmission.
From Fig.~\ref{fig:varLTEup}, it can be observed that the fluctuation of variance for LTE band 13 is much lower than the other ones. Although the mean value of LTE 12 and 41 show similar behaviour, the variance of LTE 41 is lower than LTE band 12.

\subsection{LTE Bands - Downlink}
\begin{figure}[!t]
\centering
\begin{subfigure}{0.49\columnwidth} 
\centering
\includegraphics[width=\textwidth]{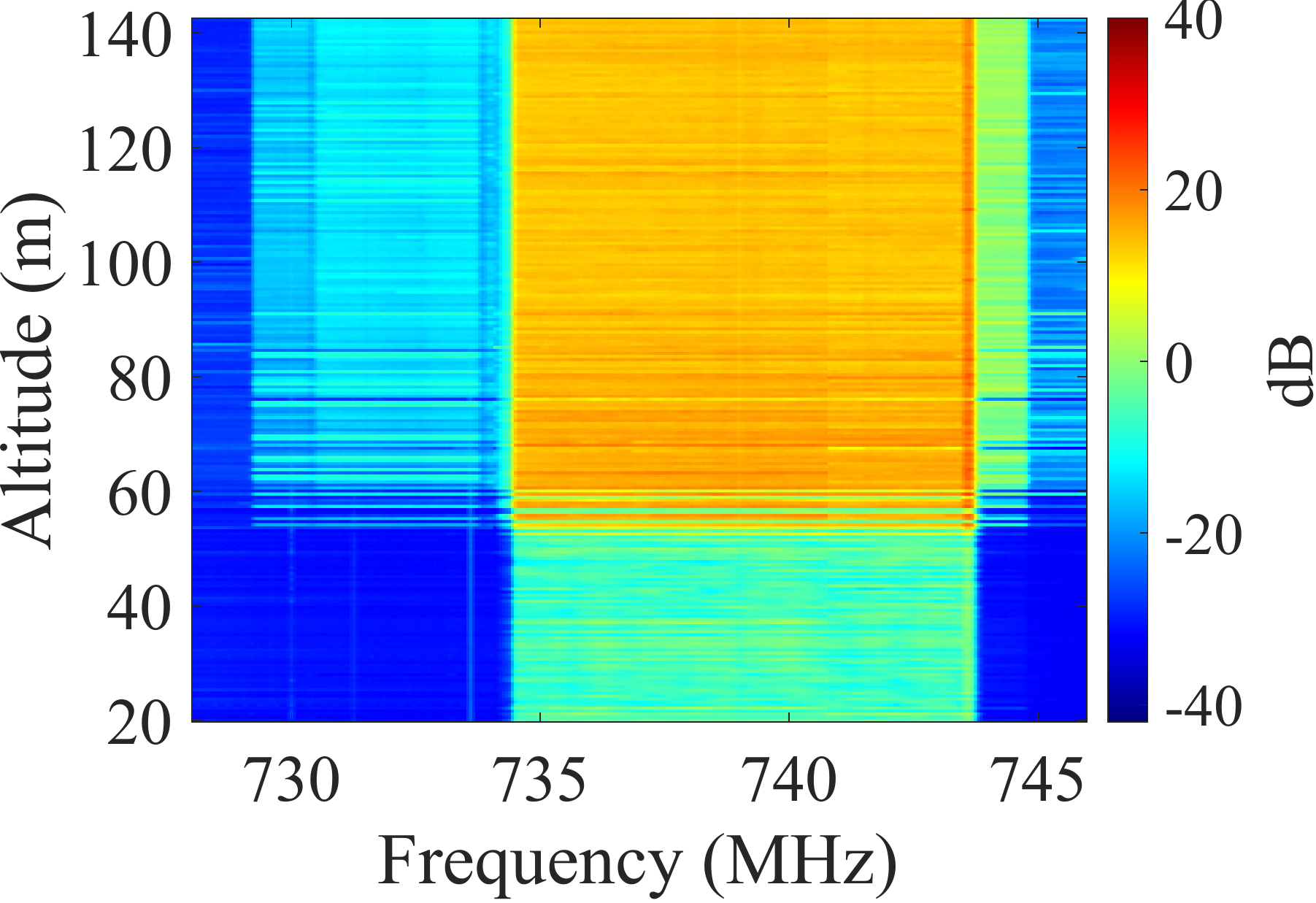} 
\caption{LTE band 12 (DL).}\label{fig:down12} 
\end{subfigure}
\begin{subfigure}{0.49\columnwidth} 
\centering
\includegraphics[width=\textwidth]{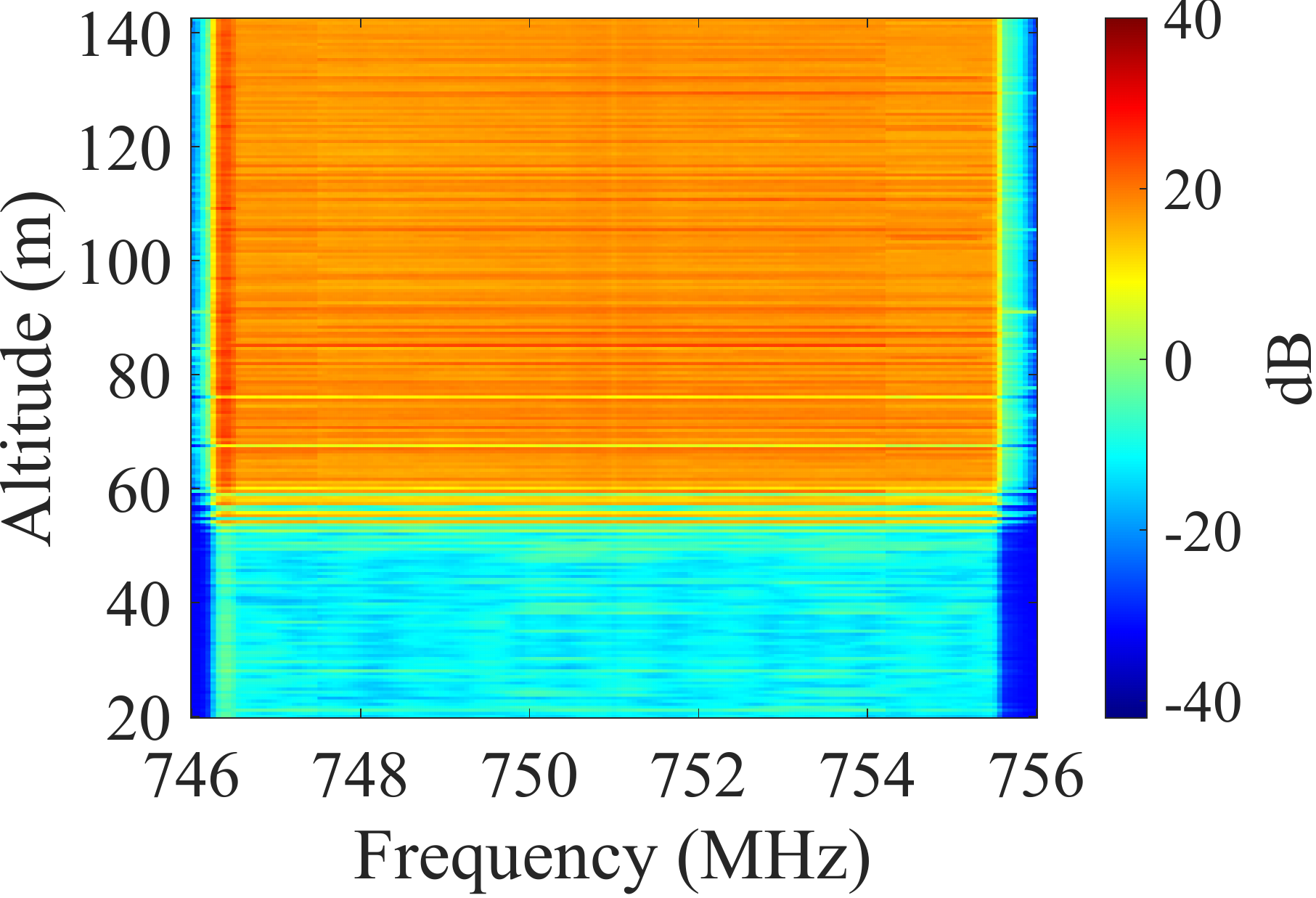}
\caption{LTE band 13 (DL).}
\label{fig:down13}
\end{subfigure}
\begin{subfigure}{0.49\columnwidth} 
\centering
\includegraphics[width=\textwidth]{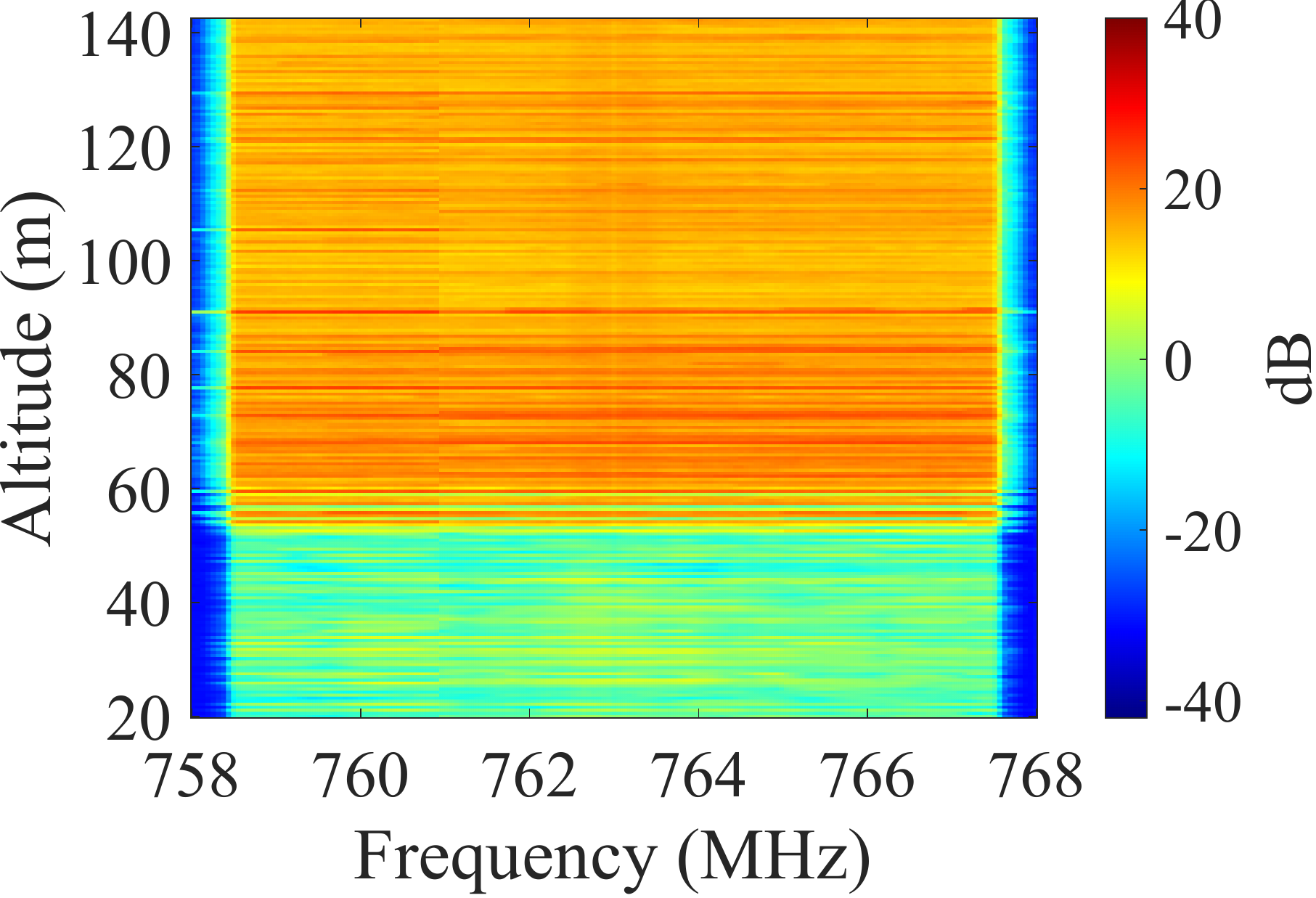} 
\caption{LTE band 14 (DL).}\label{fig:down14}
\end{subfigure} 
\begin{subfigure}{0.49\columnwidth} 
\centering
\includegraphics[width=\textwidth]{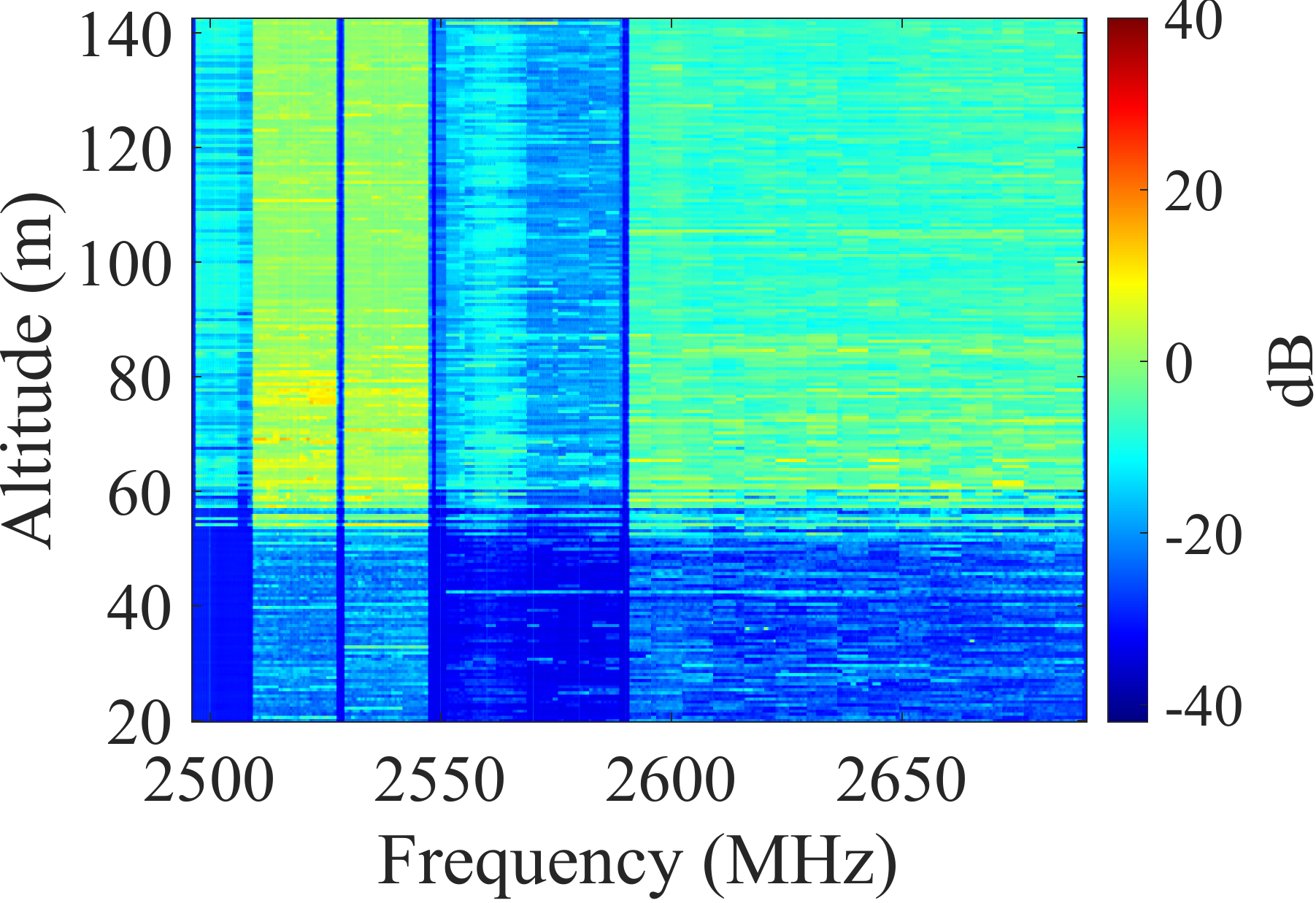} 
\caption{LTE band 41 (TDD UL/DL).}\label{fig:down41}
\end{subfigure} 
\caption{Measured LTE DL power for urban environment.}
\label{fig:LTE_power_freq_alt_down}
\end{figure}

Considering the DL frequency range for different LTE bands, Fig.~\ref{fig:LTE_power_freq_alt_down} illustrates the measured power for the bands under consideration. It can be readily checked that the spectrum of DL frequency ranges are more crowded compared with the UL ones.
Although the occupied spectrum for LTE 13 and 14 expand the whole range, the main frequency usage of LTE 12 is between 735 - 745~MHz.  

\begin{figure}[!t]
\centering
\begin{subfigure}{0.49\columnwidth} 
\centering
\includegraphics[width=\textwidth]{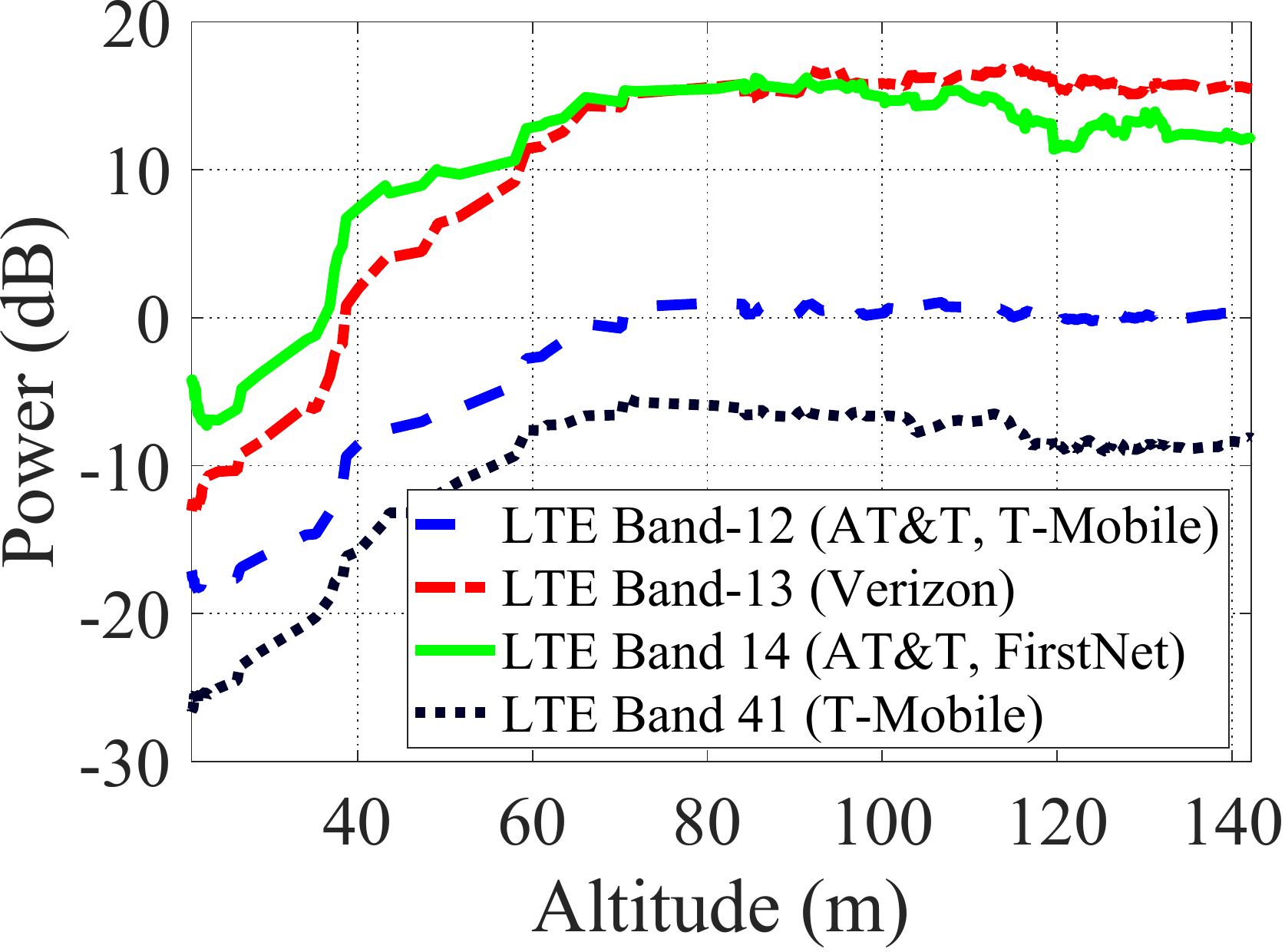} 
\caption{Mean.}\label{fig:meanLTEdown} 
\end{subfigure}
\begin{subfigure}{0.49\columnwidth} 
\centering
\includegraphics[width=\textwidth]{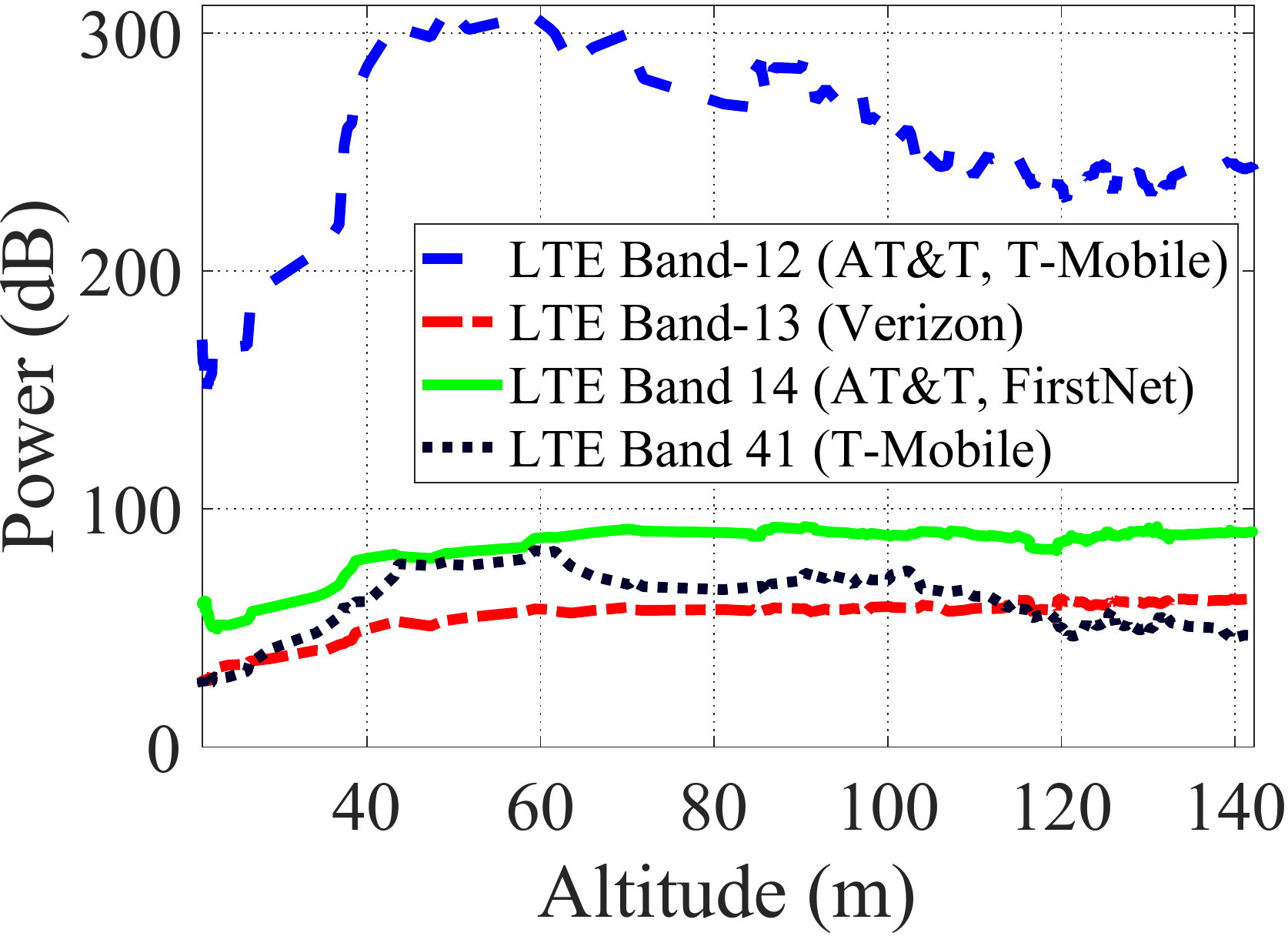}
\caption{Variance.}
\label{fig:varLTEdown}
\end{subfigure}
        \caption{Spectrum occupancy versus altitude in LTE bands 12, 13, 14 and 41 (DL) for urban environment.}
        \label{fig:mean_var_LTE_down}
\end{figure}

Fig.~\ref{fig:mean_var_LTE_down} shows the mean and variance of the measured power versus altitude. 
As it can be observed from Fig.~\ref{fig:meanLTEdown}, the mean value of the measured power increases as the altitude increases up to almost 80~m.
This is due to the fact that at high altitudes the probability of receiving signal from neighbor cells increases as the obstacles decrease, which results in the availability of the line of sight (LoS). For higher altitudes (i.e., higher than 80 m), the mean values for LTE bands under consideration remain almost constant. 
As it is shown in Fig.~\ref{fig:varLTEdown}, the variance of the measured power for LTE bands 13, 14 and 41 show relatively smaller variation over different altitudes compared to LTE band 12. The main reason for this behavior can be found by observing the measured power for LTE band 12 shown in Fig.~\ref{fig:down12}. It seems that some portion of the LTE band 12 is not fully utilized.

\subsection{5G Bands - Uplink}
\begin{figure}[!t]
\centering
\begin{subfigure}{0.49\columnwidth} 
\centering
\includegraphics[width=\textwidth]{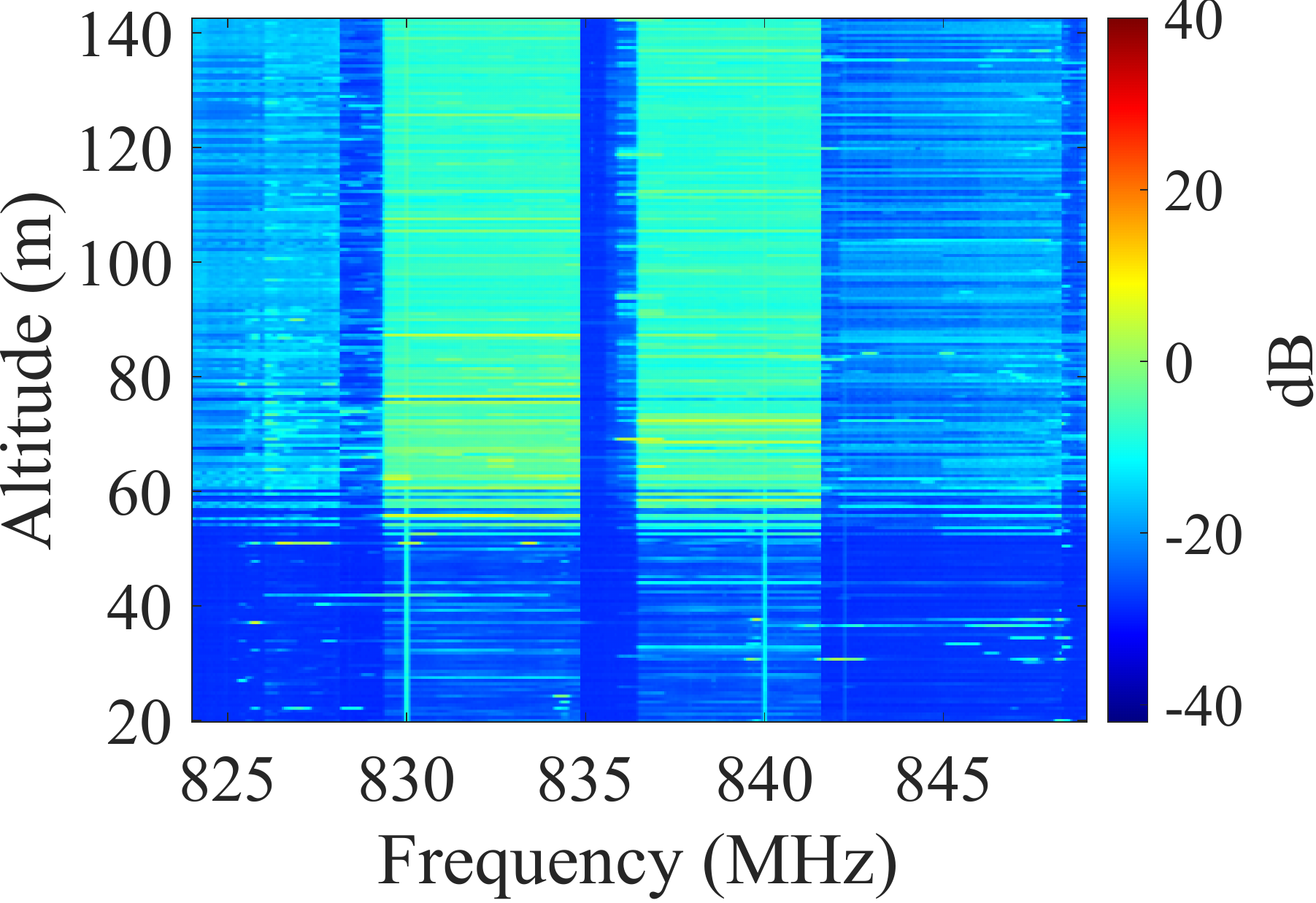} 
\caption{5G band n5 (UL).}\label{fig:upn5} 
\end{subfigure}
\begin{subfigure}{0.49\columnwidth} 
\centering
\includegraphics[width=\textwidth]{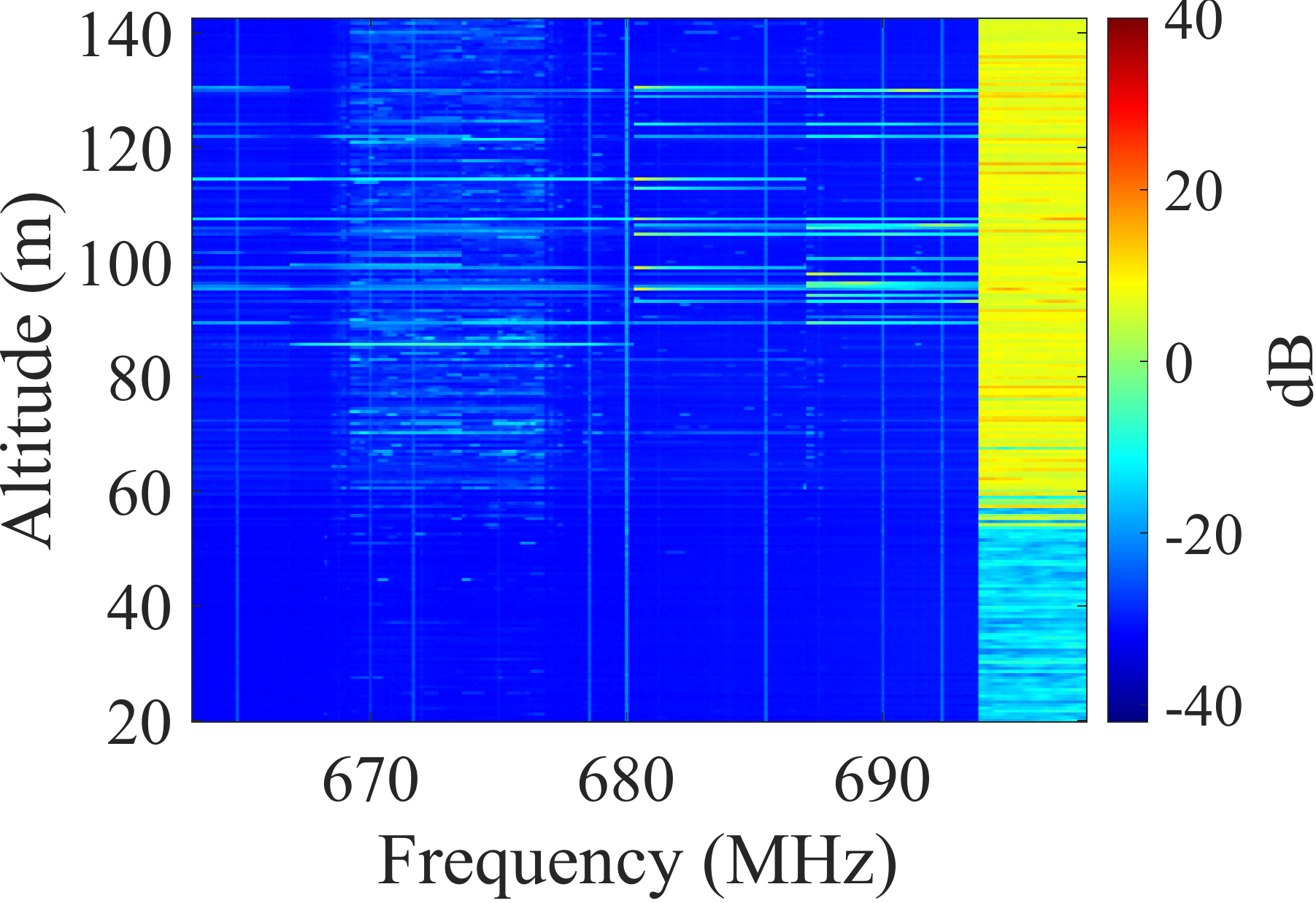}
\caption{5G band n71 (UL)}
\label{fig:upn71}
\end{subfigure}
\begin{subfigure}{0.49\columnwidth} 
\centering
\includegraphics[width=\textwidth]{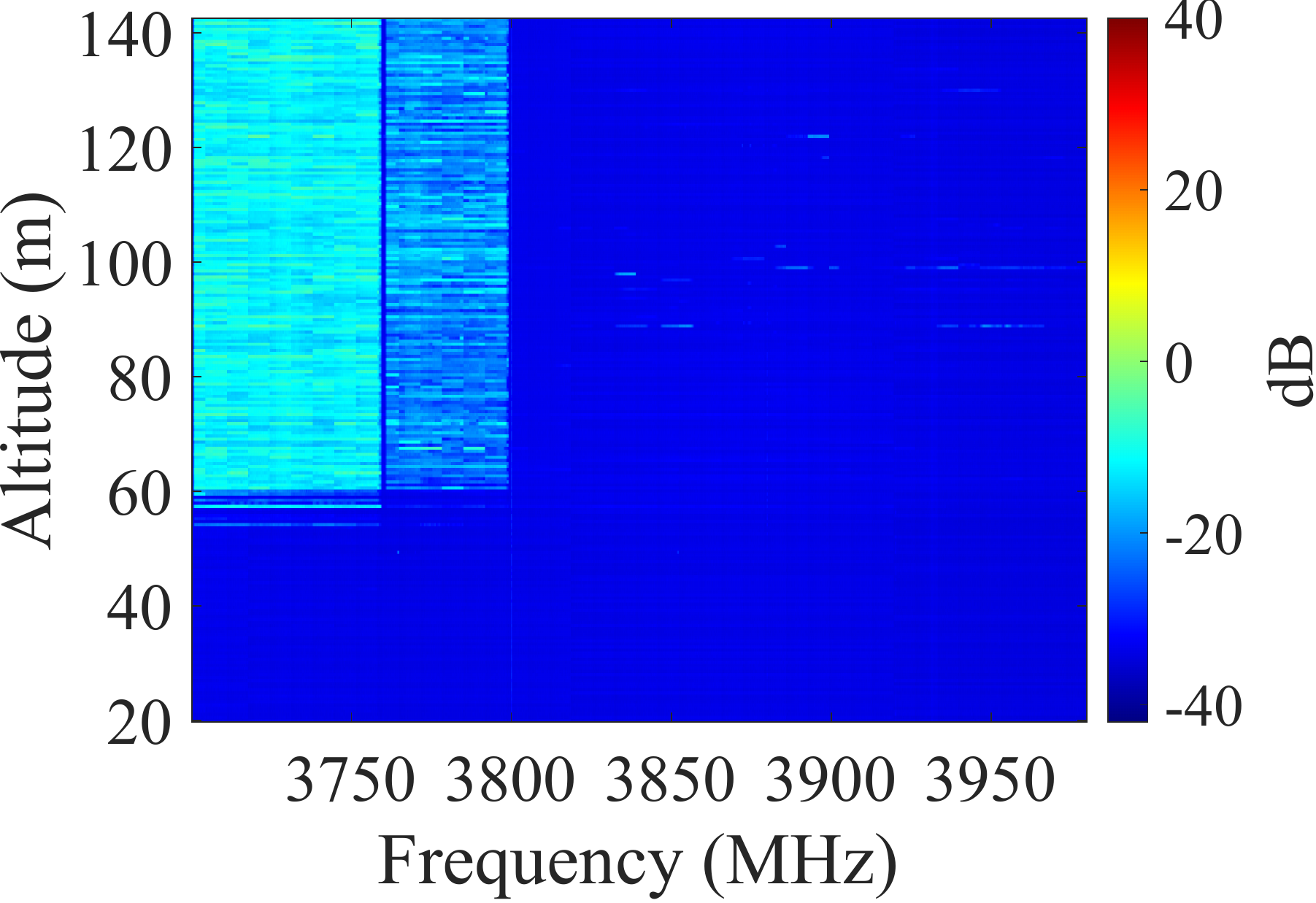} 
\caption{5G band n77 (TDD UL/DL).}\label{fig:upn77}
\end{subfigure} 
\caption{Measured 5G UL power for urban environment.}
\label{fig:5G_power_freq_alt_up}
\end{figure}

\begin{figure}[!t]
\centering
\begin{subfigure}{0.49\columnwidth} 
\centering
\includegraphics[width=\textwidth]{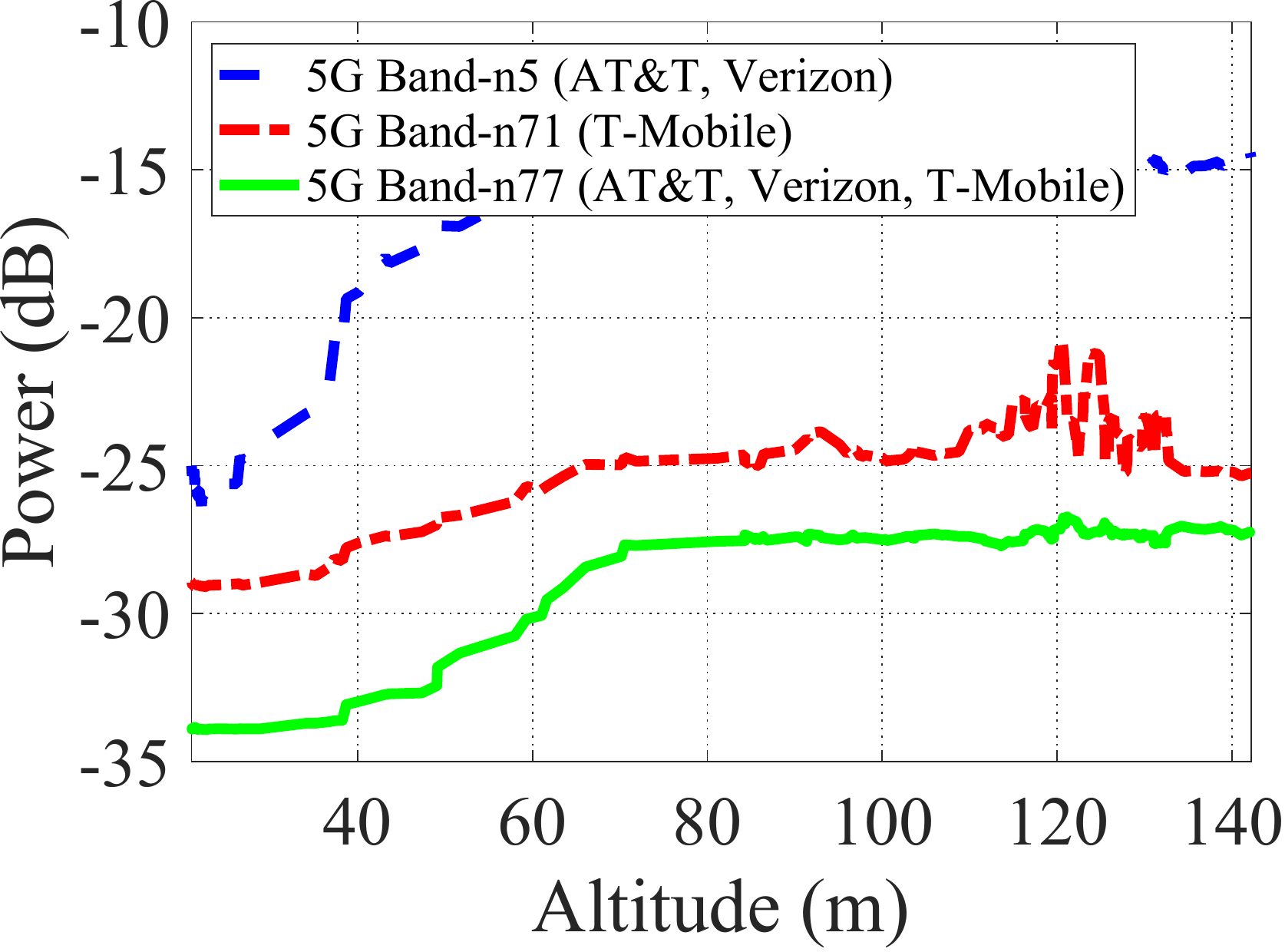} 
\caption{Mean.}\label{fig:mean5Gu} 
\end{subfigure}
\begin{subfigure}{0.49\columnwidth} 
\centering
\includegraphics[width=\textwidth]{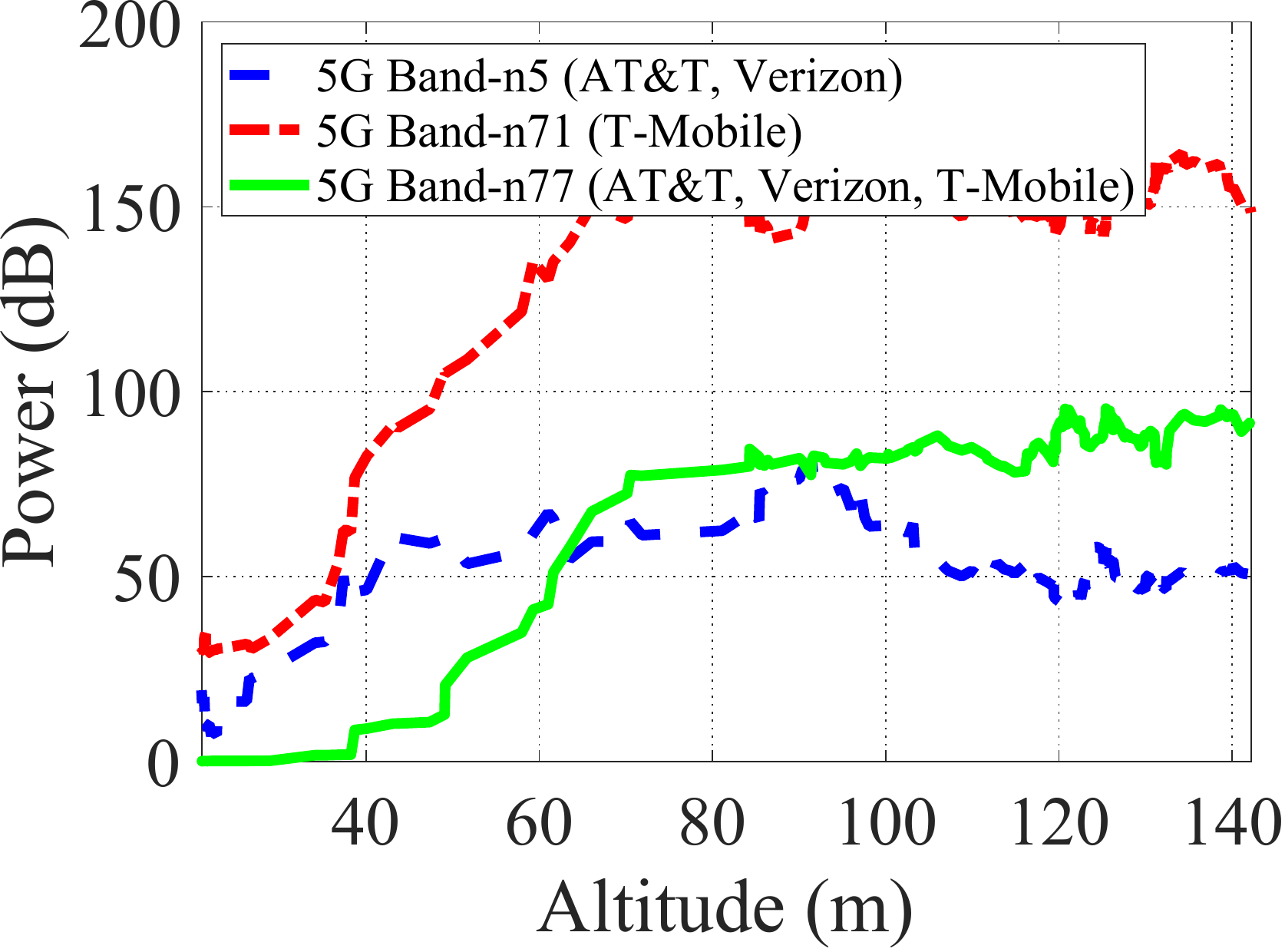}
\caption{Variance.}
\label{fig:var5Gu}
\end{subfigure}
        \caption{Spectrum occupancy versus altitude in 5G n5, n71 and n77 bands  (UL) for urban environment.}
        \label{fig:mean_var_5Gup}
\end{figure}

Fig.~\ref{fig:5G_power_freq_alt_up} presents the measured power for 5G bands n5, n71 and n77 considering the UL frequency spectrum ranges. This result reveals that the spectrum of n77 is mainly occupied between 3700-3800~MHz.
One should also note that 5G band n5 and n71 utilize the frequency-division duplexing (FDD), while 5G band n77 exploit TDD mode.
The performance of mean and variance of the measured power for 5G bands (uplink) are presented in Fig.~\ref{fig:mean_var_5Gup}. 
As it can be observed from Fig.~\ref{fig:mean5Gu}, the mean value of the measured power increases as the altitude increases up to almost 80~m due to the same argument mentioned earlier. The mean value of 5G band n5 shows higher value compared with n71 and n77. 
As it is shown in Fig.~\ref{fig:var5Gu}, the variance of the measured power for 5G bands n5 and n77 intersect with each other around the altitude of 60~m. The variance of n77 band keeps increasing as the altitude increases.

\begin{figure}[!t]
\centering
\begin{subfigure}{0.49\columnwidth} 
\centering
\includegraphics[width=\textwidth]{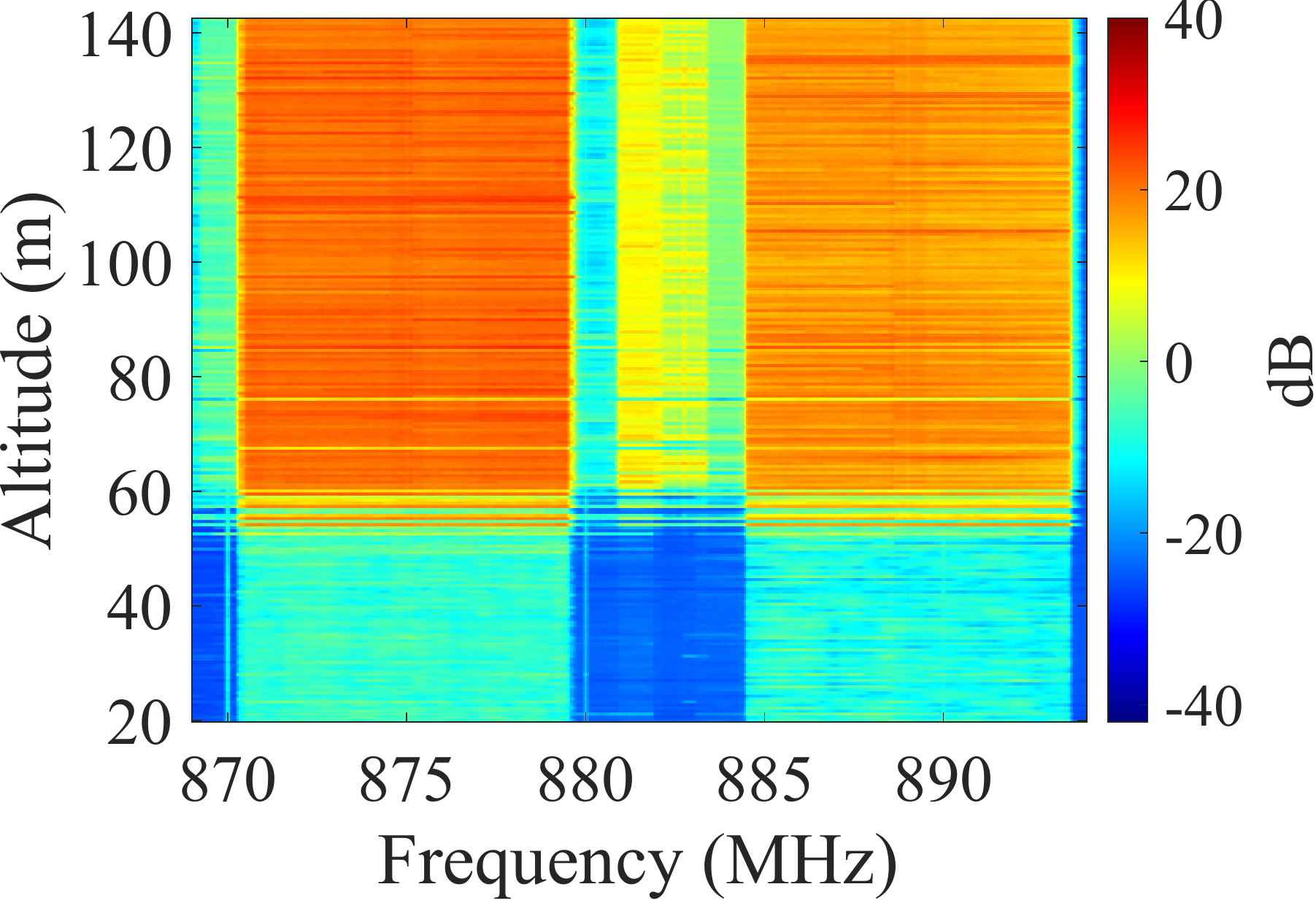} 
\caption{5G band n5 (DL).}\label{fig:n5_downlink} 
\end{subfigure}
\begin{subfigure}{0.49\columnwidth} 
\centering
\includegraphics[width=\textwidth]{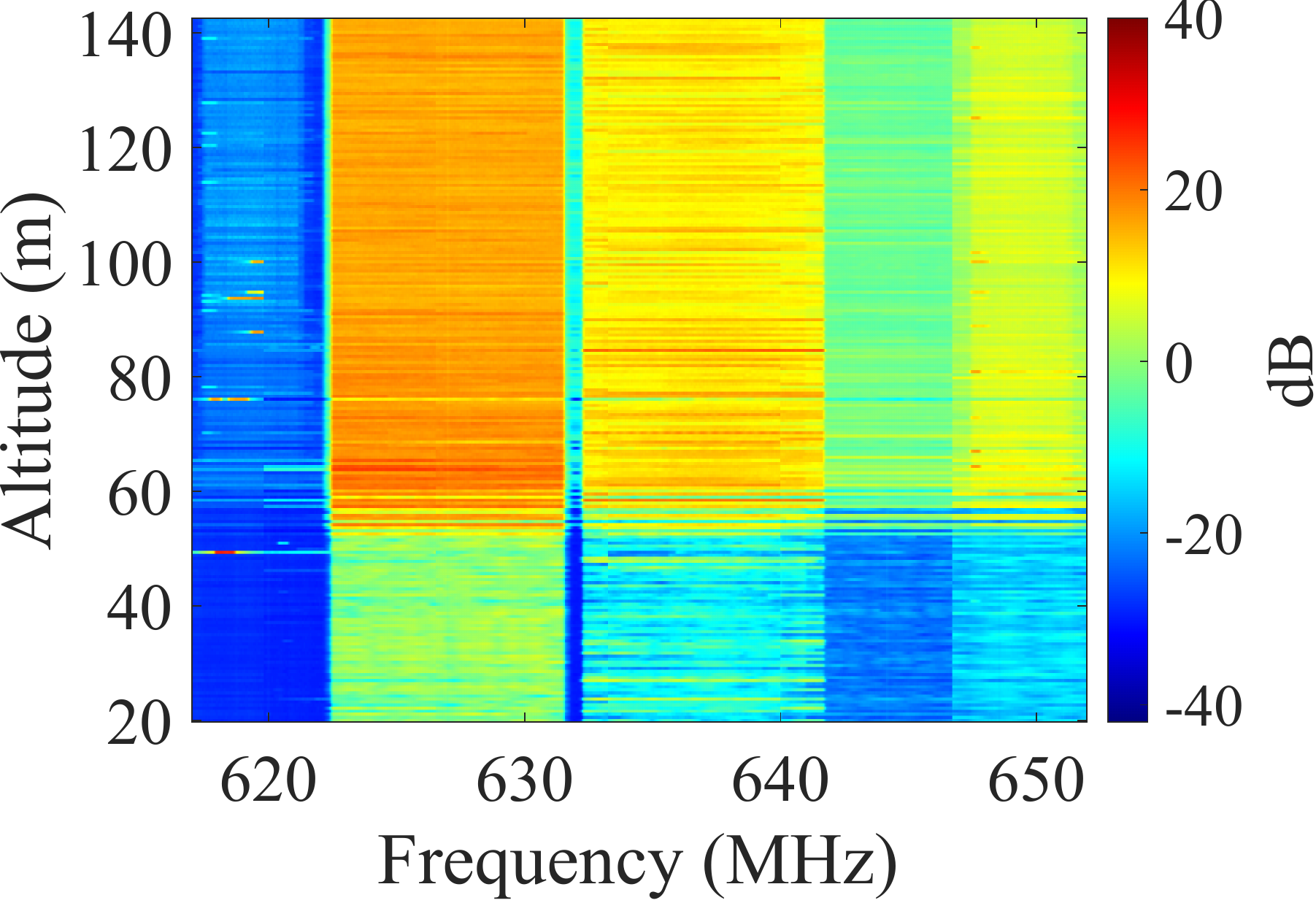}
\caption{5G band n71 (DL).}
\label{fig:n71_downlink}
\end{subfigure}
\caption{Measured 5G DL power for urban environment.}
\label{fig:n5_n71_downlink}
\end{figure}

\begin{figure}[!t]
\centering
\begin{subfigure}{0.49\columnwidth} 
\centering
\includegraphics[width=\textwidth]{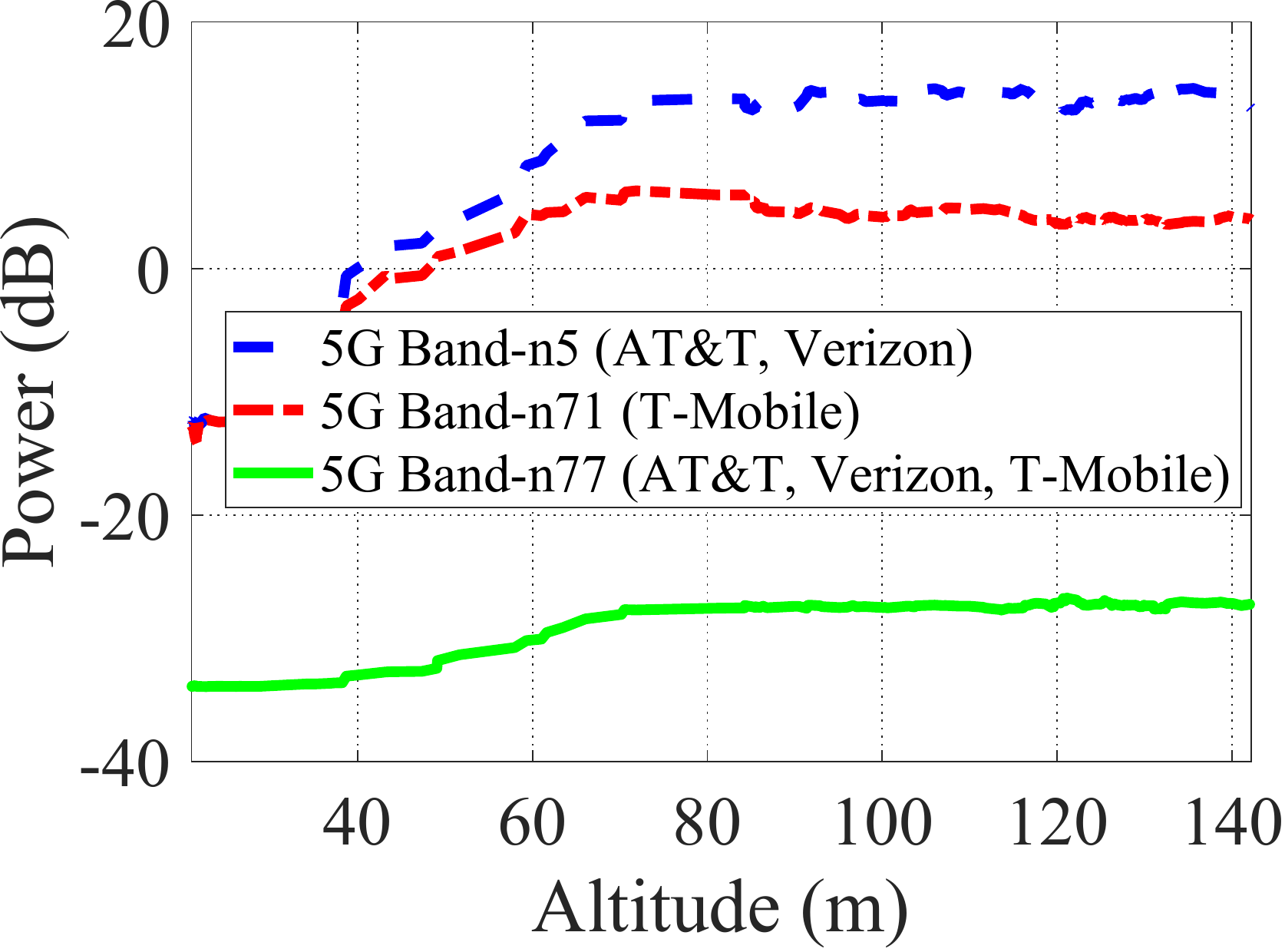} 
\caption{Mean.}\label{fig:mean5Gd} 
\end{subfigure}
\begin{subfigure}{0.49\columnwidth} 
\centering
\includegraphics[width=\textwidth]{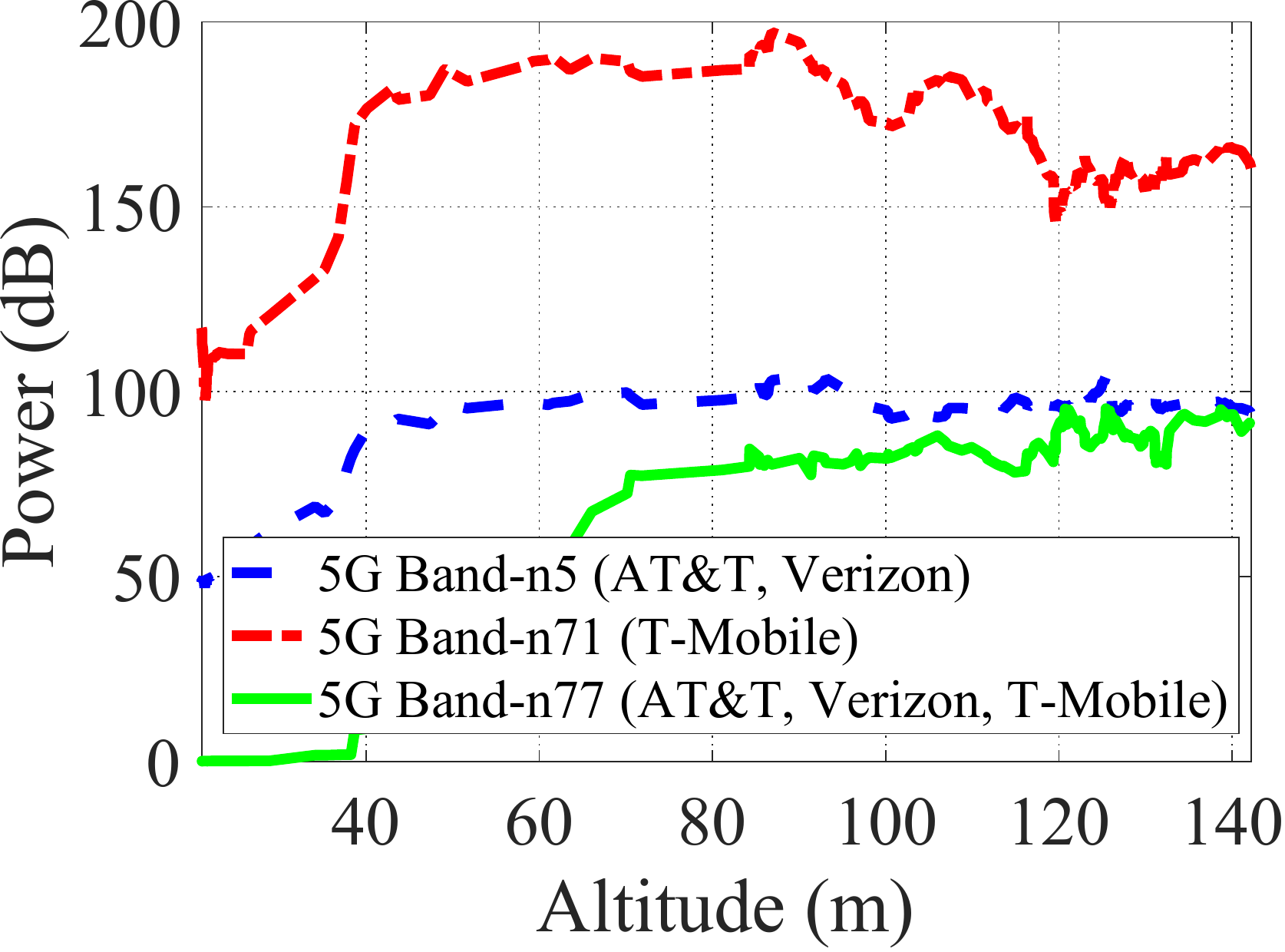}
\caption{Variance.}
\label{fig:var5Gd}
\end{subfigure}
        \caption{Spectrum occupancy versus altitude in 5G bands n5 and n77 (DL) for urban environment.}
        \label{fig:mean_var_5Gdown}
\end{figure}

\subsection{5G Bands - Downlink}
Fig.~\ref{fig:n5_n71_downlink} illustrates the measured power for 5G n5 and n71 bands by considering the DL frequency range. It can be seen that the measured power for 870 - 880~MHz and 885-894~MHz are higher than the rest of spectrum.
Fig.~\ref{fig:mean_var_5Gdown} shows the mean and variance of the measured power versus altitude. 
As it can be observed from Fig.~\ref{fig:mean5Gd}, the mean value of the measured power for n5 and n71 are similar and significantly higher than n77. For the bands under consideration, the mean value increases as the altitude increases up to almost 80~m. 
As it is shown in Fig.~\ref{fig:var5Gd}, the variance of the measured power for n77 starts with a small value, while it climes up to near those of n5 values as the altitude increases. The variance of n71 band depicts a higher value for all the measured altitudes compared with those others 5G bands.

\subsection{CBRS Band}
\begin{figure}[!t]
\centering
\begin{subfigure}{0.49\columnwidth} 
\centering
\includegraphics[width=\textwidth]{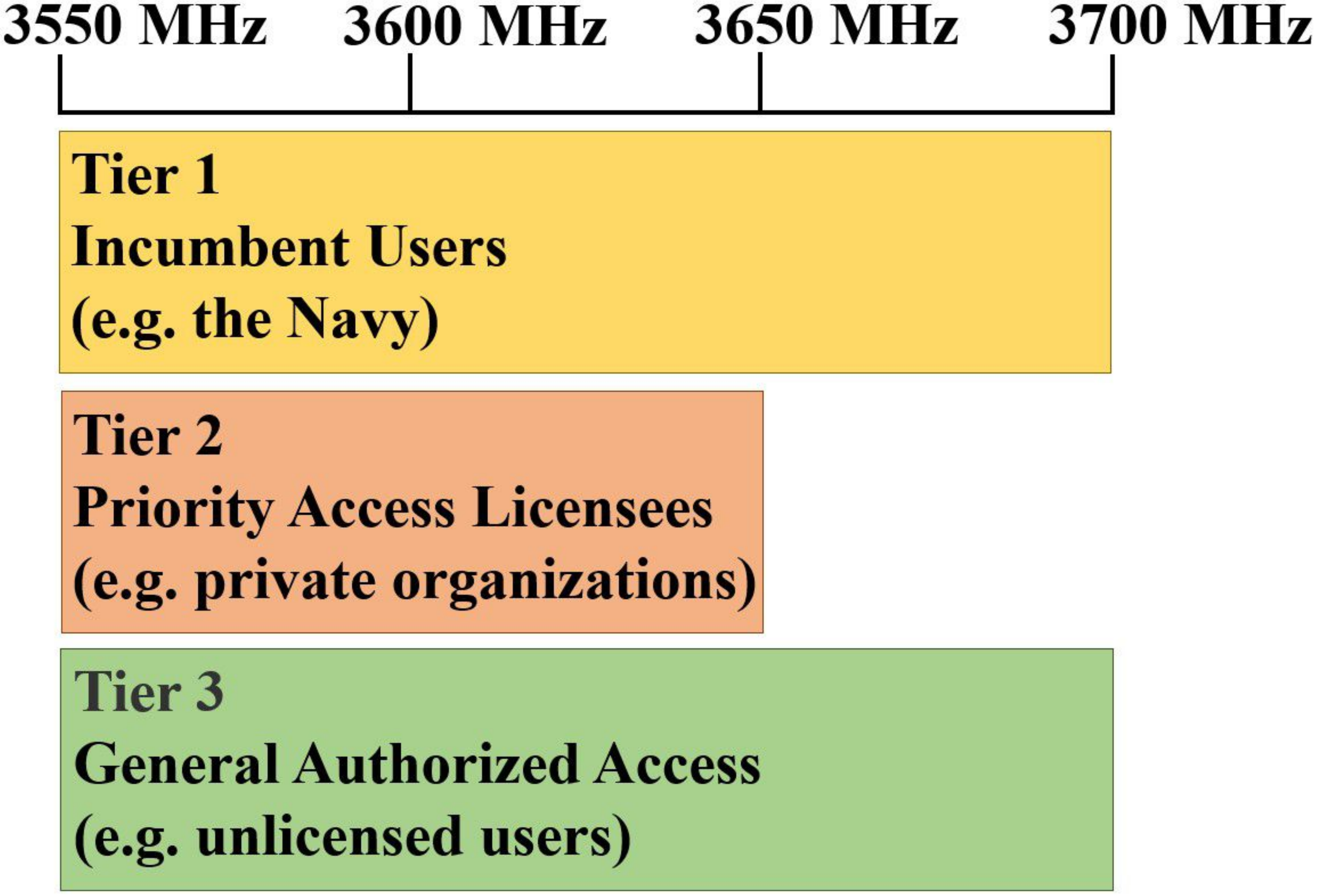} 
\caption{}\label{fig:cbrs} 
\end{subfigure}
\begin{subfigure}{0.49\columnwidth} 
\centering
\includegraphics[width=\textwidth]{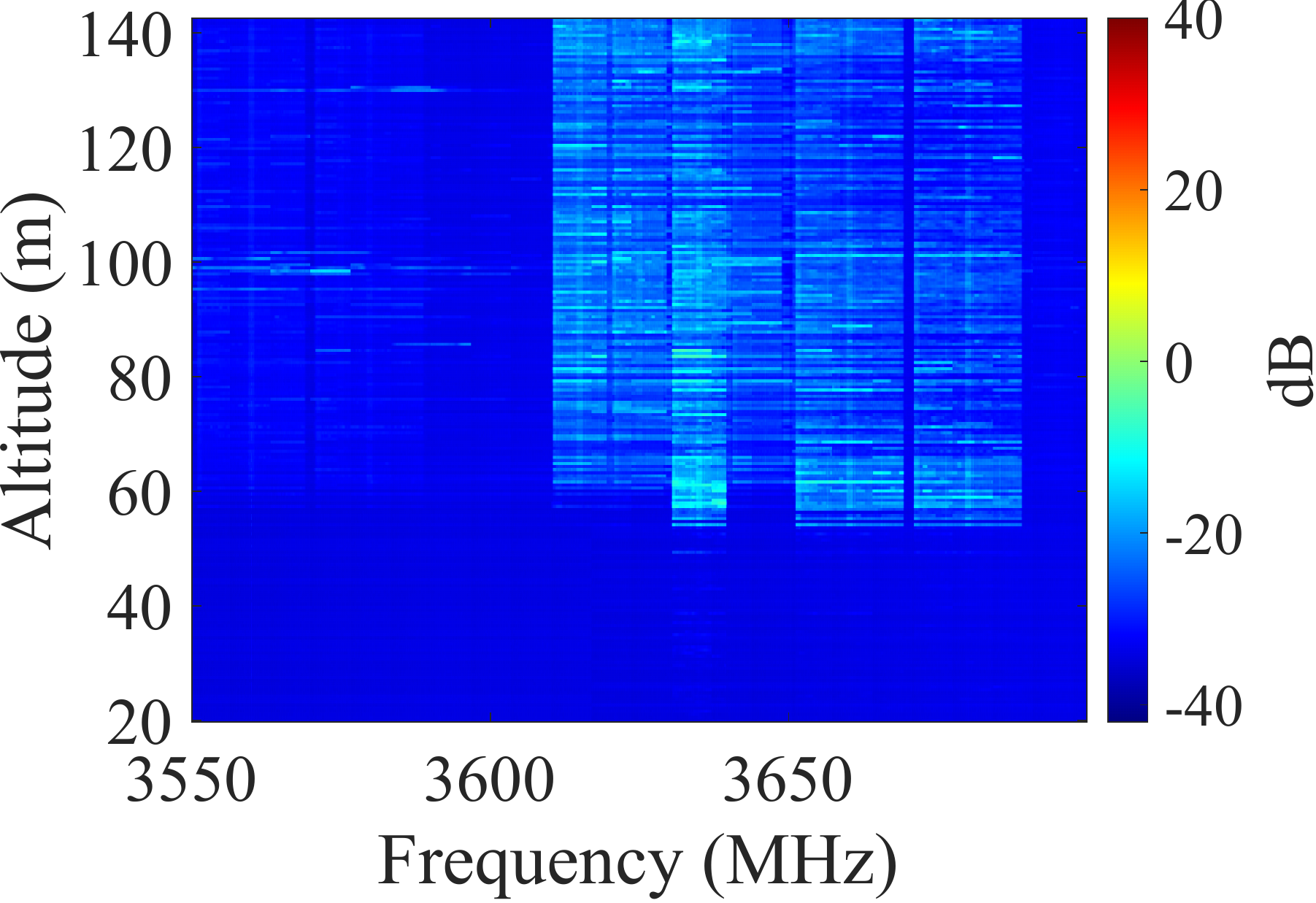}
\caption{}
\label{fig:n48}
\end{subfigure}
        \caption{ \textbf{(a)} CBRS spectrum and tiers; and \textbf{(b)} Measured CBRS band n48 power for urban environment (TDD UL/DL).}
        \label{fig:cbrs_n48}
\end{figure}

\begin{figure}[!t]
\centering
\begin{subfigure}{0.49\columnwidth} 
\centering
\includegraphics[width=\textwidth]{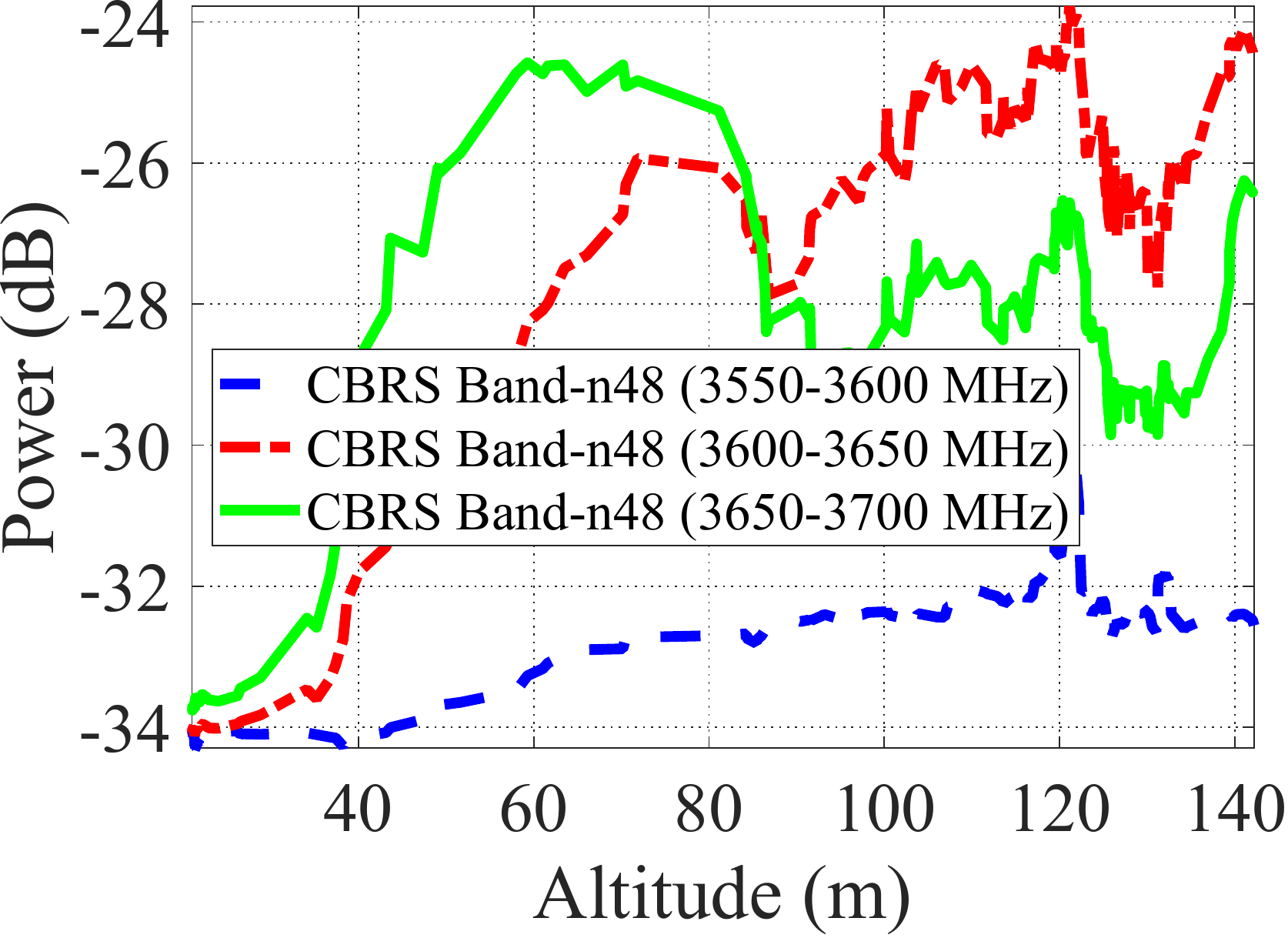} 
\caption{Mean.}\label{fig:meancbrs} 
\end{subfigure}
\begin{subfigure}{0.49\columnwidth} 
\centering
\includegraphics[width=\textwidth]{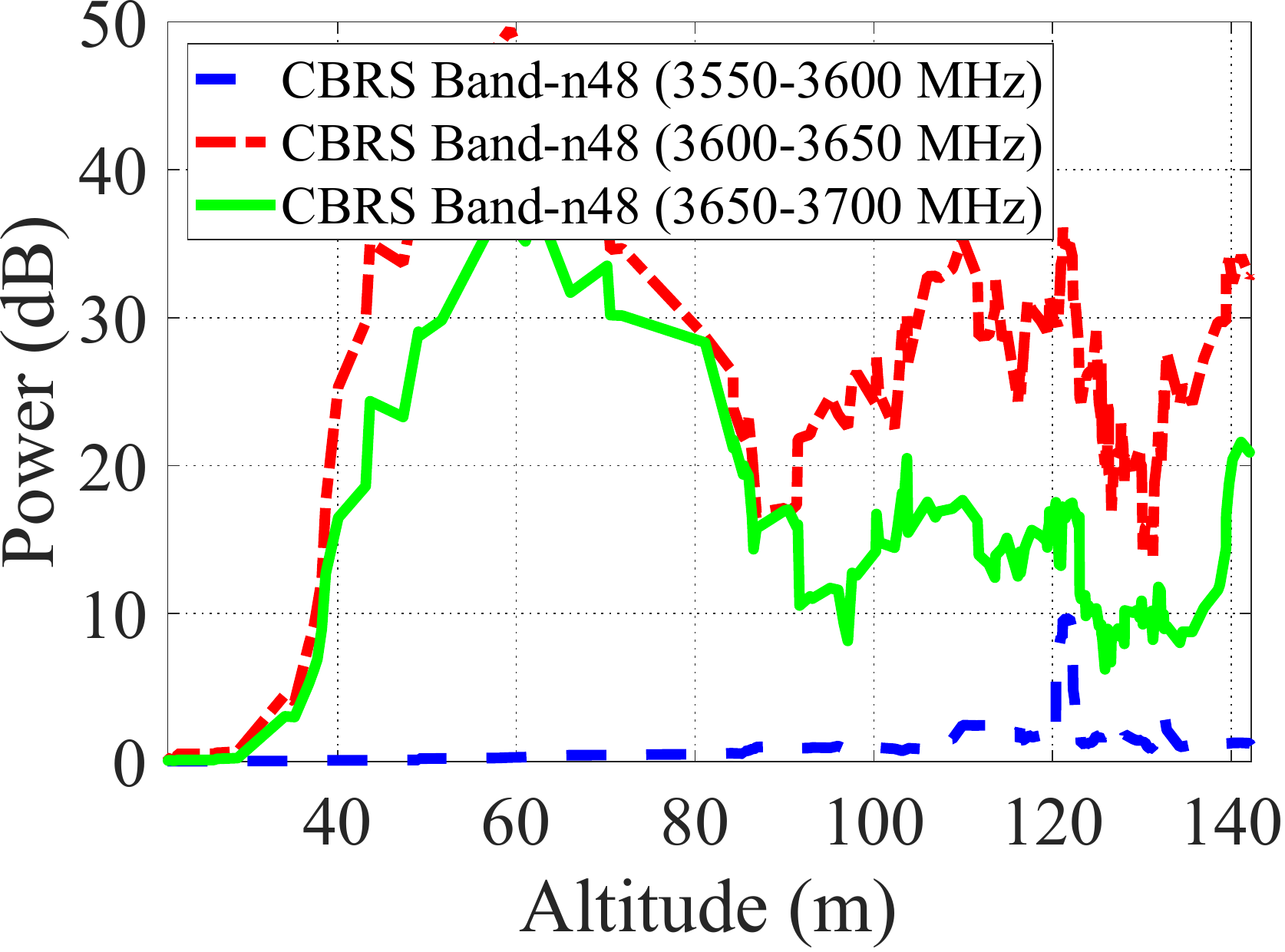}
\caption{Variance.}
\label{fig:varcbrs}
\end{subfigure}
        \caption{Spectrum occupancy versus altitude in CBRS band for urban environment.}
        \label{fig:mean_var_CBRS}
\end{figure}
Fig.~\ref{fig:cbrs} illustrates the CBRS spectrum which it lays out three tiers of users.
Fig.~\ref{fig:n48} presents the measured power for CBRS n48 band. Similar to LTE 41 and 5G n77 bands, n48 also exploits TDD mode. As it can be seen, the spectrum is mainly occupied within the range of 3610-3690~MHz. In Fig.~\ref{fig:mean_var_CBRS}, we study the mean and variance of the measured power versus altitude whereas the CBRS band is divided into three equal portions. As it can be observed, the mean and variance of the measured power for the first portion (i.e., 3550-3600~MHz) are lower than the other parts. The mean value of the third portion (i.e., 3650-3700~MHz) increases as the altitude increases up to 60~m and then it drops afterwards. However, the man value of the second part (i.e., 3600-3650~MHz) keeps increasing as the altitude increases.

\section{Rural Spectrum Occupancy Results}\label{results_rural}   
In this section, we study the spectrum occupancy and its characteristic for the similar bands as previous section by considering the experimental results for the rural environment.

\begin{figure}[!t]
\centering
\begin{subfigure}{0.49\columnwidth} 
\centering
\includegraphics[width=\textwidth]{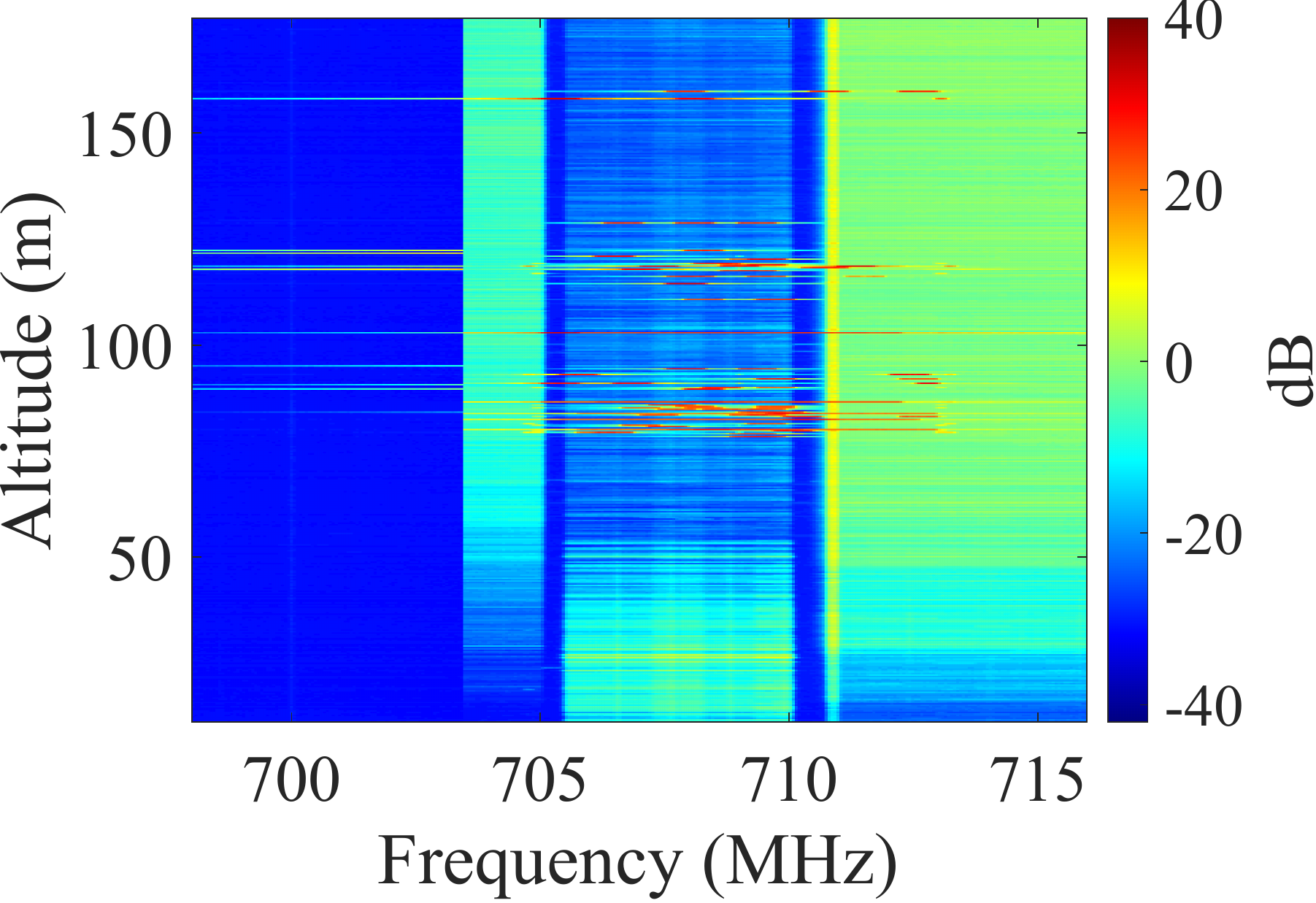} 
\caption{LTE band 12 (UL).}\label{fig:up12_wheeler} 
\end{subfigure}
\begin{subfigure}{0.49\columnwidth} 
\centering
\includegraphics[width=\textwidth]{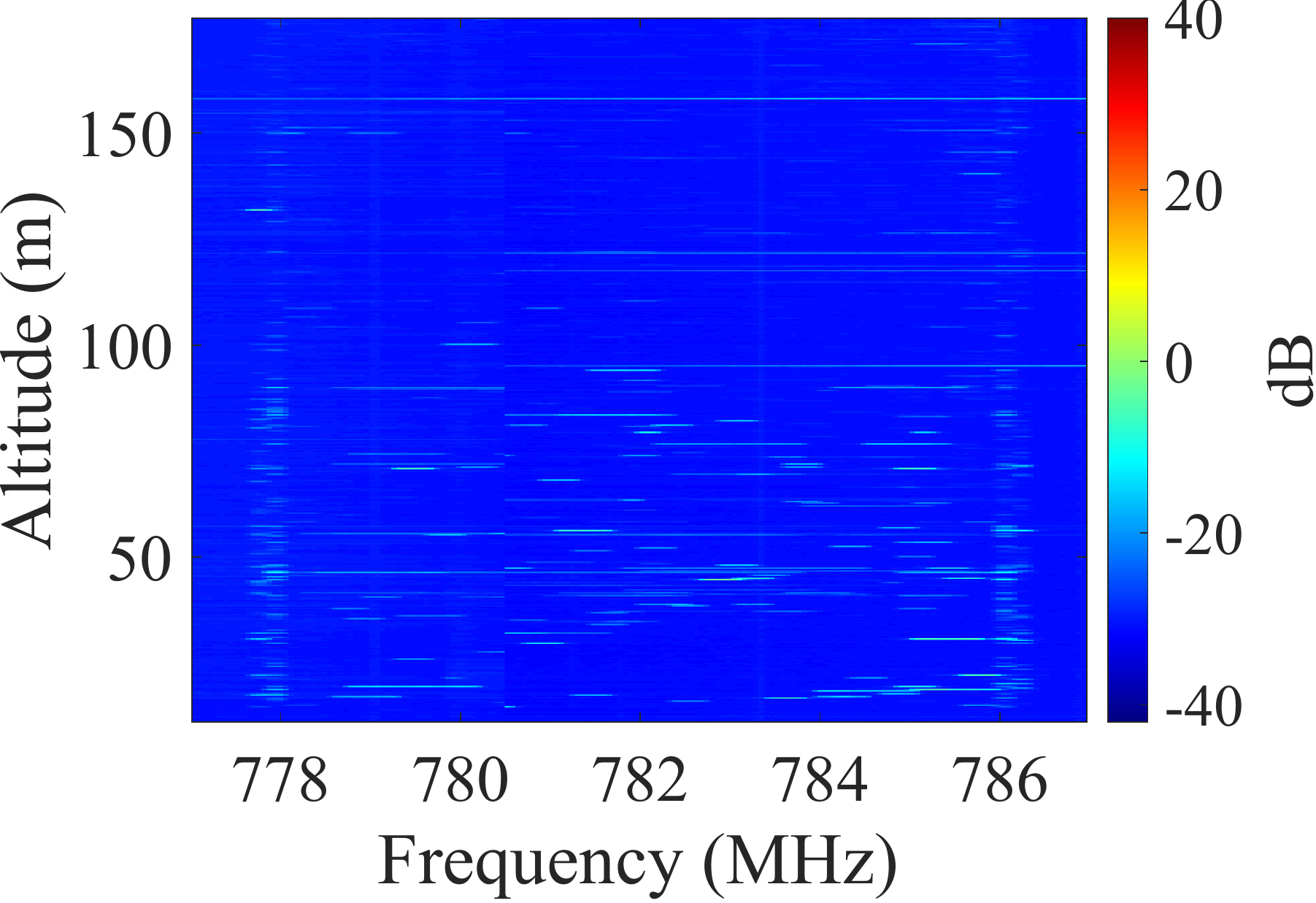}
\caption{LTE band 13 (UL).}
\label{fig:up13_wheeler}
\end{subfigure}
\begin{subfigure}{0.49\columnwidth} 
\centering
\includegraphics[width=\textwidth]{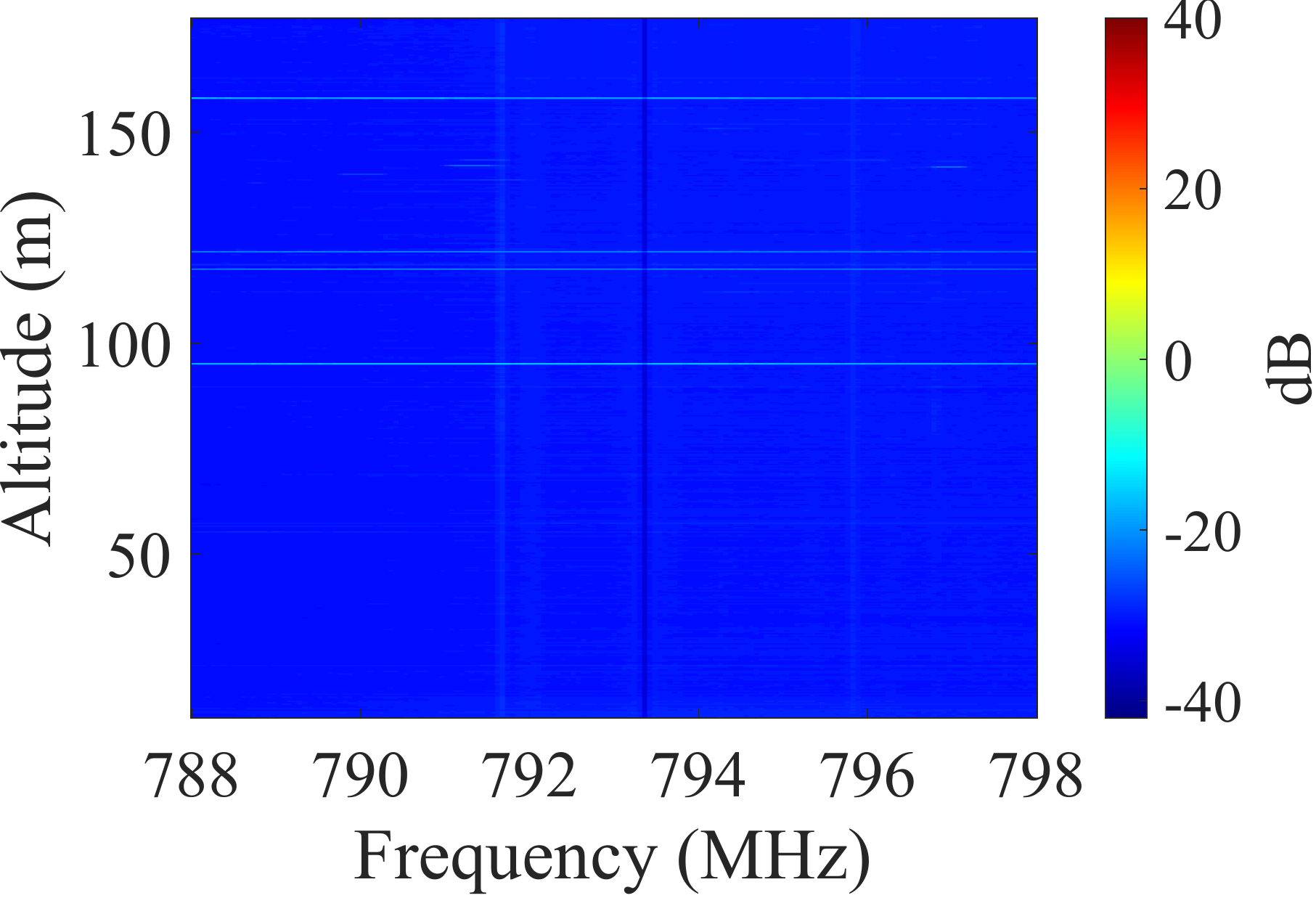} 
\caption{LTE band 14 (UL).}\label{fig:up14_wheeler}
\end{subfigure} 
\begin{subfigure}{0.49\columnwidth} 
\centering
\includegraphics[width=\textwidth]{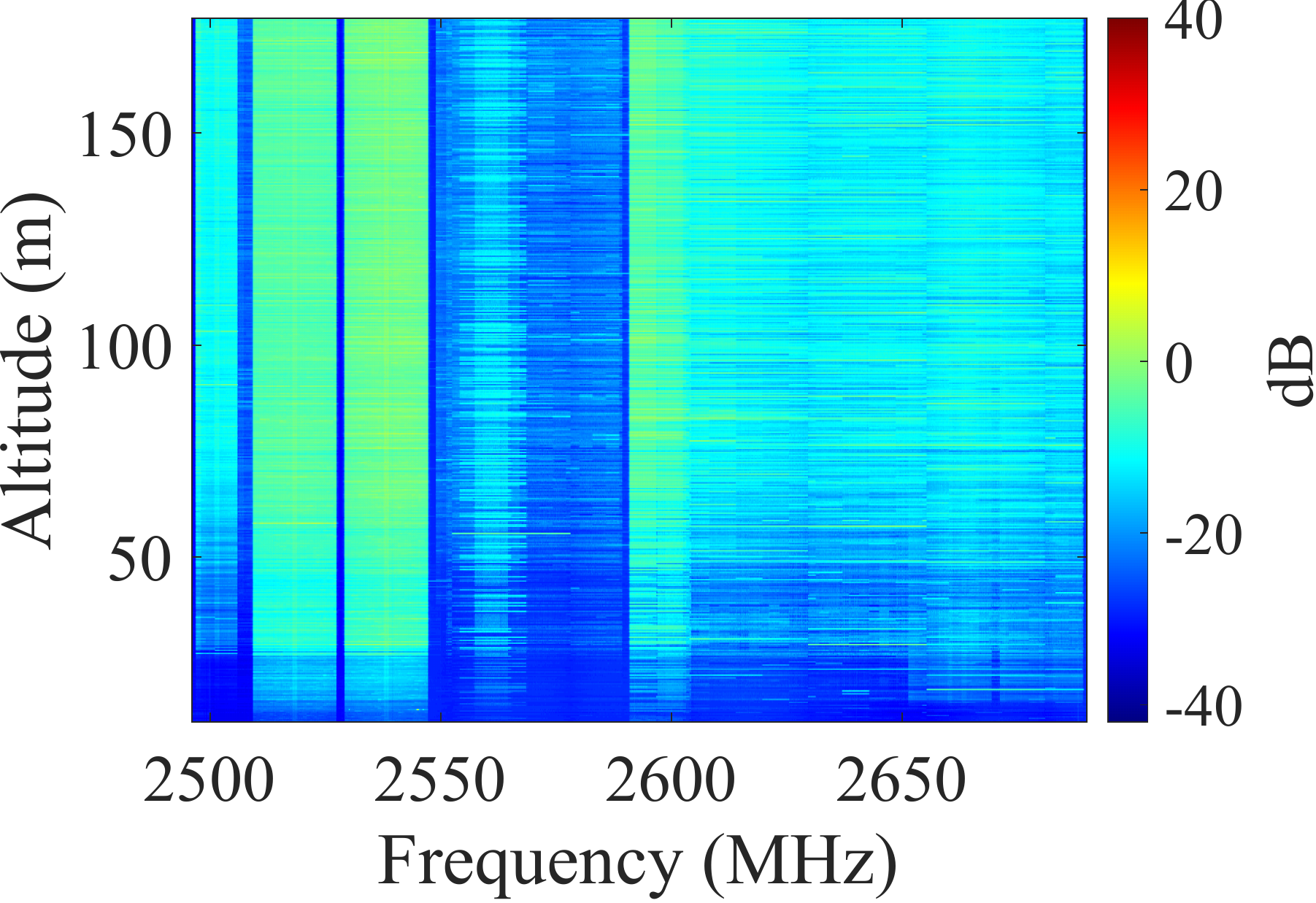} 
\caption{LTE band 41 (TDD UL/DL).}\label{fig:up41_wheeler}
\end{subfigure} 
\caption{Measured LTE UL power for rural environment.}
\label{fig:LTE_power_freq_alt_up_wheeler}
\end{figure}
\subsection{LTE Bands - Uplink}
\begin{figure}[!t]
\centering
\begin{subfigure}{0.49\columnwidth} 
\centering
\includegraphics[width=\textwidth]{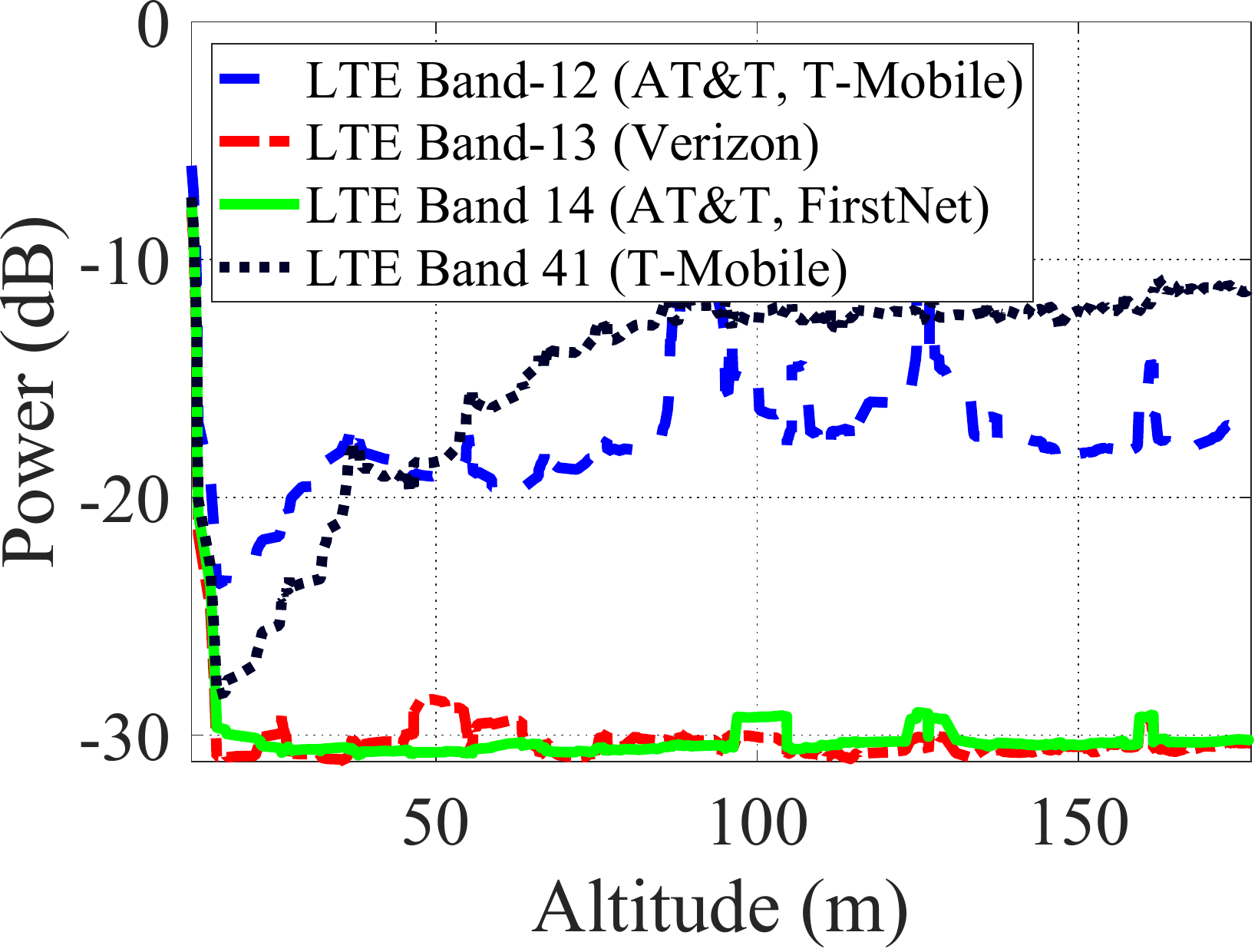} 
\caption{Mean.}\label{fig:meanLTEup_wheeler} 
\end{subfigure}
\begin{subfigure}{0.49\columnwidth} 
\centering
\includegraphics[width=\textwidth]{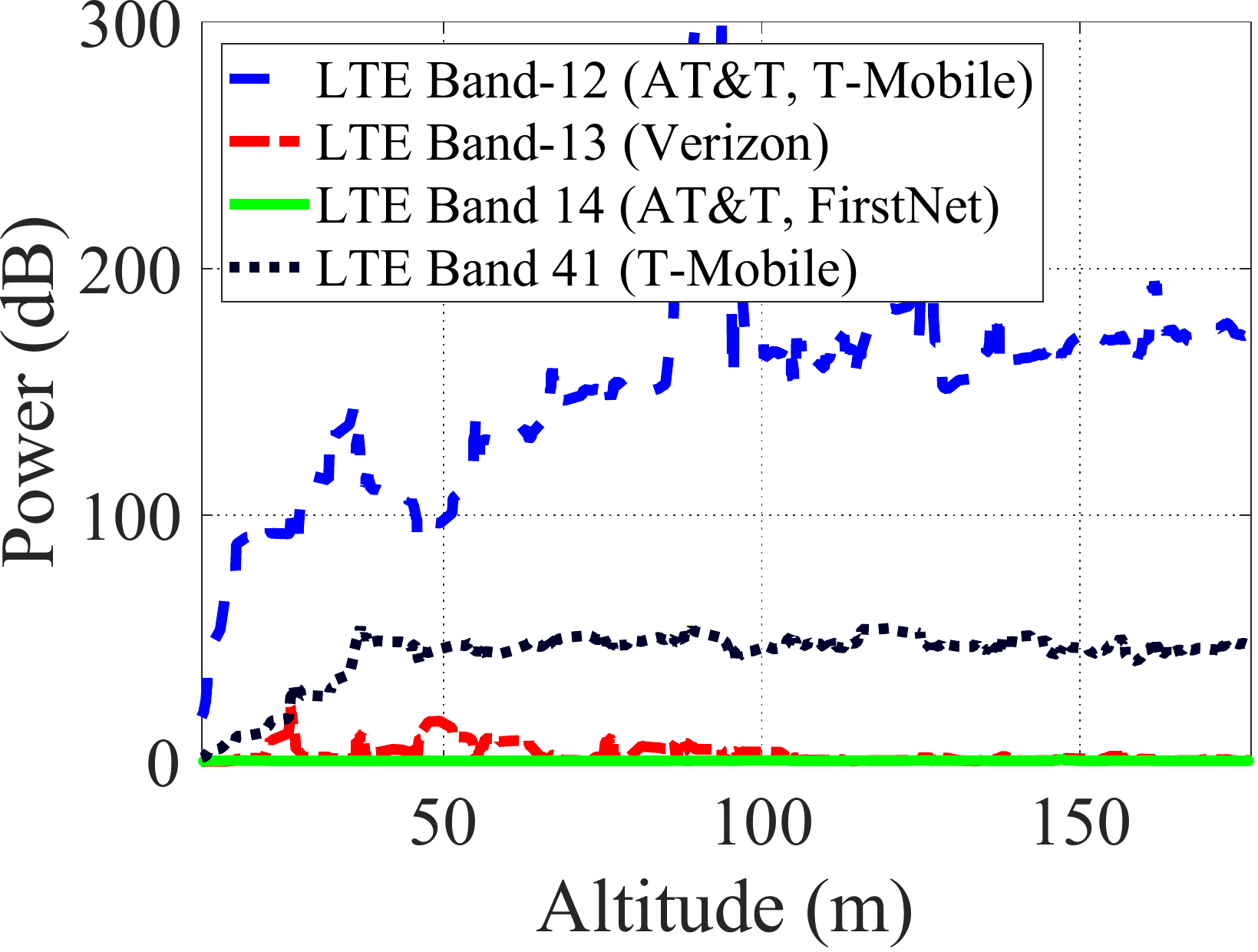}
\caption{Variance.}
\label{fig:varLTEup_wheeler}
\end{subfigure}
        \caption{Spectrum occupancy versus altitude in LTE bands 12, 13, 14 and 41 (UL) for rural environment.}
        \label{fig:mean_var_LTEup_wheeler}
\end{figure}

Fig.~\ref{fig:LTE_power_freq_alt_up_wheeler} illustrates the measured power for for LTE bands 13,
14, 15 and 41 considering the UL frequency spectrum. 
As it can be seen, LTE bands 12 and 41 show more crowded spectrum compared with LTE bands 13 and 14. 
The mean and variance of the measured
power for various LTE bands are presented in Fig.~\ref{fig:mean_var_LTEup_wheeler}.  As opposed to the urban environment (cf. Fig.~\ref{fig:meanLTEup}), the
mean value for LTE bands 13 and 14 are much higher than the other two bands
under consideration.

\begin{figure}[!t]
\centering
\begin{subfigure}{0.49\columnwidth} 
\centering
\includegraphics[width=\textwidth]{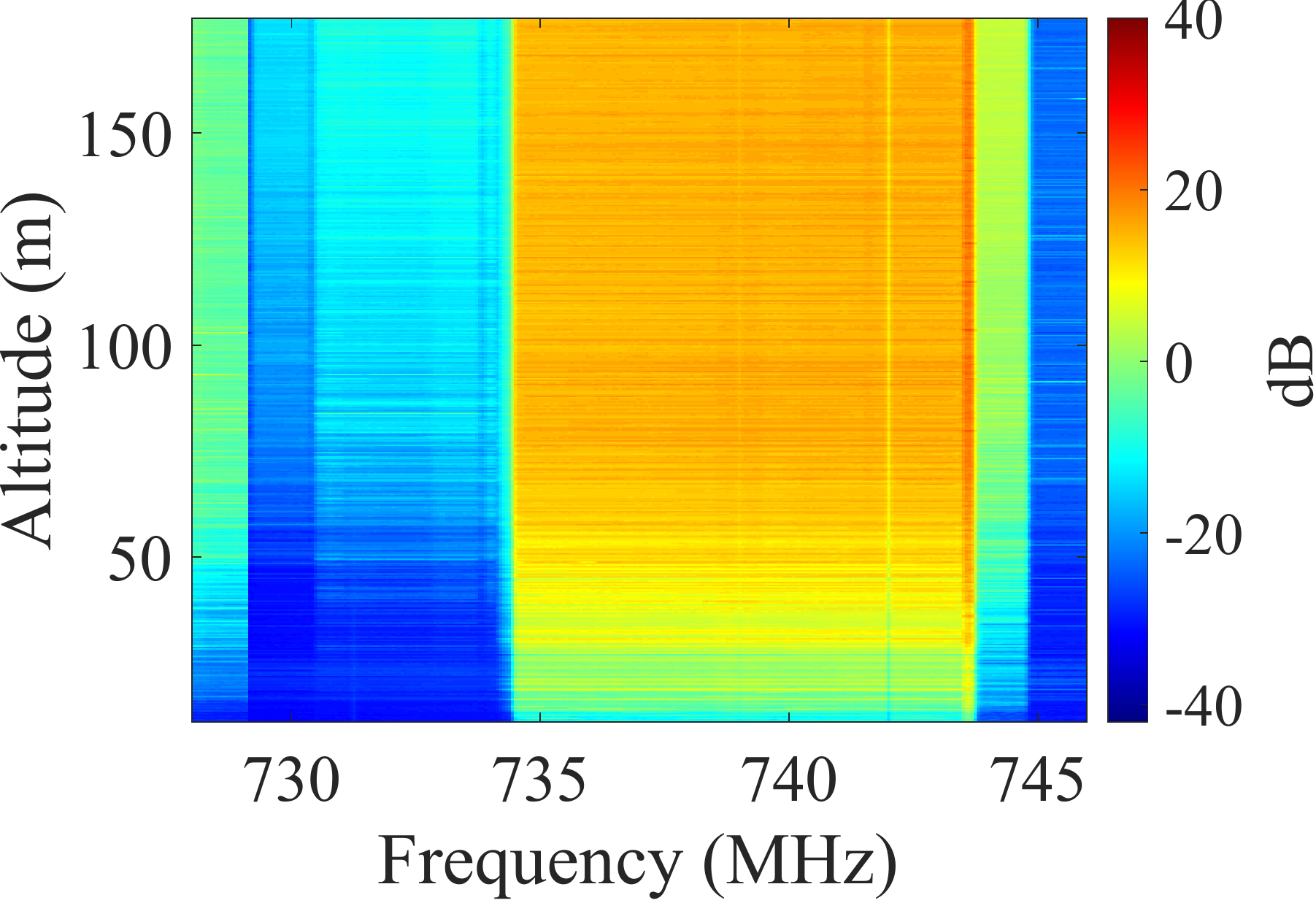} 
\caption{LTE band 12 (DL).}\label{fig:down12_wheeler} 
\end{subfigure}
\begin{subfigure}{0.49\columnwidth} 
\centering
\includegraphics[width=\textwidth]{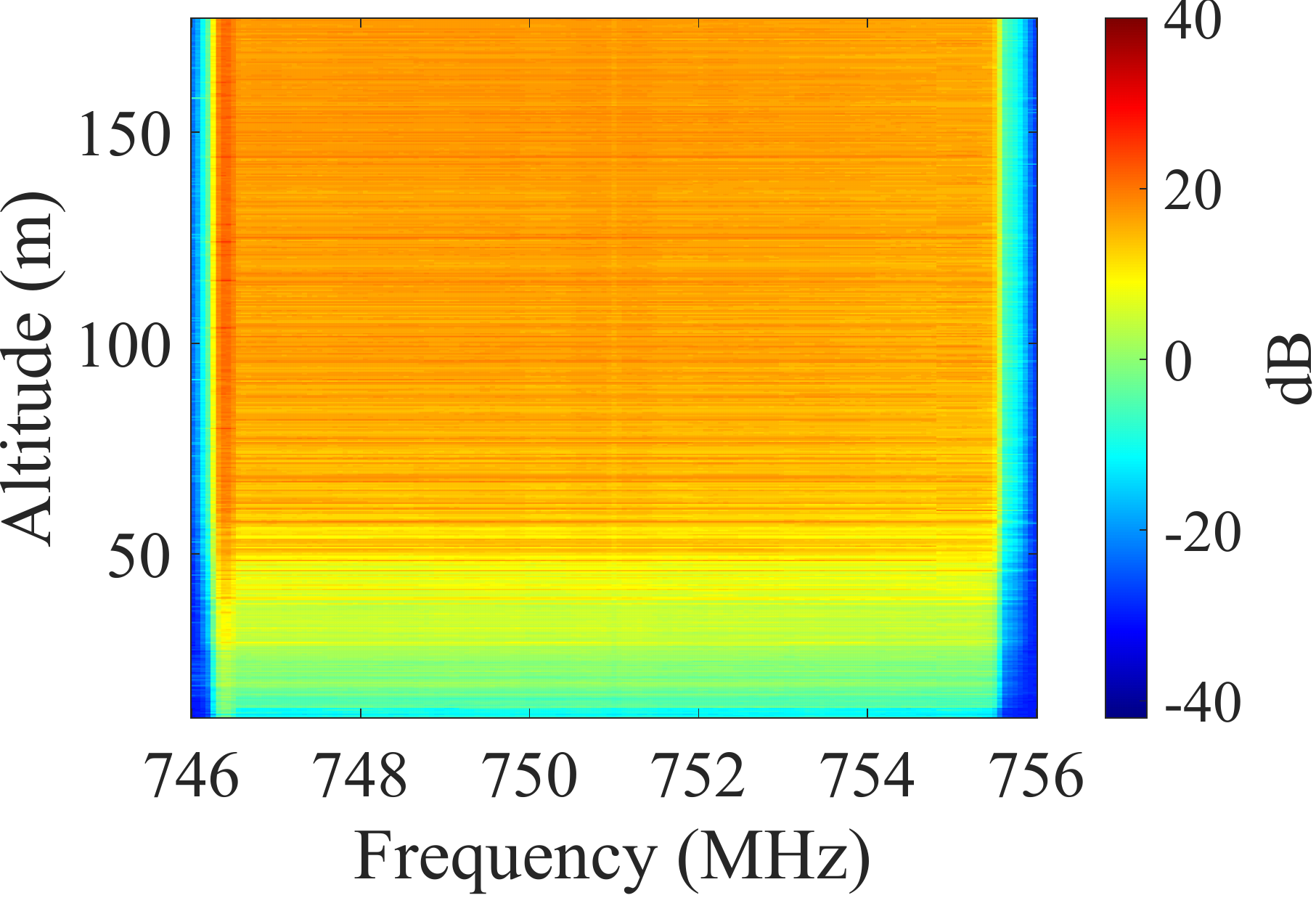}
\caption{LTE band 13 (DL).}
\label{fig:down13_wheeler}
\end{subfigure}
\begin{subfigure}{0.49\columnwidth} 
\centering
\includegraphics[width=\textwidth]{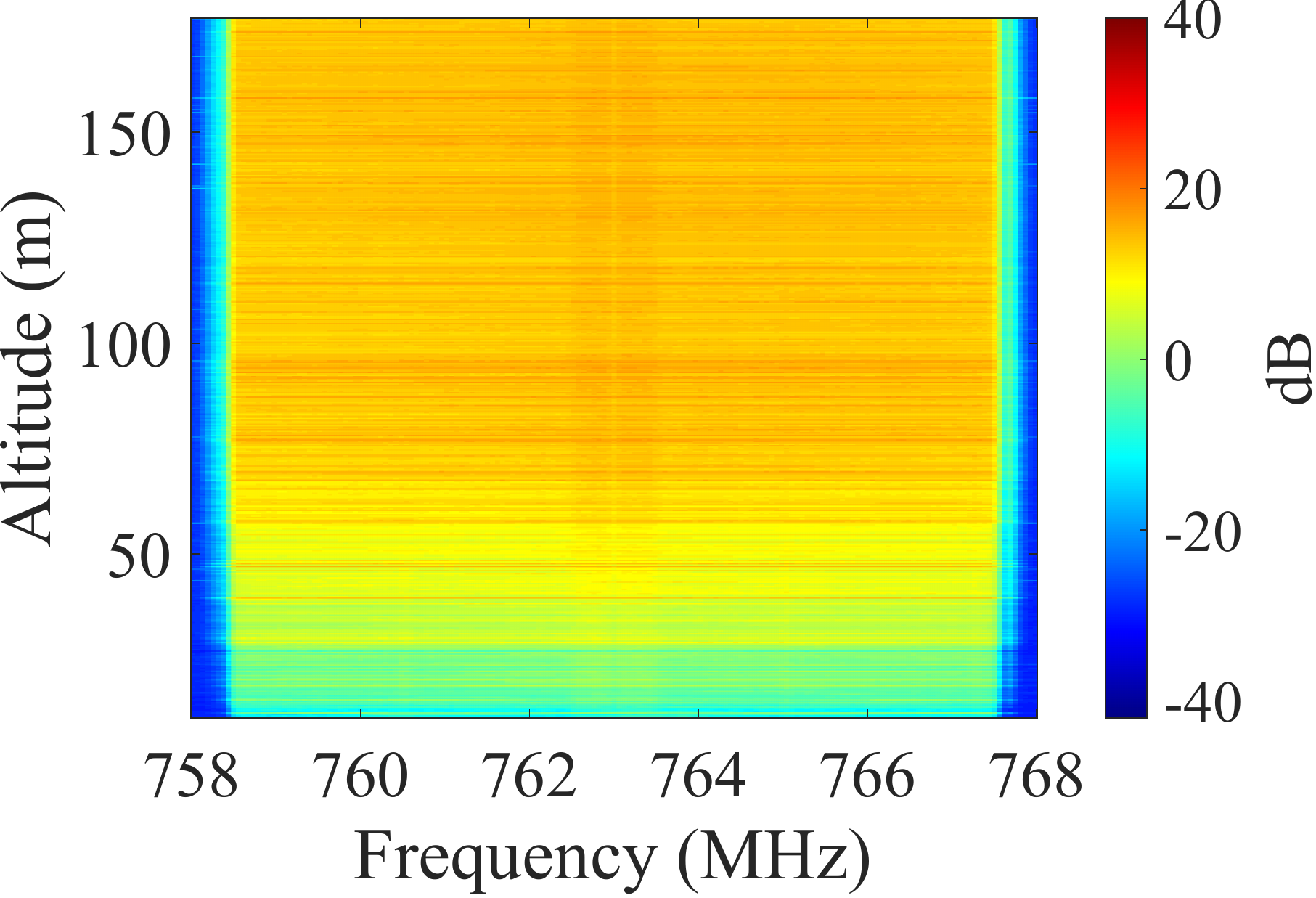} 
\caption{LTE band 14 (DL).}\label{fig:down14_wheeler}
\end{subfigure} 
\begin{subfigure}{0.49\columnwidth} 
\centering
\includegraphics[width=\textwidth]{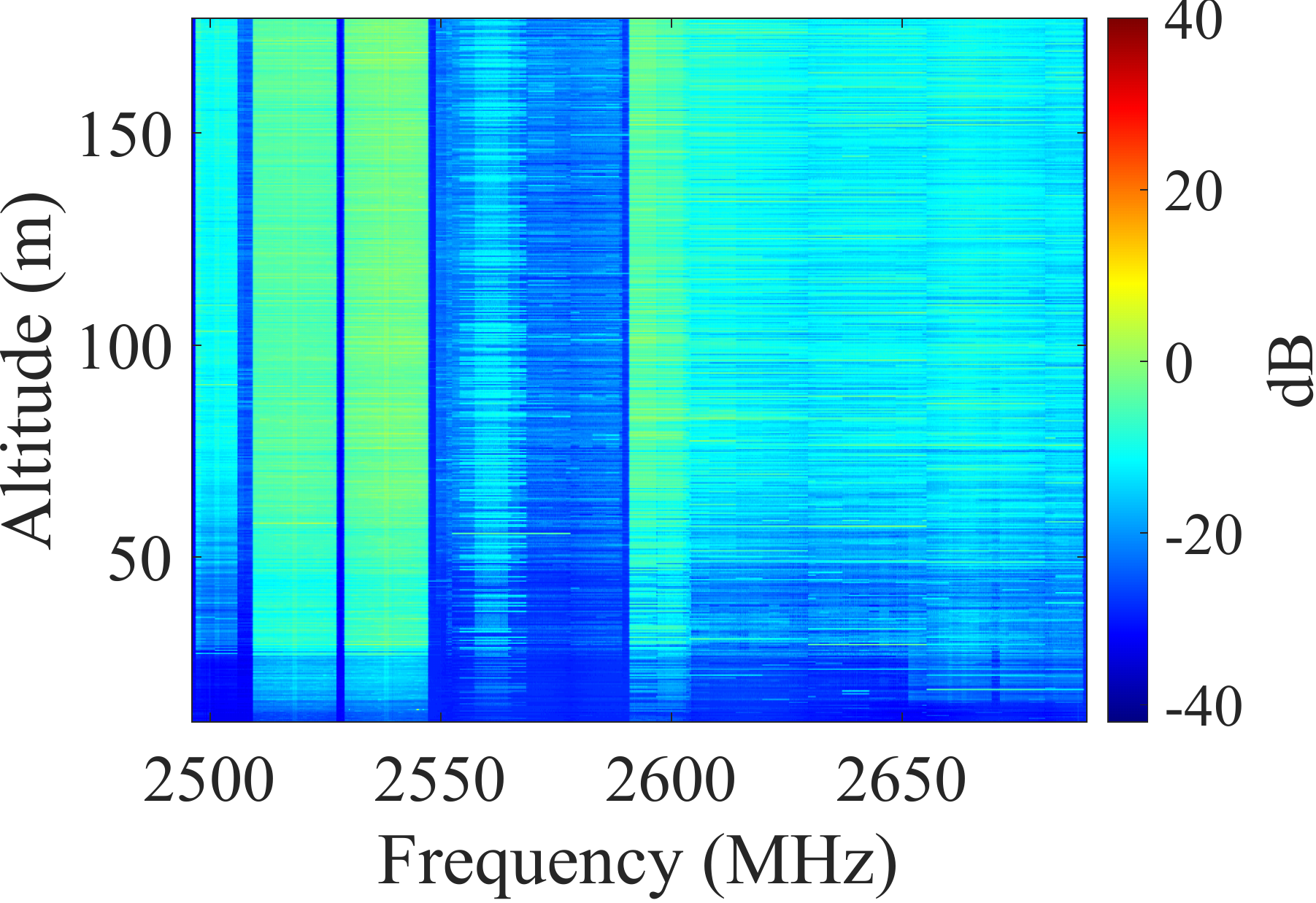} 
\caption{LTE band 41 (TDD UL/DL).}\label{fig:down41_wheeler}
\end{subfigure} 
\caption{Measured LTE DL power for rural environment.}
\label{fig:LTE_power_freq_alt_down_wheeler}
\end{figure}

\subsection{LTE Bands - Downlink}
\begin{figure}[!t]
\centering
\begin{subfigure}{0.49\columnwidth} 
\centering
\includegraphics[width=\textwidth]{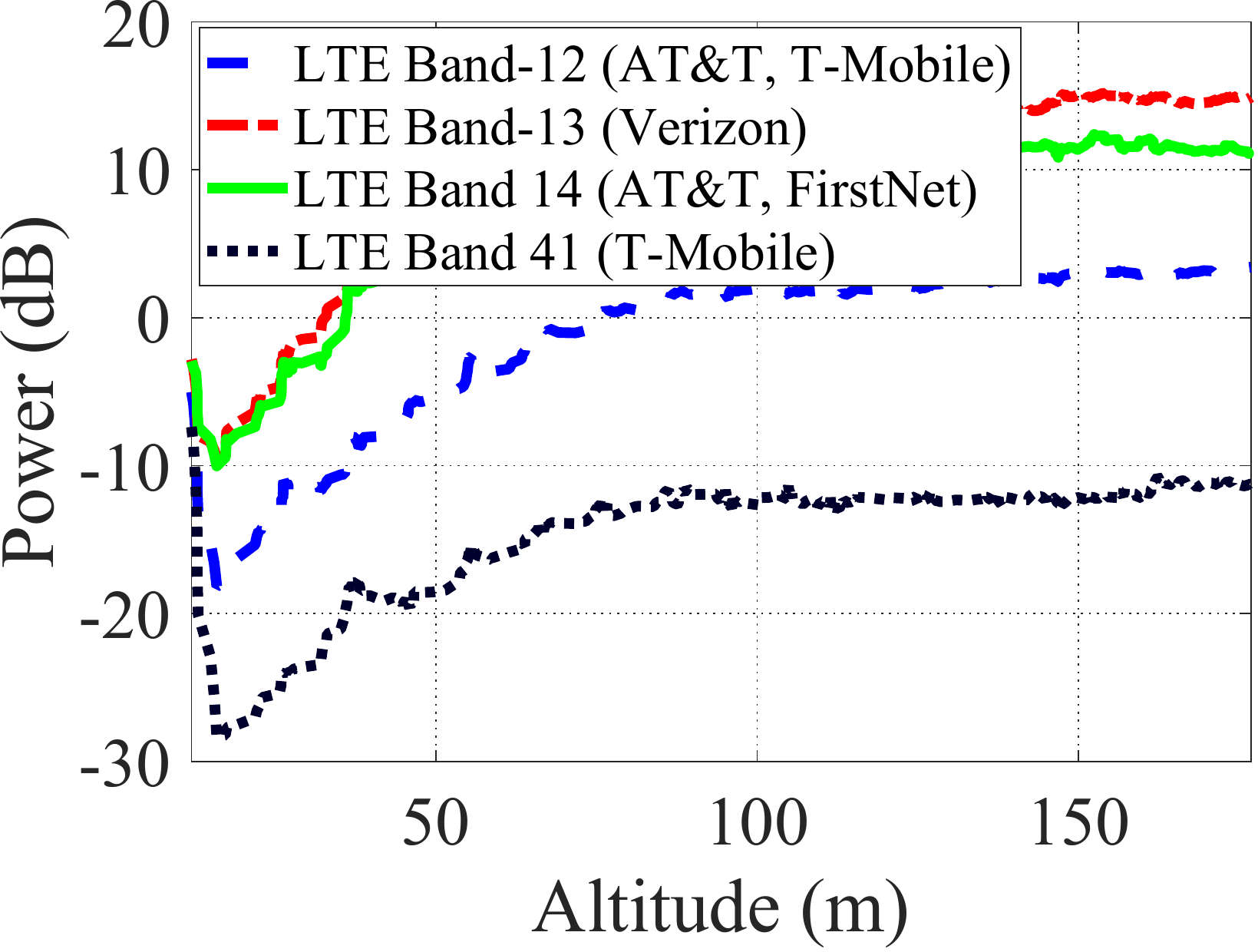} 
\caption{Mean.}\label{fig:meanLTEdown_wheeler} 
\end{subfigure}
\begin{subfigure}{0.49\columnwidth} 
\centering
\includegraphics[width=\textwidth]{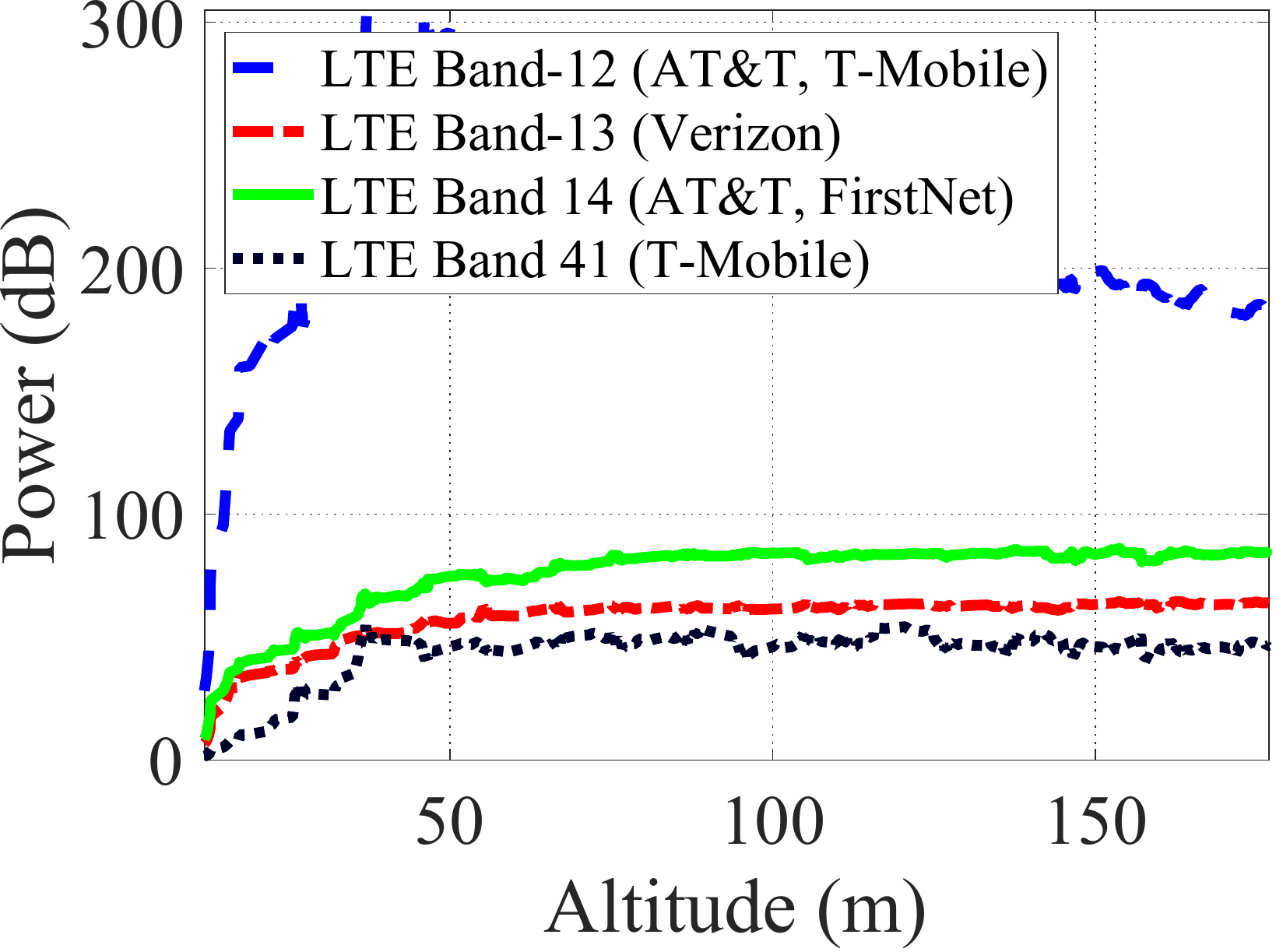}
\caption{Variance.}
\label{fig:varLTEdown_wheeler}
\end{subfigure}
        \caption{Spectrum occupancy versus altitude in LTE bands 12, 13, 14 and 41 (DL) for rural environment.}
        \label{fig:mean_var_LTE_down_wheeler}
\end{figure}

Considering the DL frequency range for different
LTE bands, Fig.~\ref{fig:LTE_power_freq_alt_down_wheeler} illustrates the measured power for the bands under consideration. Same as the urban results, the spectrum of DL frequency range are more crowded compared with the UL ones in the rural environment. 
Fig.~\ref{fig:mean_var_LTE_down_wheeler} shows the mean and variance of the measured
power versus altitude. 
As it can be observed from Fig.~\ref{fig:meanLTEdown_wheeler}, the mean value of
the measured power increases as the altitude increases up to 80~m and it remains almost constant for the higher altitudes.
The variance of LTE bands 13, 14, and 41 show similar behaviour, while the corresponded plot for LTE band 12 starts with increasing for the altitude up to 40~m and then it drops afterwards.

\subsection{5G Bands - Uplink}
\begin{figure}[!t]
\centering
\begin{subfigure}{0.49\columnwidth} 
\centering
\includegraphics[width=\textwidth]{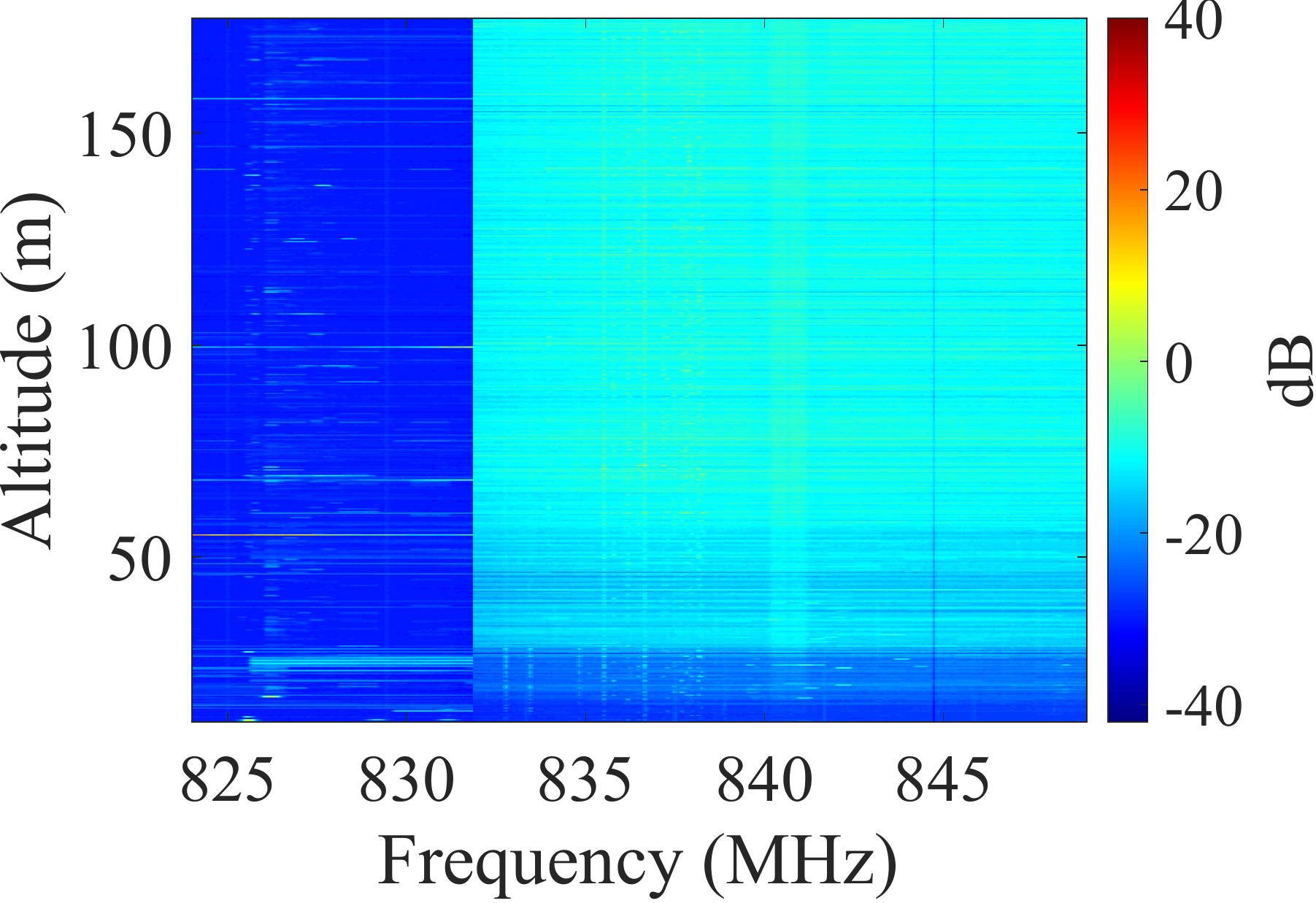} 
\caption{5G band n5 (UL).}\label{fig:upn5_wheeler} 
\end{subfigure}
\begin{subfigure}{0.49\columnwidth} 
\centering
\includegraphics[width=\textwidth]{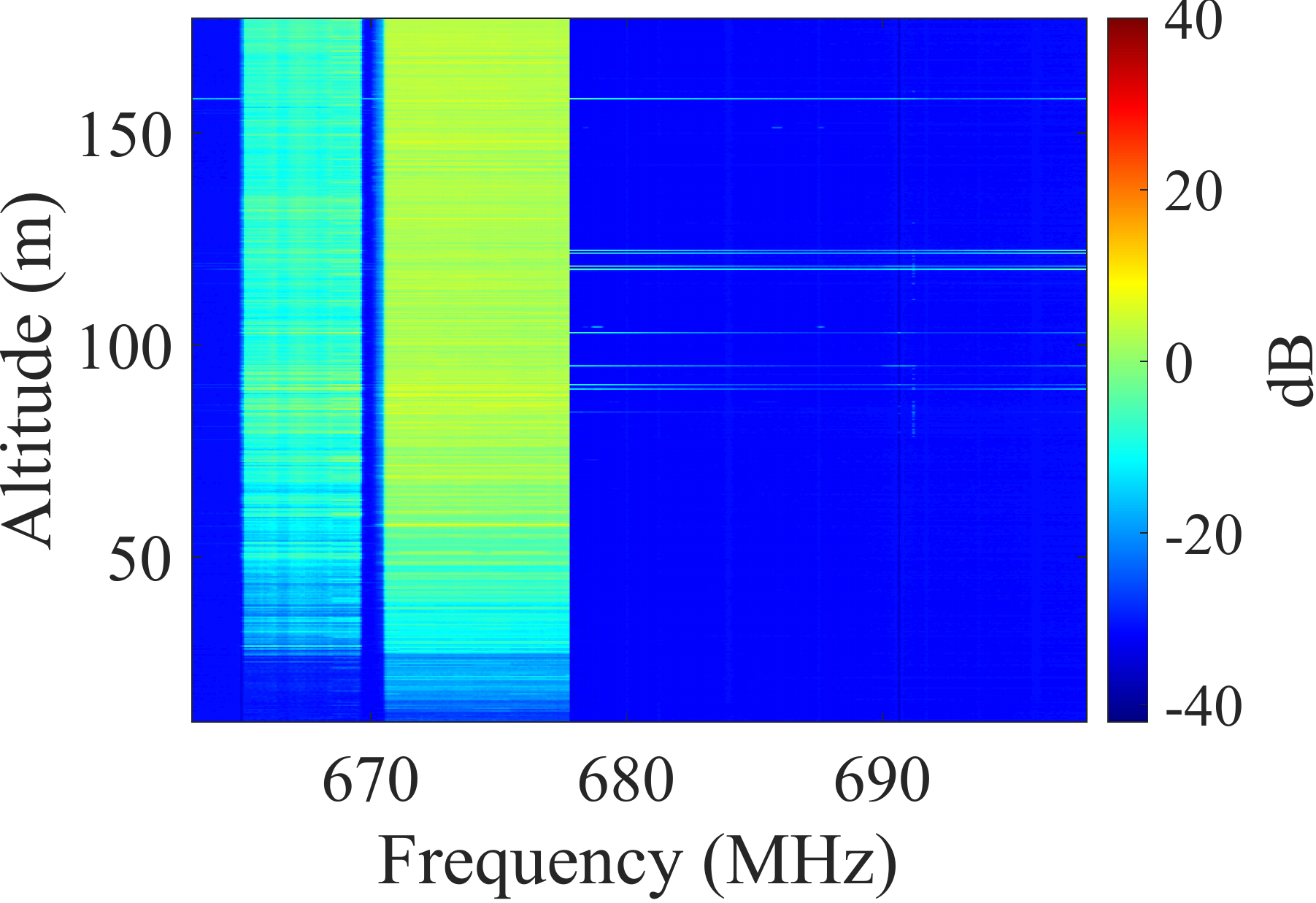}
\caption{5G band n71 (UL).}
\label{fig:upn71_wheeler}
\end{subfigure}
\begin{subfigure}{0.49\columnwidth} 
\centering
\includegraphics[width=\textwidth]{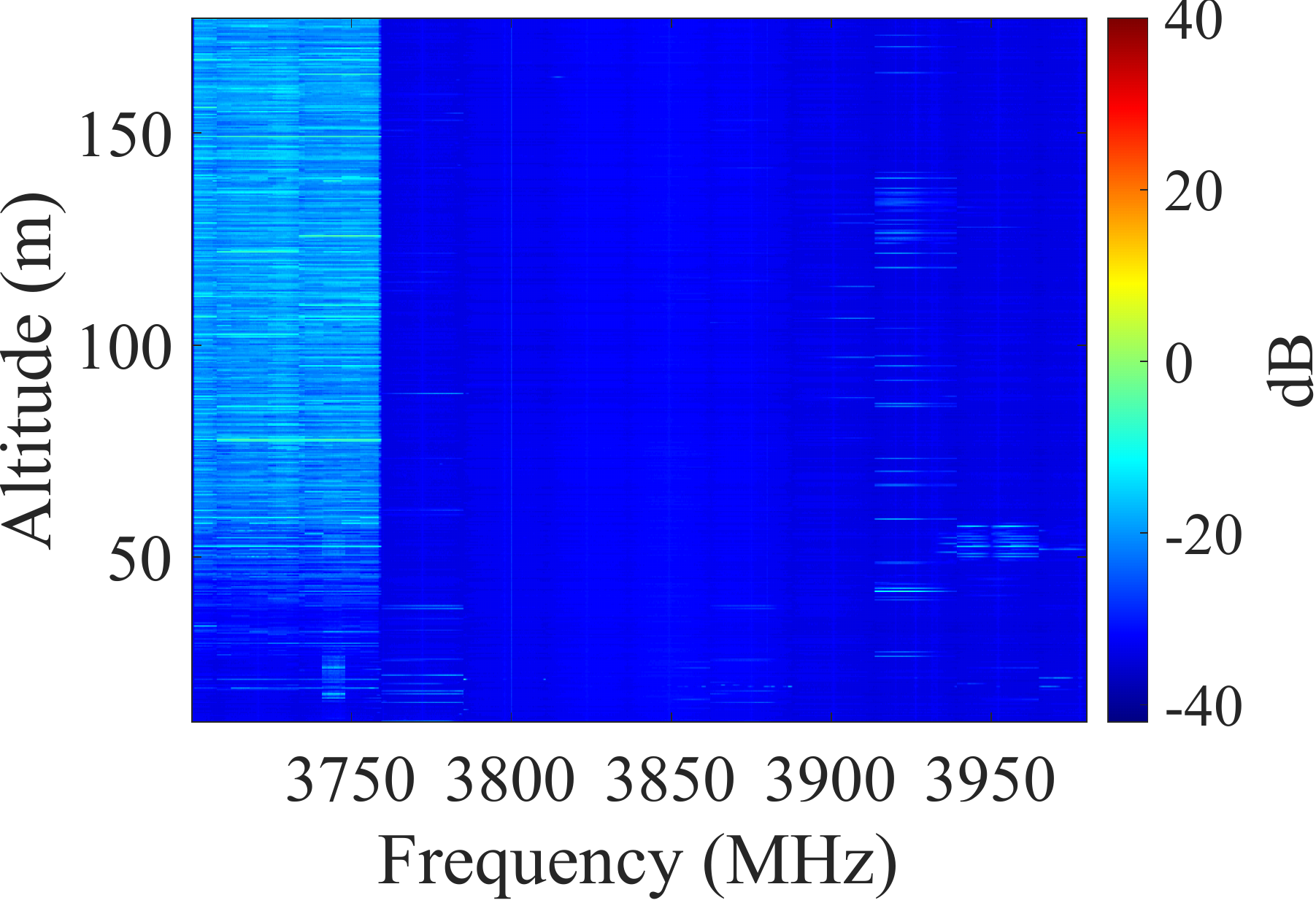} 
\caption{5G band n77 (TDD UL/DL).}\label{fig:upn77_wheeler}
\end{subfigure} 
\caption{Measured 5G UL power for rural environment.}
\label{fig:5G_power_freq_alt_up_wheeler}
\end{figure}

\begin{figure}[!t]
\centering
\begin{subfigure}{0.49\columnwidth} 
\centering
\includegraphics[width=\textwidth]{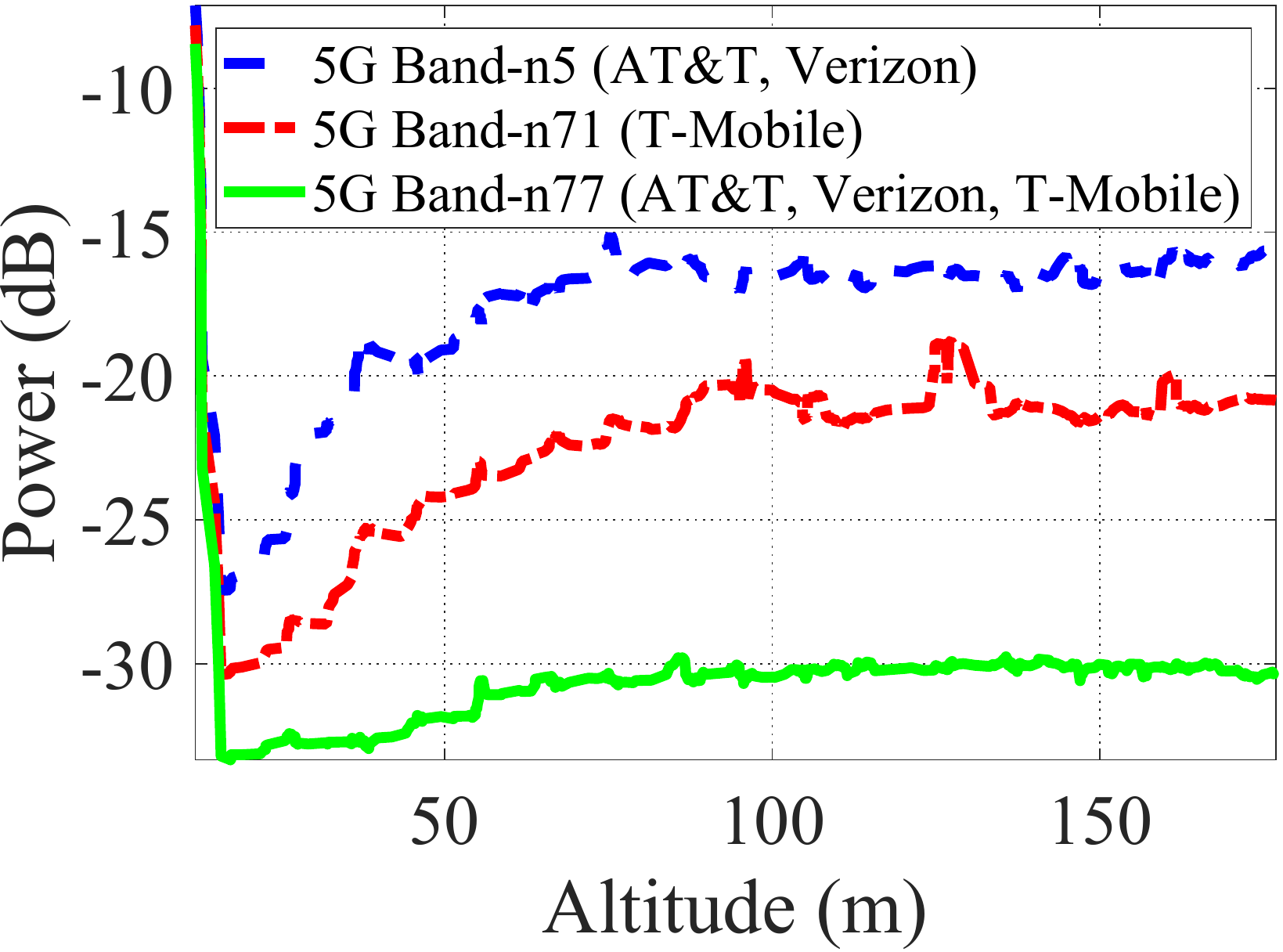} 
\caption{Mean.}\label{fig:mean5Gu_wheeler} 
\end{subfigure}
\begin{subfigure}{0.49\columnwidth} 
\centering
\includegraphics[width=\textwidth]{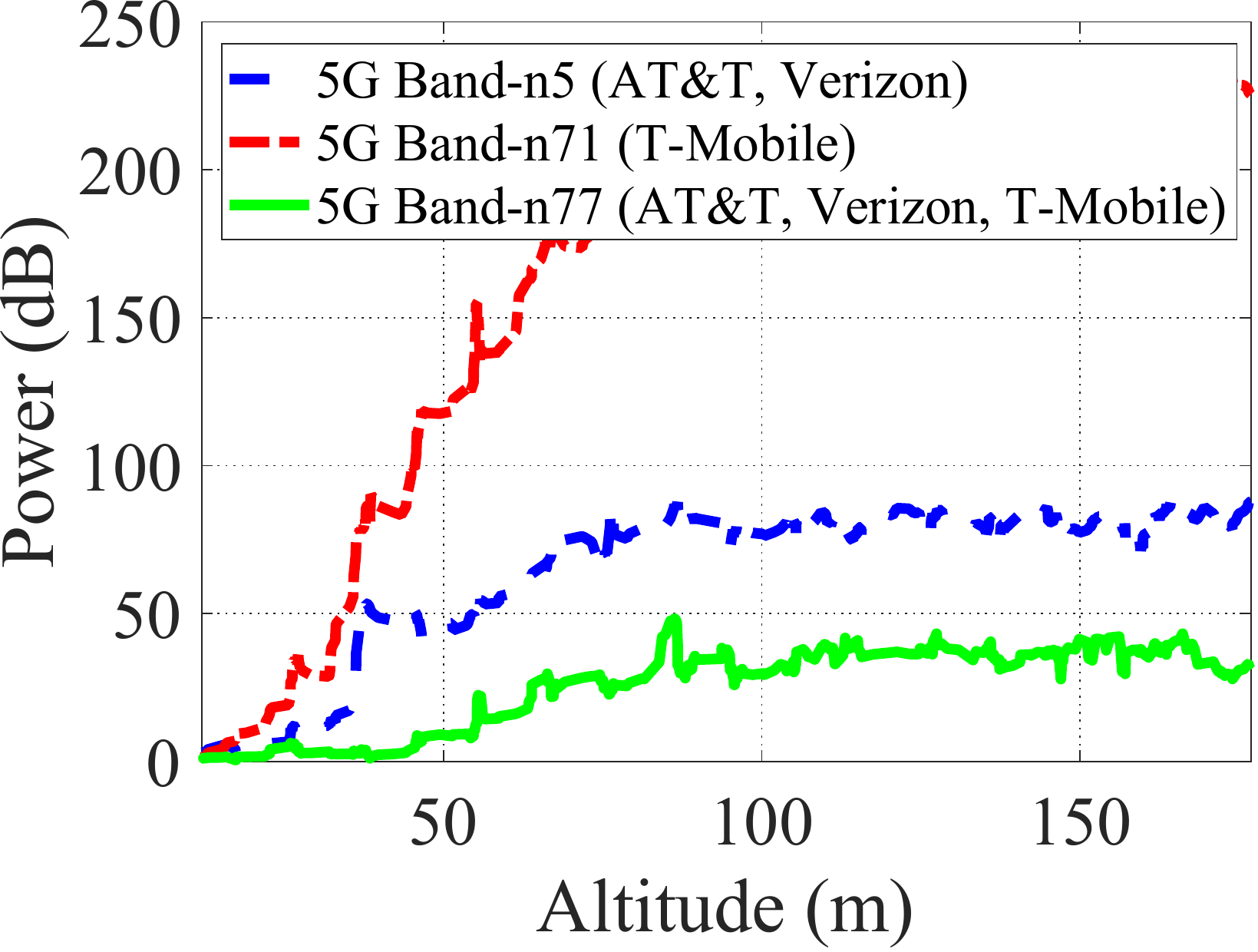}
\caption{Variance.}
\label{fig:var5Gu_wheeler}
\end{subfigure}
        \caption{Spectrum occupancy versus altitude in 5G n5 and n77 bands  (UL) for rural environment.}
        \label{fig:mean_var_5Gup_wheeler}
\end{figure}

Fig.~\ref{fig:5G_power_freq_alt_up_wheeler} illustrates the measured power for 5G bands n5, n71 and n77 considering the UL frequency spectrum ranges. This result reveals that the spectrum of n77 is less crowded than those of n5 and n71.
The performance of mean and variance of the measured power for 5G bands (uplink) are presented in Fig.~\ref{fig:mean_var_5Gup_wheeler}. As it can be observed from Fig.~\ref{fig:mean5Gu_wheeler}, while the mean value of the measured power for n77 is almost independent of the altitude, it increases for n5 and n71 bands as the altitude increases. 
As it is shown in Fig.~\ref{fig:var5Gu_wheeler}, the variance of the measured power for n71 depicts higher value compared with the other 5G bands.

\subsection{5G Bands - Downlink}
\begin{figure}[!t]
\centering
\begin{subfigure}{0.49\columnwidth} 
\centering
\includegraphics[width=\textwidth]{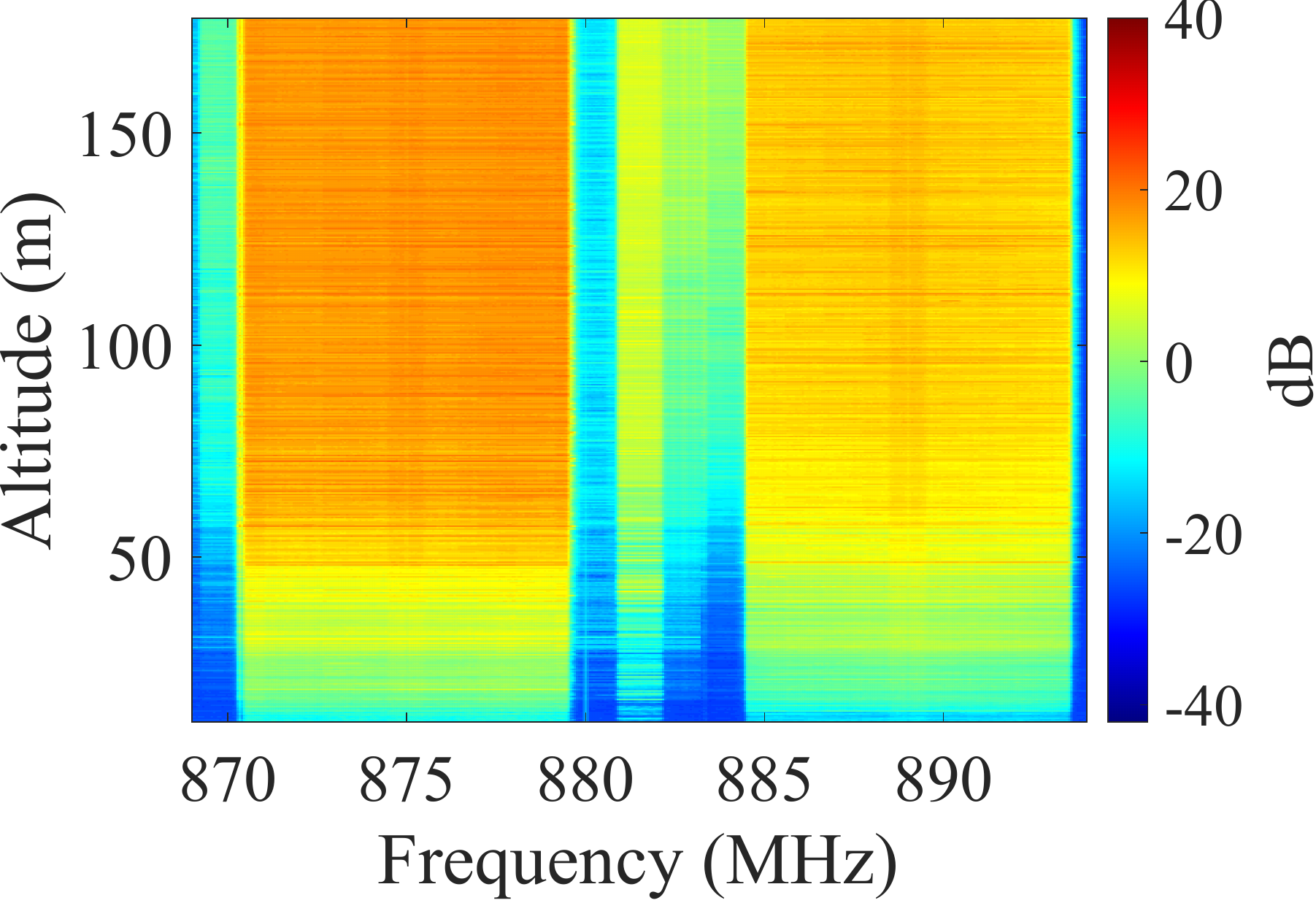} 
\caption{5G band n5 (DL).}\label{fig:n5_downlink_wheeler} 
\end{subfigure}
\begin{subfigure}{0.49\columnwidth} 
\centering
\includegraphics[width=\textwidth]{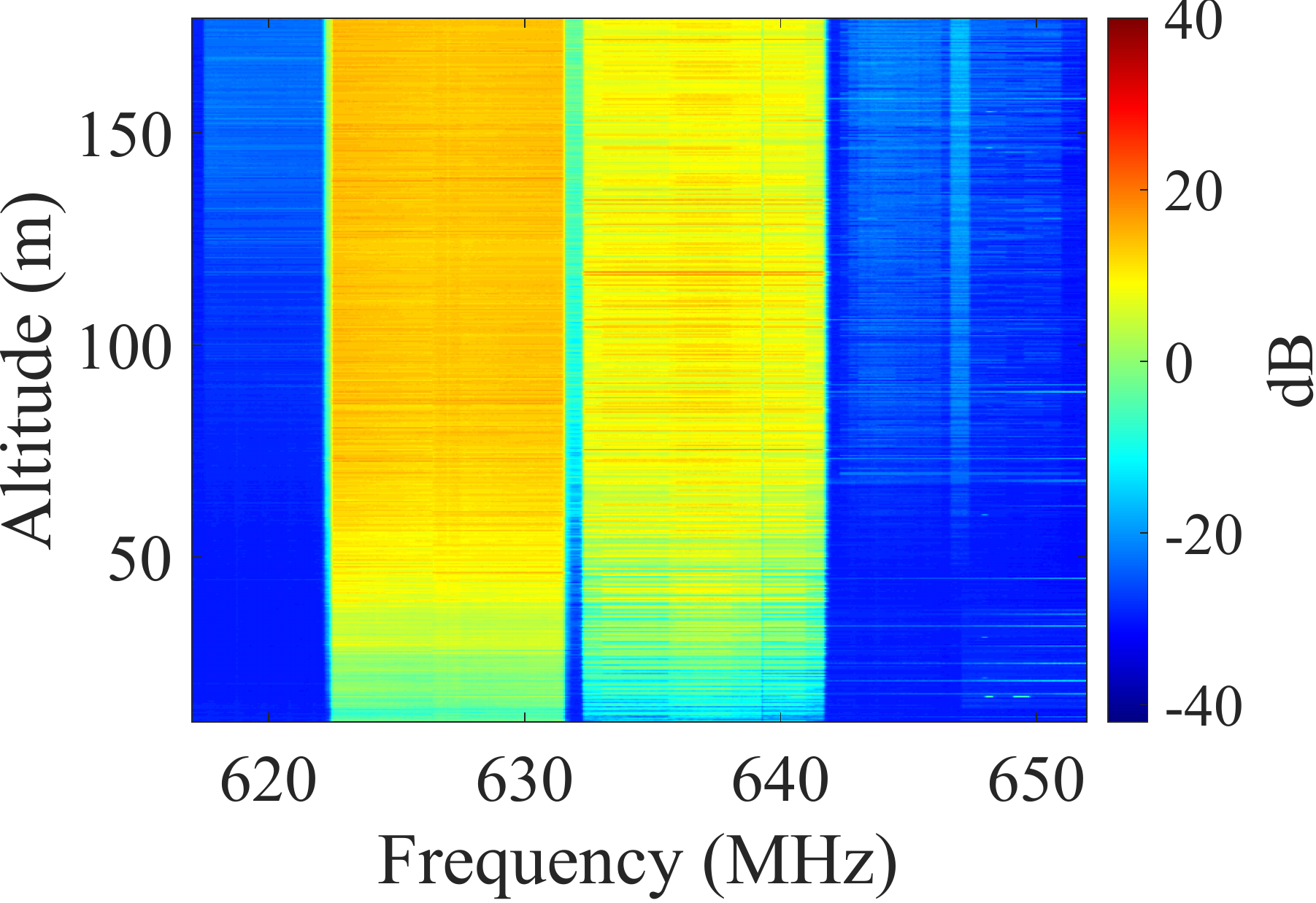}
\caption{5G band n71 (DL).}
\label{fig:n71_downlink_wheeler}
\end{subfigure}
\caption{Measured 5G DL power for rural environment.}
\label{fig:n5_n71_downlink_wheeler}
\end{figure}

\begin{figure}[!t]
\centering
\begin{subfigure}{0.49\columnwidth} 
\centering
\includegraphics[width=\textwidth]{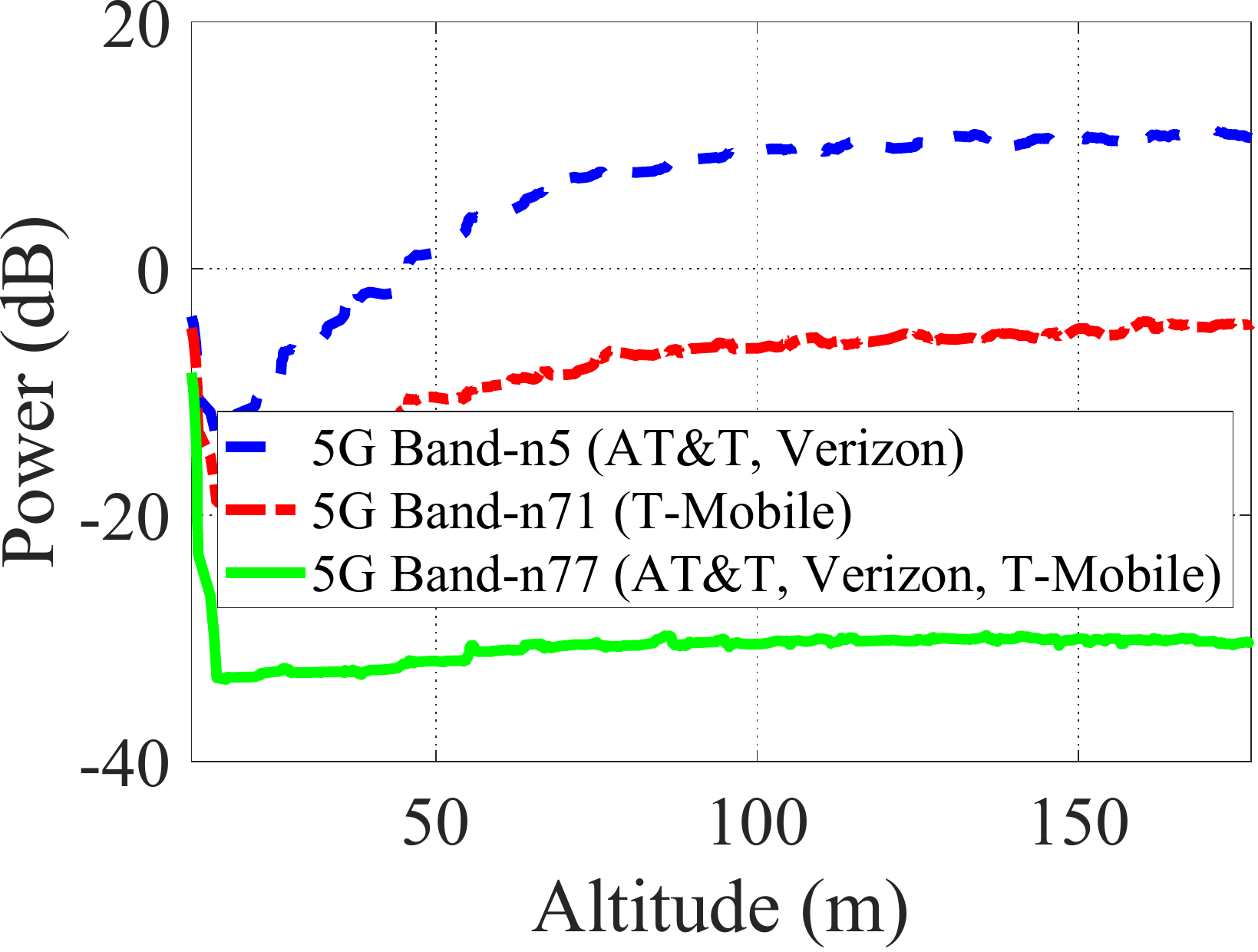} 
\caption{Mean.}\label{fig:mean5Gd_wheeler} 
\end{subfigure}
\begin{subfigure}{0.49\columnwidth} 
\centering
\includegraphics[width=\textwidth]{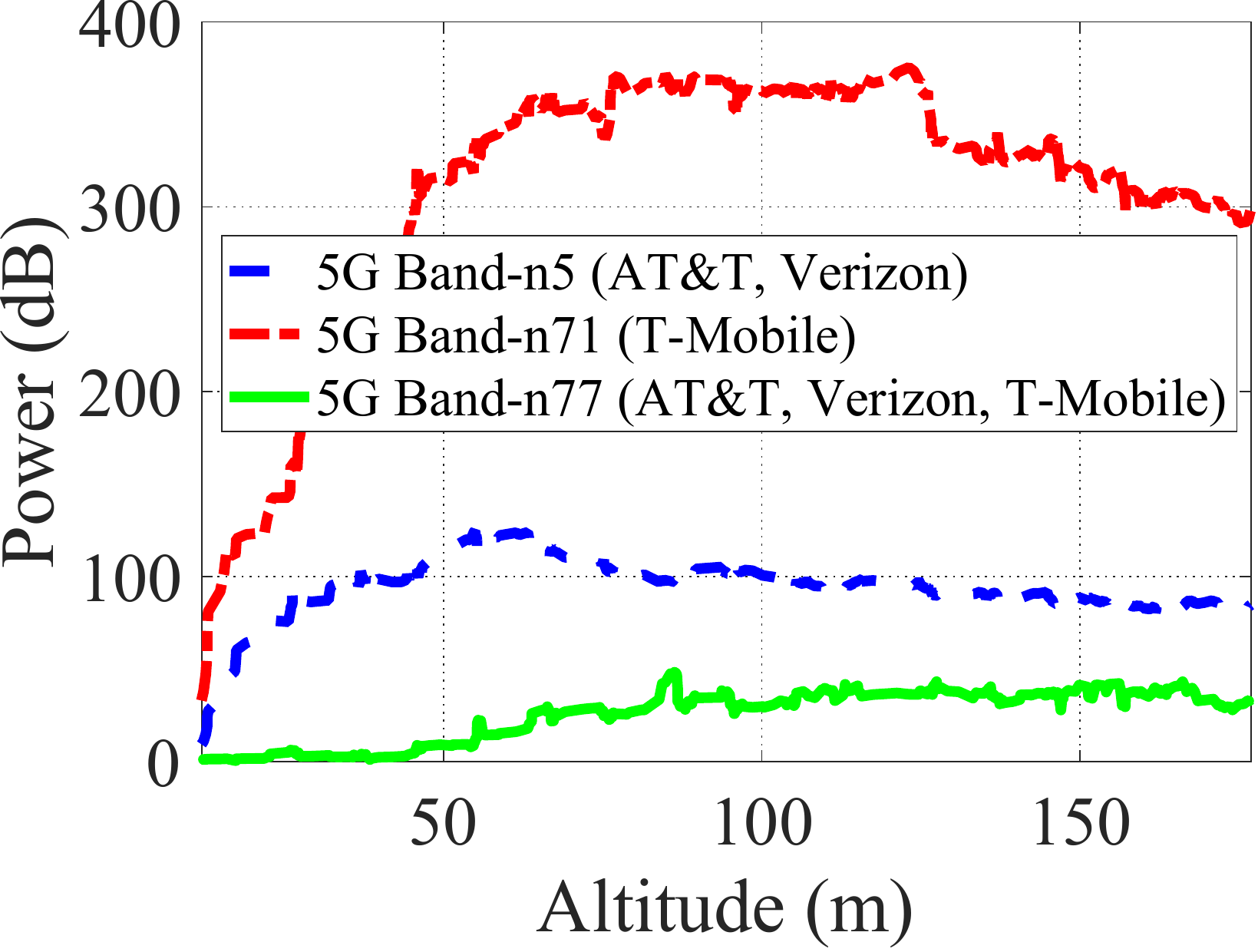}
\caption{Variance.}
\label{fig:var5Gd_wheeler}
\end{subfigure}
        \caption{Spectrum occupancy versus altitude in 5G bands n5 and n77 (DL) for rural environment.}
        \label{fig:mean_var_5Gdown_wheeler}
\end{figure}

Fig.~\ref{fig:n5_n71_downlink_wheeler} illustrates the measured power for 5G n5 and n71 bands by considering the DL frequency range. Similar to the urban case, it can be seen that the measured power for 870 - 880~MHz and 885-894~MHz are higher than the rest of spectrum in the rural environment.
Fig.~\ref{fig:mean_var_5Gdown_wheeler} depicts the mean and variance of the measured power versus altitude. 
As it can be observed from Fig.~\ref{fig:mean5Gd_wheeler}, the mean value of the measured power for n77 band remains almost constant for different altitudes, while it increases as the altitude increases up to almost 80~m for n5 and n71 bands.
As it is shown in Fig.~\ref{fig:var5Gd_wheeler}, the variance of the measured power for 5G band n71 shows higher values compared with n5 and n77.

\subsection{CBRS Band}

\begin{figure}[t!]
\centering
\includegraphics[width=0.49\linewidth]{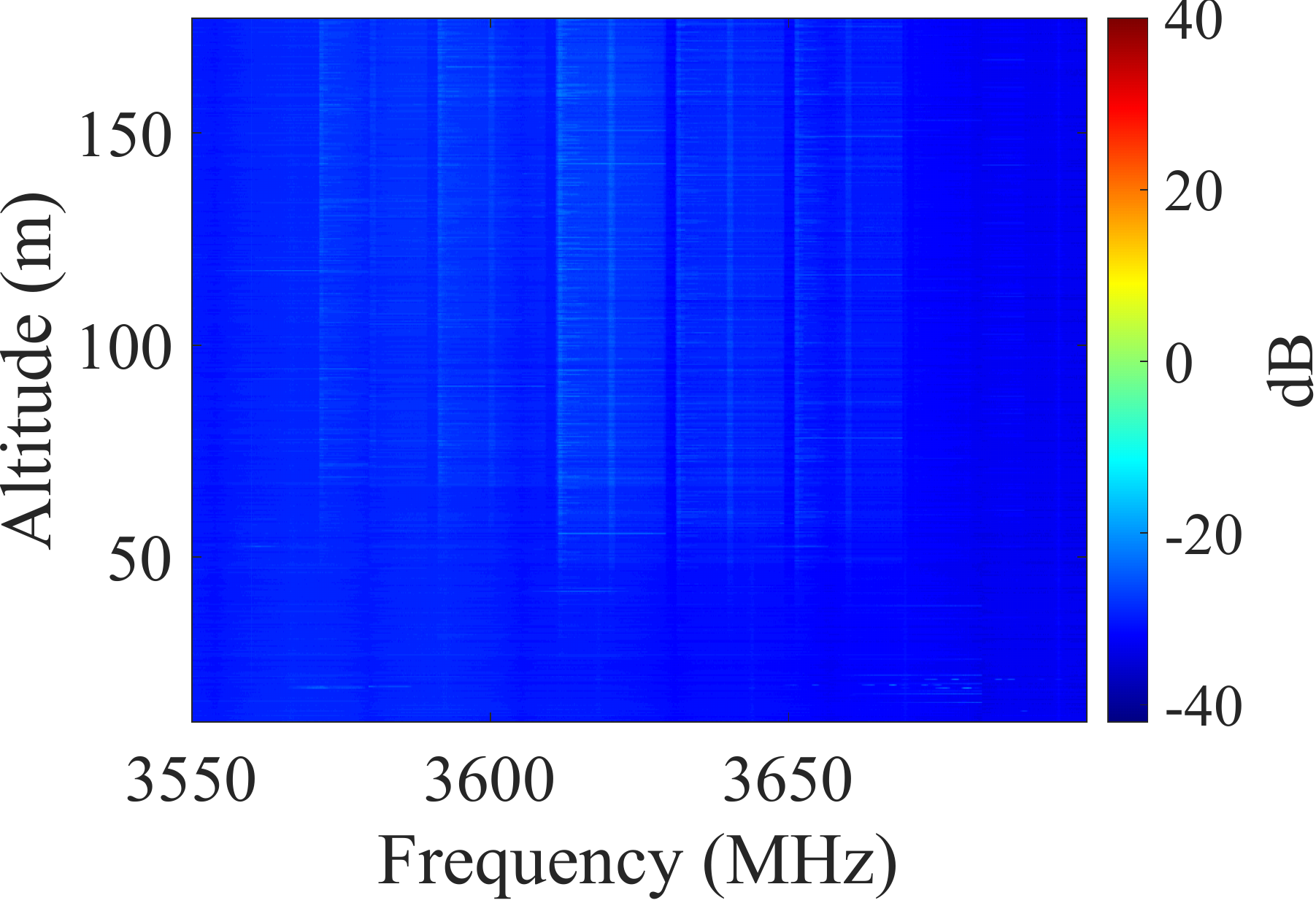}
\caption{Measured power during Helikite operation over rural
environment for CBRS band n48 (TDD UL/DL).}
\label{fig:n48_wheeler}
\end{figure}

\begin{figure}[!t]
\centering
\begin{subfigure}{0.49\columnwidth} 
\centering
\includegraphics[width=\textwidth]{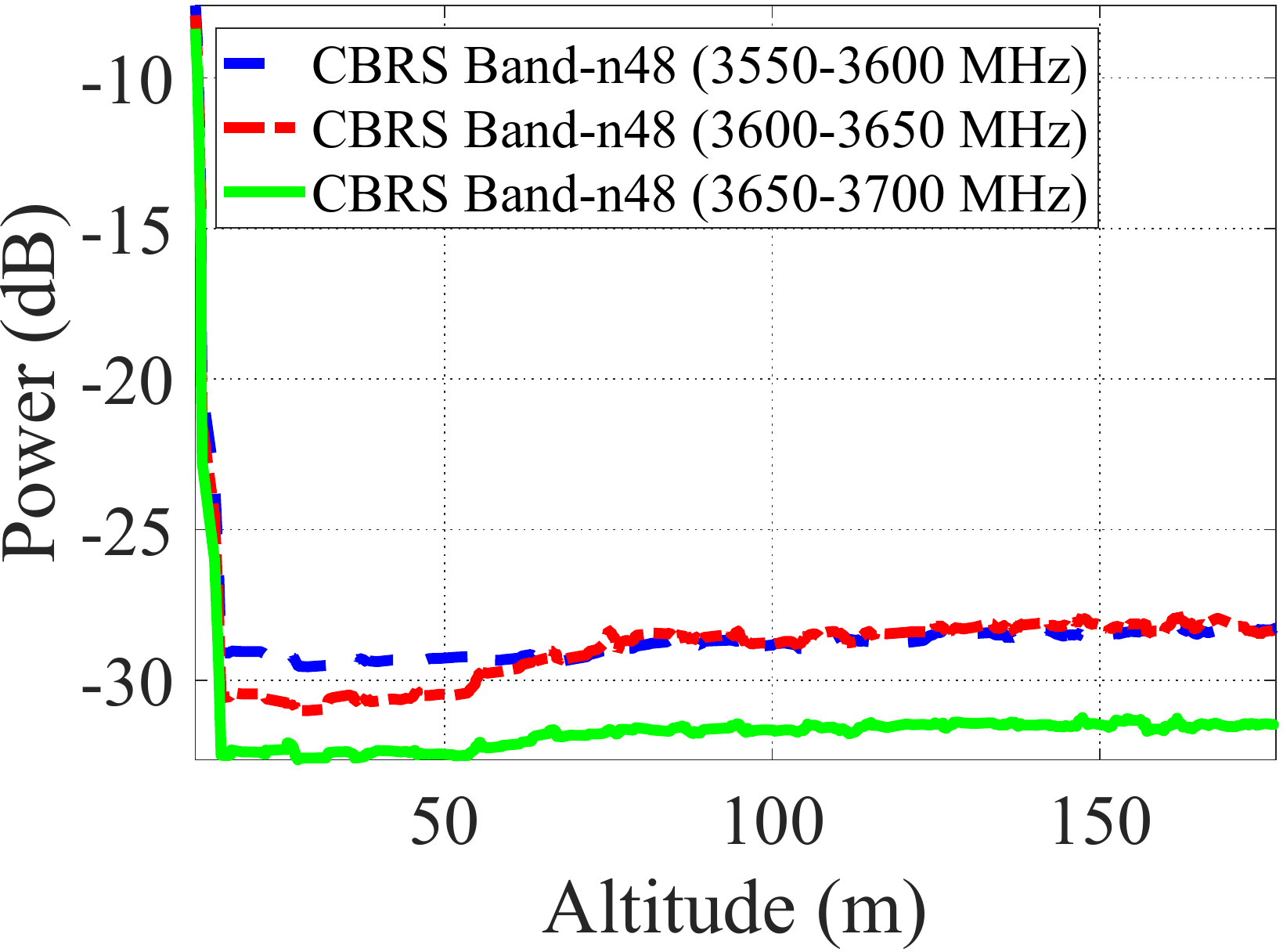} 
\caption{Mean.}\label{fig:meancbrs_wheeler} 
\end{subfigure}
\begin{subfigure}{0.49\columnwidth} 
\centering
\includegraphics[width=\textwidth]{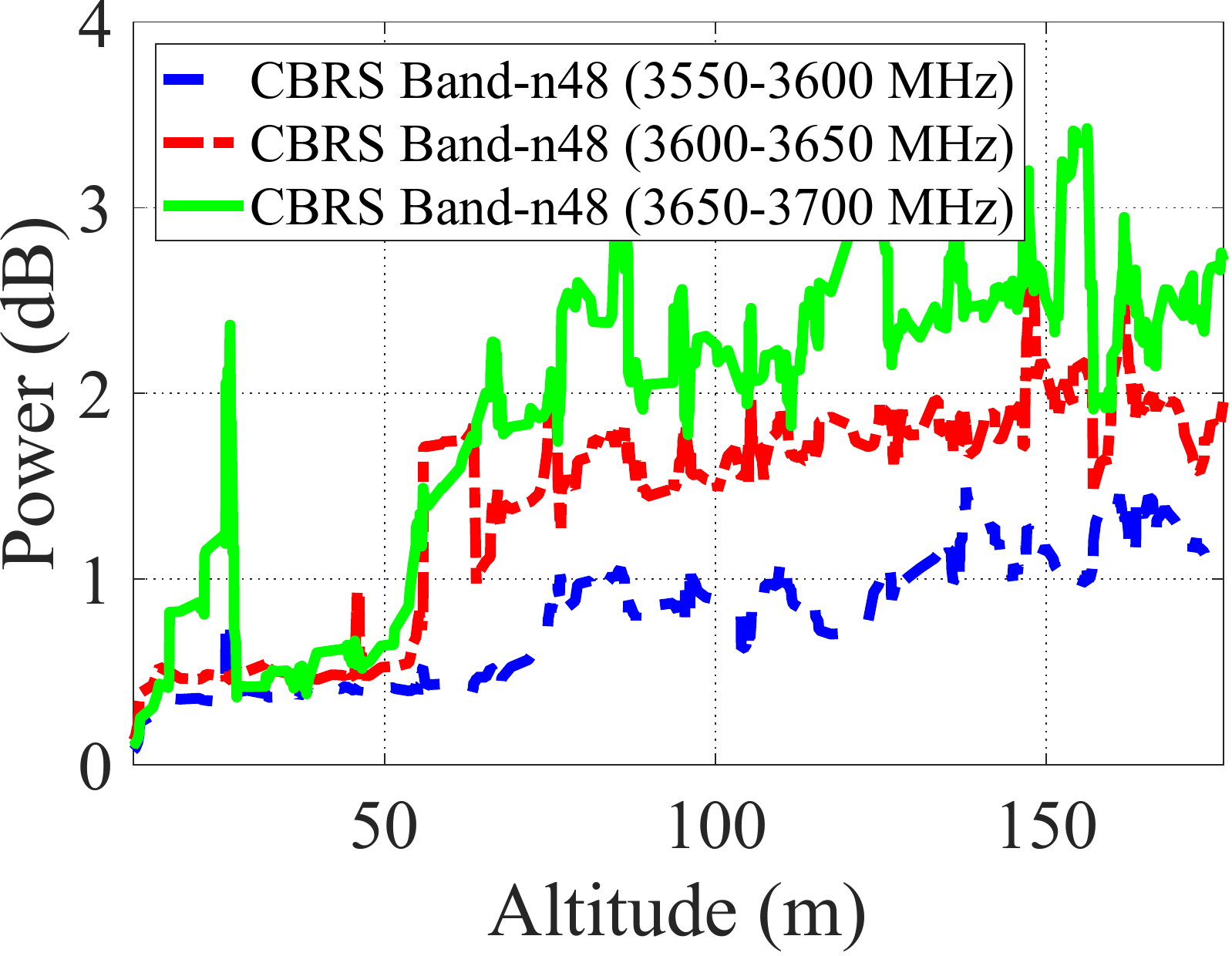}
\caption{Variance.}
\label{fig:varcbrs_wheeler}
\end{subfigure}
        \caption{Spectrum occupancy versus altitude in CBRS band for rural environment.}
        \label{fig:mean_var_CBRS_wheeler}
\end{figure}
Fig.~\ref{fig:n48_wheeler} present the measured power for CBRS n48 band for rural environment. As it can be seen, the spectrum is less crowded compared with the rural environment. In Fig.~\ref{fig:mean_var_CBRS_wheeler}, we study the mean and variance of the measured power versus altitude. As it can be observed, the mean value of the measured power for all three considered portions are almost similar and remain constant as the altitude increases. In addition, the variance also shows slight fluctuations compared to the other bands under consideration.

\section{Time Domain Analysis of Spectrum Occupency}\label{sec:spec_time}
\begin{figure*}[t!]
	\centering
	\subfloat[700~MHz - 800~MHz.]{\includegraphics[width=0.33\textwidth]{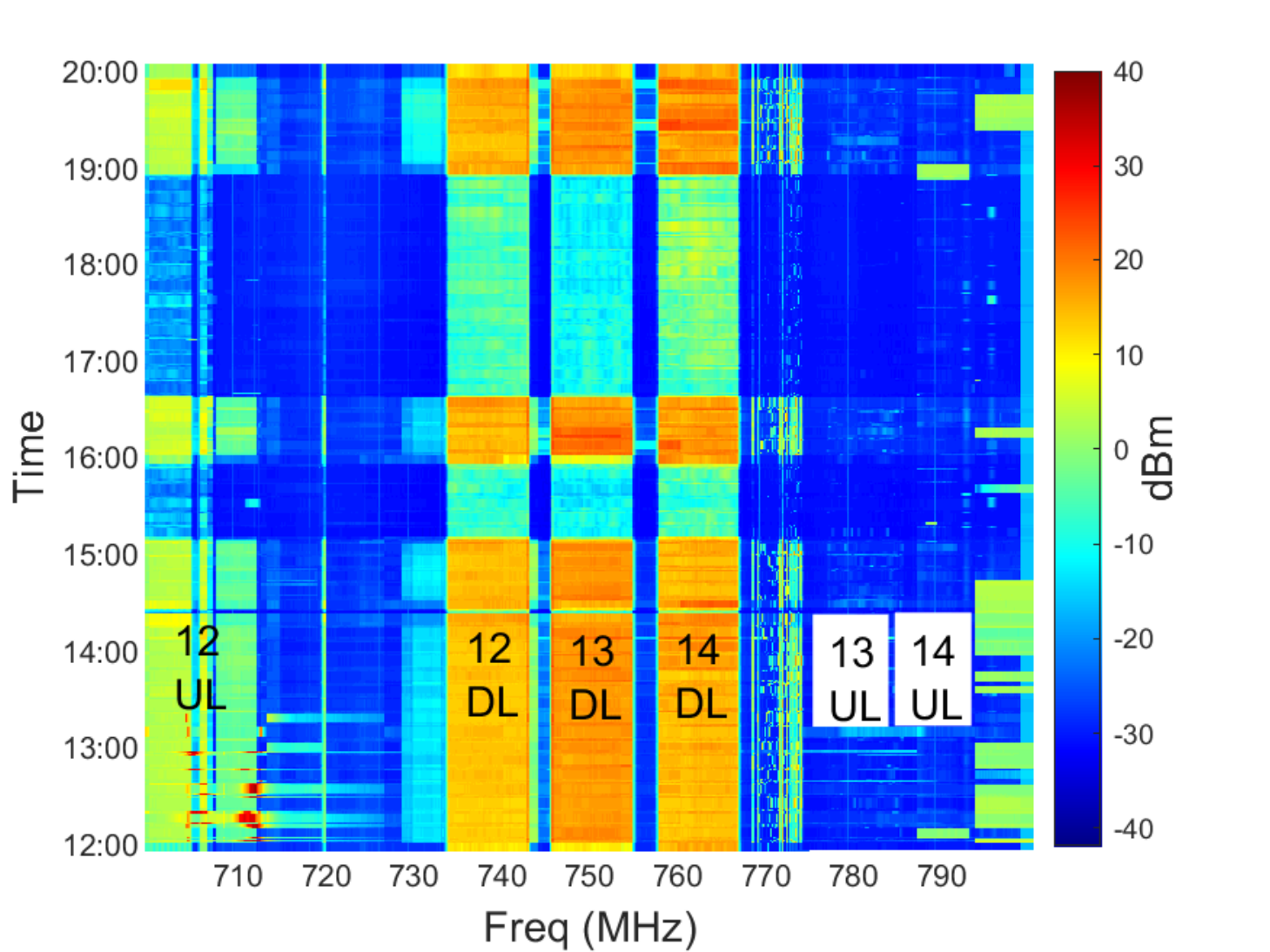}\label{fig:time_spec_1}}~
	\subfloat[2500~MHz - 2700~MHz.]{\includegraphics[width=0.33\textwidth]{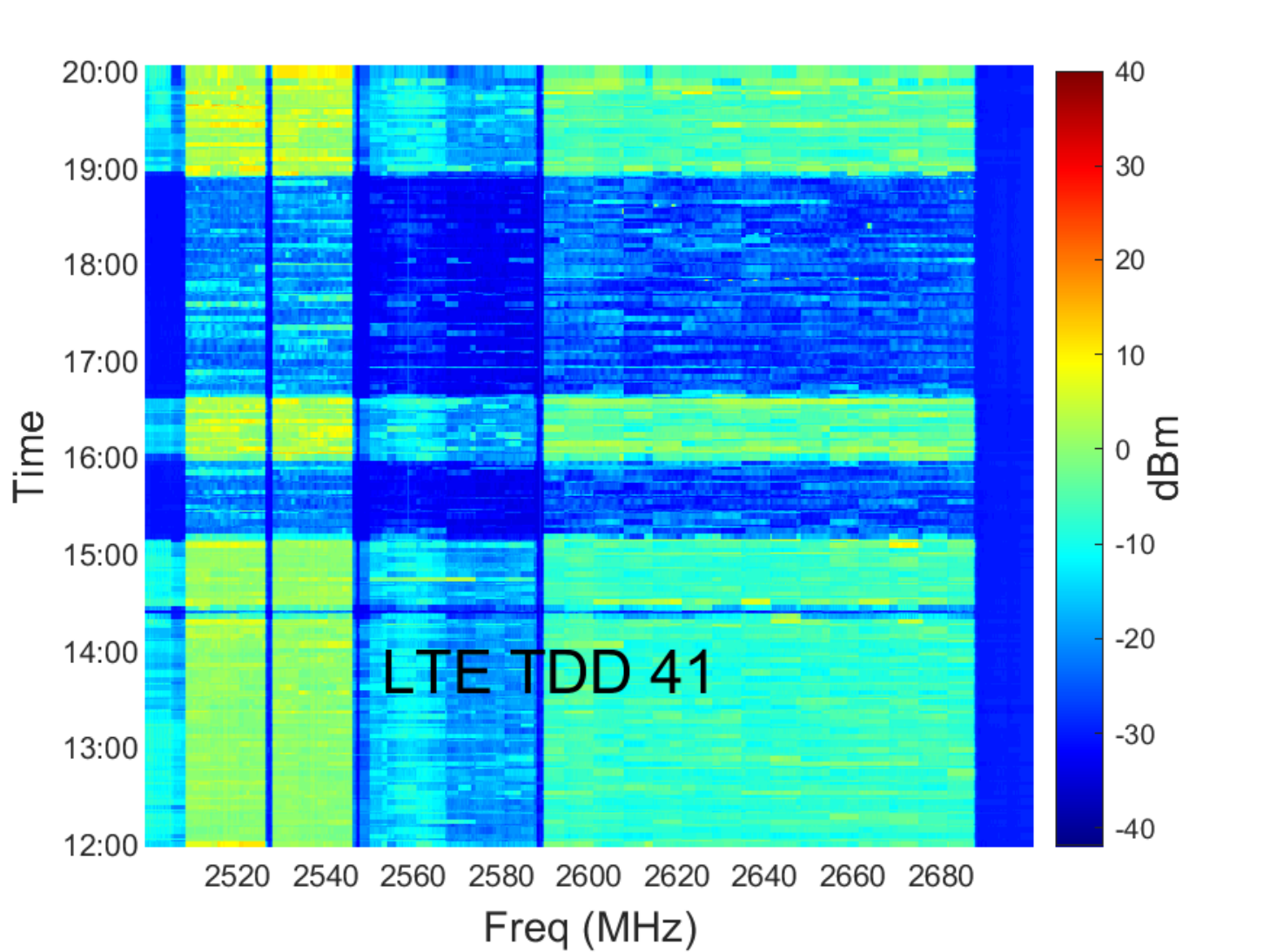}\label{fig:time_spec_2}}~
	\subfloat[3700~MHz - 3800~MHz.]{\includegraphics[width=0.33\textwidth]{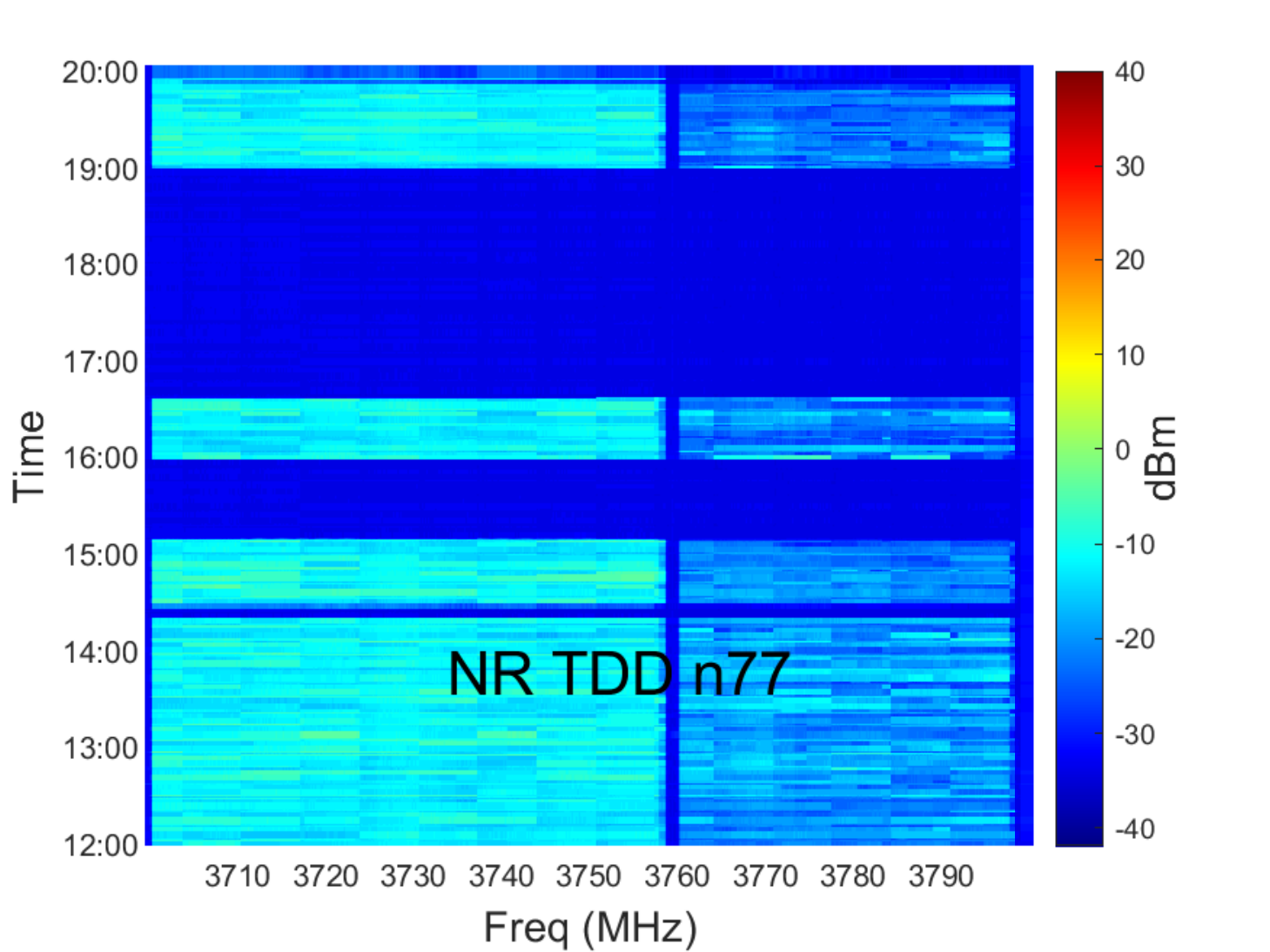}\label{fig:time_spec_3}}
 \vspace{-1mm}
	\caption{Spectrum monitoring during the measurement time. We observe different LTE and NR bands' occupancy and the received signal strength is strong when the Helikite floats at a high altitude.}\label{fig:time_spec}
  \vspace{-1mm}
\end{figure*}

\begin{figure*}[t!]
	\centering
 \vspace{-5mm}
	\subfloat[LTE FDD 12 band.]{\includegraphics[width=0.33\textwidth]{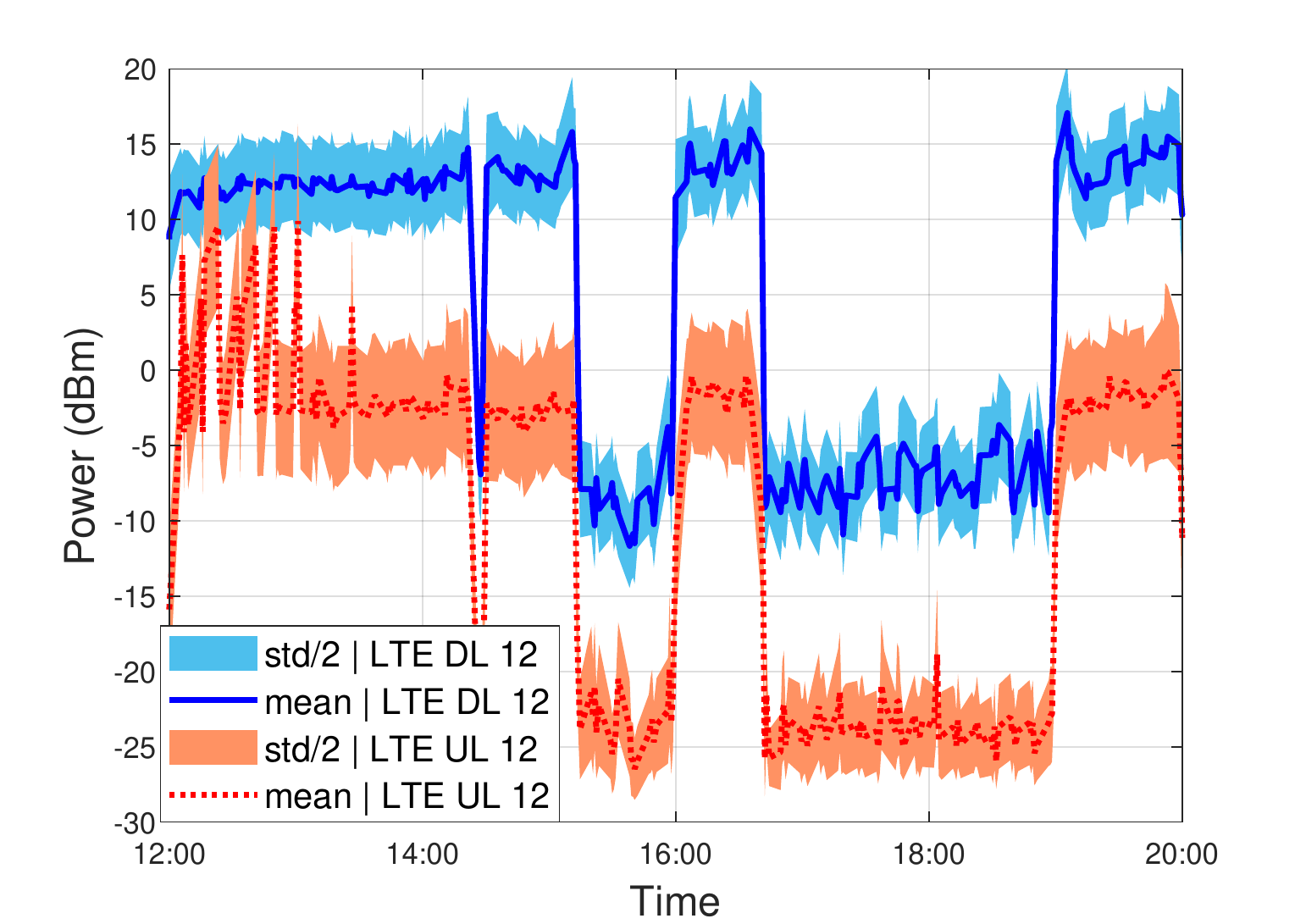}\label{fig:time_pw_1}}~
	\subfloat[LTE TDD 41 band.]{\includegraphics[width=0.33\textwidth]{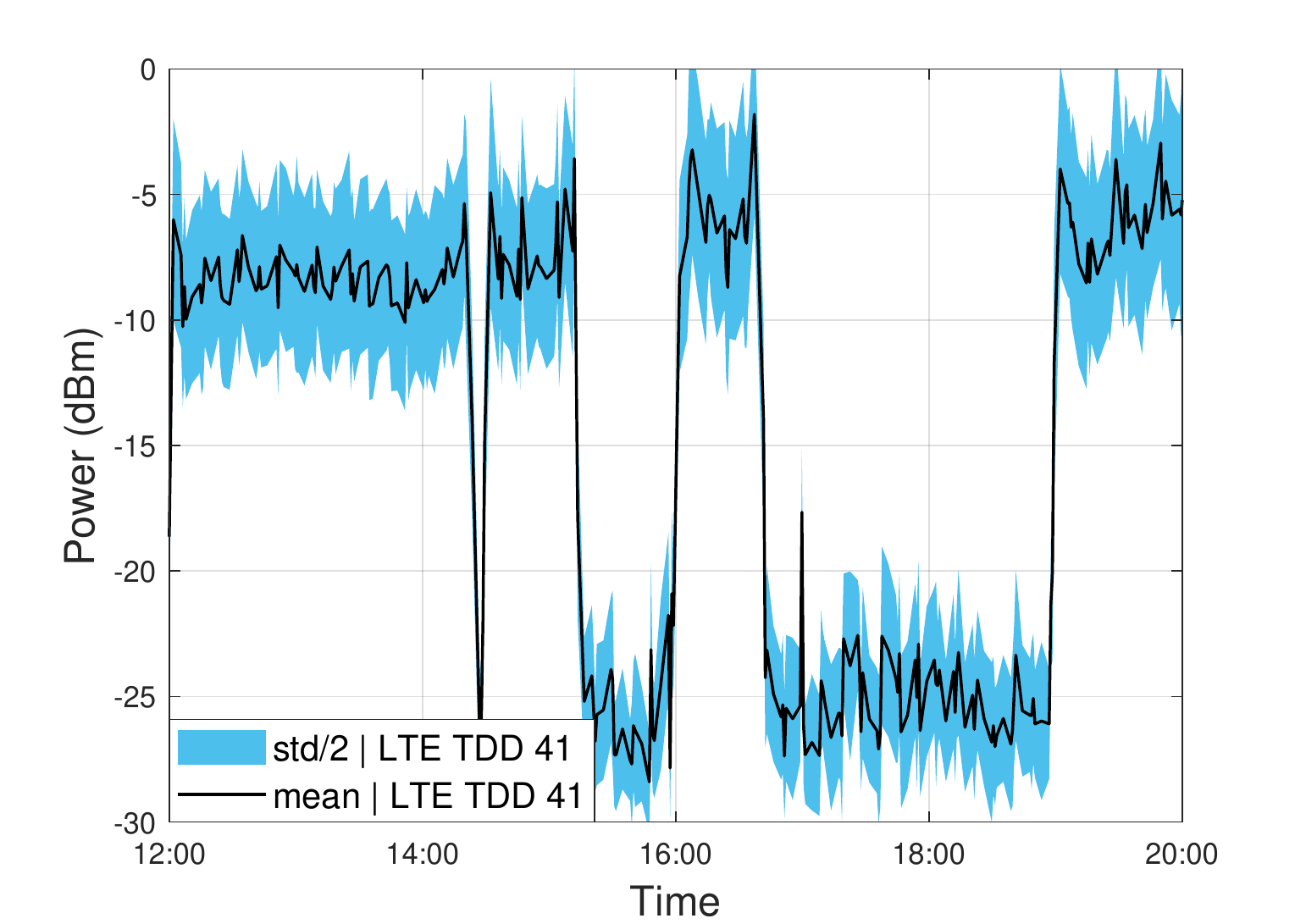}\label{fig:time_pw_2}}~
	\subfloat[NR TDD 77 band.]{\includegraphics[width=0.33\textwidth]{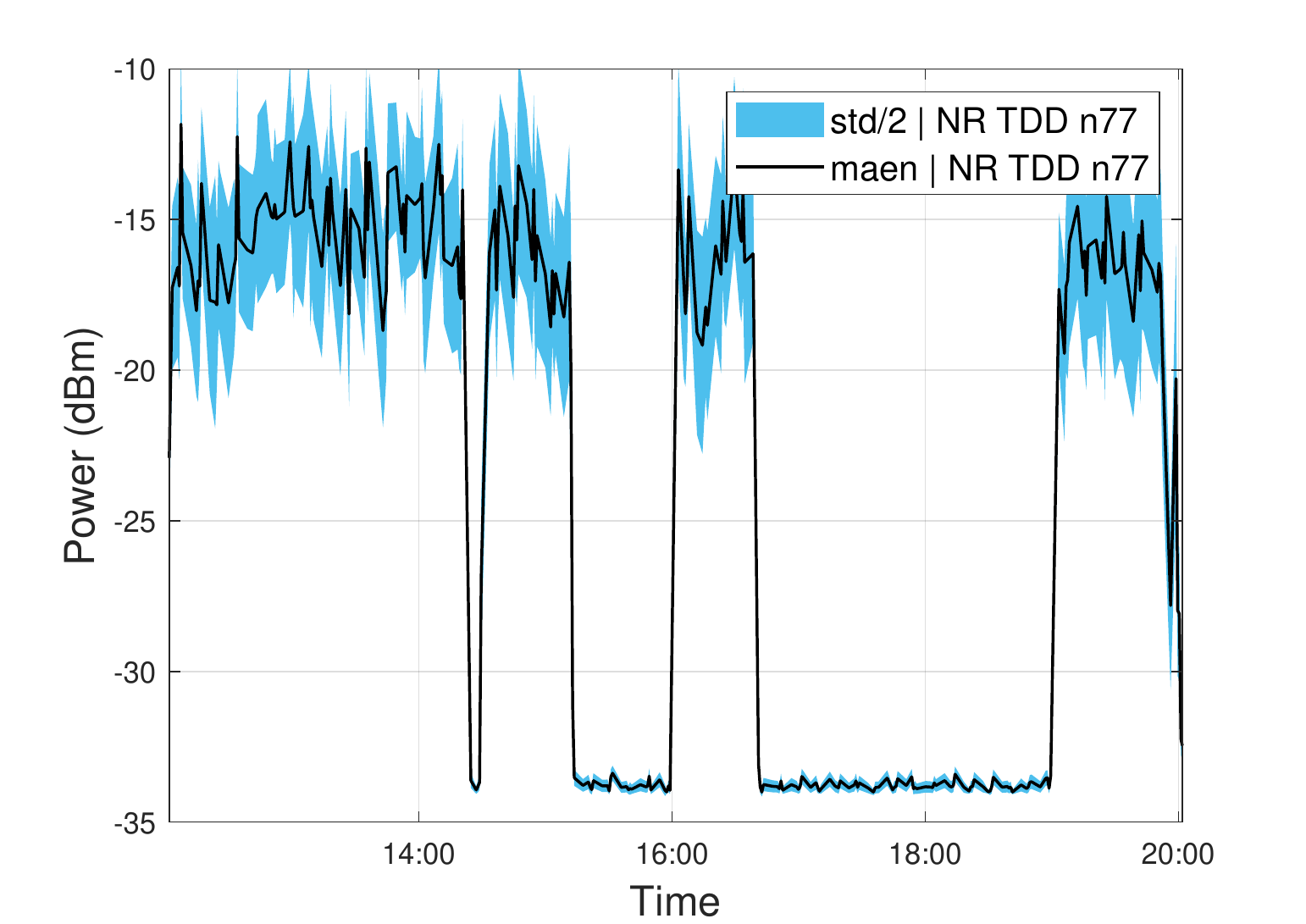}\label{fig:time_pw_3}}
 \vspace{-1mm}
	\caption{Received power of different LTE and NR bands during the measurement time. The solid lines represent the mean value of signal power and shaded areas indicate half of the standard deviation (std) of signal strength, which shows the variation of signal strength inside the specific bands.}\label{fig:time_pw}
 \vspace{-5mm}
\end{figure*}

In this section, we focus on the spectrum occupancy of LTE and NR signals in time, while we describe the altitude dependency of the spectrum in the previous section. For around 8 hours of measurement duration by the Helikite in the urban environment, we observe signal strength changes. This section focuses exclusively on those urban environment measurements.

Fig.~\ref{fig:time_spec} shows the spectrum monitoring results by the Helikite. The x-axis is the monitored spectrum range and the y-axis is the measured time stamp, which is indicated by hours and minutes. In Fig.~\ref{fig:time_spec_1}, we capture the frequency range from 700~MHz to 800~MHz, which contains LTE FDD bands 12, 13, 14 (see Table~\ref{table1}). First of all, we can clearly observe a series of occupied 10~MHz bandwidth 12, 13, and, 14 DL bands. On the other hand, the signal strength of UL bands is lower than DL bands, and UL bands 13 and 14 are scarcely occupied. We also observe that there are time periods when signal strength becomes low for the whole observed frequency range, which coincides with the periods where the altitude of the Helikite stays low in Fig.~\ref{fig:alt_time}. It implies that received signal strength is abruptly reduced by the blockage when the altitude of the Helikite is lower than a certain height. In addition, this tendency is observed in other frequency bands as well in Fig.~\ref{fig:time_spec_2} and Fig.~\ref{fig:time_spec_3}. In Fig~\ref{fig:time_spec_2}, we capture the frequency range 2500~MHz - 2700~MHz, which contains LTE TDD 41 band. Since carrier frequency is higher than Fig.~\ref{fig:time_spec_1}, we observe that this LTE band covers wider bandwidth: 20~MHz, 40~MHz, and 100~MHz. It is also observed that the received signal strength is lower than the frequency range in Fig.~\ref{fig:time_spec_1}. This is due to the fact that as carrier frequency increases a received signal suffers higher path loss, which is also observed in a much higher carrier frequency range in Fig.~\ref{fig:time_spec_3}. In particular, Fig.~\ref{fig:time_spec_3} shows spectrum occupancy of NR TDD n77 band, 3700~MHz - 3800~MHz. We can observe 40~MHz and 60~MHz bandwidth signals.

Fig.~\ref{fig:time_pw} shows the received signal strength changes during the measurement time for the captured LTE and NR bands. In Fig.~\ref{fig:time_pw_1}, we observe the LTE FDD UL/DL 12 band shown in Fig.~\ref{fig:time_spec_1}. Mean value of the received signal strength across the frequency band is represented by lines and half of the standard deviation (std) of signal strength is described by the shaded area around lines. It is observed that the signal strength of UL is lower than DL, while the variation of the signal strength of UL inside the band is higher than DL, which can be observed from higher std values. Fig.~\ref{fig:time_pw_2} and Fig.~\ref{fig:time_pw_2} show the received signal strength changes of LTE TDD 41 and NR TDD 77 bands which can be shown in Fig.~\ref{fig:time_spec_2} and Fig.~\ref{fig:time_spec_3}. It is observed that the signal strength fluctuation of NR TDD 77 band is higher than other bands such as LTE 12 and 41 bands.

\section{conclusion}\label{conclusion}

Using the data measured by a Helikite flying over an urban and rural environments, in this paper we studied spectrum measurements in various sub-6~GHz 4G, 5G and CBRS bands.
Both UL and DL spectrum occupancy has been investigated.
Our results revealed that generally the mean value of measured power tends to increase as the altitude increases due to higher probability of line-of-sight, at least for the considered maximum altitude range. Further, the spectrum of DL frequency ranges showed to be more crowded compared with the uplink ones for both environments. It has been also seen that for the rural environment the
mean value for LTE bands 13 and 14 are much higher than the other two bands
under consideration, as opposed to the urban environment. Furthermore, the performance of CBRS band for urban environment indicates more activity compared with the rural condition.

\bibliographystyle{IEEEtran}
\bibliography{reference}

\begin{thebibliography}{10}
\providecommand{\url}[1]{#1}
\csname url@samestyle\endcsname
\providecommand{\newblock}{\relax}
\providecommand{\bibinfo}[2]{#2}
\providecommand{\BIBentrySTDinterwordspacing}{\spaceskip=0pt\relax}
\providecommand{\BIBentryALTinterwordstretchfactor}{4}
\providecommand{\BIBentryALTinterwordspacing}{\spaceskip=\fontdimen2\font plus
\BIBentryALTinterwordstretchfactor\fontdimen3\font minus
  \fontdimen4\font\relax}
\providecommand{\BIBforeignlanguage}[2]{{%
\expandafter\ifx\csname l@#1\endcsname\relax
\typeout{** WARNING: IEEEtran.bst: No hyphenation pattern has been}%
\typeout{** loaded for the language `#1'. Using the pattern for}%
\typeout{** the default language instead.}%
\else
\language=\csname l@#1\endcsname
\fi
#2}}
\providecommand{\BIBdecl}{\relax}
\BIBdecl

\bibitem{islam2008spectrum}
M.~H. Islam, C.~L. Koh, S.~W. Oh, X.~Qing, Y.~Y. Lai, C.~Wang, Y.-C. Liang,
  B.~E. Toh, F.~Chin, G.~L. Tan \emph{et~al.}, ``Spectrum survey in
  {S}ingapore: {O}ccupancy measurements and analyses,'' in \emph{Proc. IEEE
  Int. conf. Cognitive Radio Oriented Wireless Networks and Communications},
  2008, pp. 1--7.

\bibitem{adelantado2017understanding}
F.~Adelantado, X.~Vilajosana, P.~Tuset-Peiro, B.~Martinez, J.~Melia-Segui, and
  T.~Watteyne, ``Understanding the limits of {LoRaWAN},'' \emph{IEEE Commun.
  Mag.}, vol.~55, no.~9, pp. 34--40, 2017.

\bibitem{ericsson}
\BIBentryALTinterwordspacing
Ericsson, ``Frequency reuse in limited spectrum networks,'' Oct. 2022,
  accessed: 2023-01-04. [Online]. Available:
  \url{https://www.ericsson.com/en/reports-and-papers/microwave-outlook/articles/maximizing-capacity-in-spectrum-limited-networks}
\BIBentrySTDinterwordspacing

\bibitem{1}
{D. Roberson}, ``The 5{G}/aviation crisis that never should have happened,''
  \url{https://www.linkedin.com/pulse/5g-aviation-crisis-never-should-have-happened-dennis-roberson/},
  2022, accessed: 2023-01-04.

\bibitem{CBand_Ref}
\BIBentryALTinterwordspacing
{W. Bellamy}, ``{US Airlines Begin Installing 5G C-Band Filter for Radio
  Altimeters on Airbus A320s},'' Sep. 2022, accessed: 2023-01-04. [Online].
  Available:
  \url{https://www.aviationtoday.com/2022/09/14/us-airlines-begin-installing-5g-c-band-filter-radio-altimeters-airbus-a320s/}
\BIBentrySTDinterwordspacing

\bibitem{GPS}
\BIBentryALTinterwordspacing
{K. Dennehy}, ``Ligado scraps {5G} network launch amid interference flap,''
  2022, accessed: 2023-01-04. [Online]. Available:
  \url{https://locationbusinessnews.substack.com/p/ligado-scraps-5g-network-launch-amid}
\BIBentrySTDinterwordspacing

\bibitem{chen2014survey}
Y.~Chen and H.-S. Oh, ``A survey of measurement-based spectrum occupancy
  modeling for cognitive radios,'' \emph{IEEE Commun. Surv. Tuts.}, vol.~18,
  no.~1, pp. 848--859, 2014.

\bibitem{al2020machine}
B.~Al~Homssi, A.~Al-Hourani, Z.~Krusevac, and W.~S. Rowe, ``Machine learning
  framework for sensing and modeling interference in {IoT} frequency bands,''
  \emph{IEEE Internet Things J.}, vol.~8, no.~6, pp. 4461--4471, 2020.

\bibitem{homssi2022artificial}
B.~A. Homssi, K.~Dakic, K.~Wang, T.~Alpcan, B.~Allen, S.~Kandeepan,
  A.~Al-Hourani, and W.~Saad, ``Artificial intelligence techniques for
  next-generation mega satellite networks,'' \emph{arXiv preprint
  arXiv:2207.00414}, 2022.

\bibitem{maeng2022analysis}
S.~J. Maeng, J.~Park, and I.~Guvenc, ``Analysis of {UAV} radar and
  communication network coexistence with different multiple access protocols,''
  \emph{arXiv preprint arXiv:2211.16614}, 2022.

\bibitem{azari2018uplink}
M.~M. Azari, F.~Rosas, A.~Chiumento, A.~Ligata, and S.~Pollin, ``Uplink
  performance analysis of a drone cell in a random field of ground
  interferers,'' in \emph{Proc. IEEE Wireless Commun. and Netw. Conf. (WCNC)},
  2018, pp. 1--6.

\bibitem{marojevic2020advanced}
V.~Marojevic, I.~Guvenc, R.~Dutta, M.~L. Sichitiu, and B.~A. Floyd, ``Advanced
  wireless for unmanned aerial systems: {5G} standardization, research
  challenges, and {AERPAW} architecture,'' \emph{IEEE Vehicular Technology
  Magazine}, vol.~15, no.~2, pp. 22--30, 2020.

\bibitem{Dataset1}
\BIBentryALTinterwordspacing
AERPAW, ``Helikite spectrum measurements (packapalooza),'' Aug. 2022, accessed:
  2023-01-04. [Online]. Available:
  \url{https://sites.google.com/ncsu.edu/aerpaw-wiki/aerpaw-user-manual/4-sample-experiments-repository/4-4-data-repository/aerpaw-nrdz-research/august-2022-helikite-spectrum-measurements-packapalooza}
\BIBentrySTDinterwordspacing

\bibitem{Dataset2}
\BIBentryALTinterwordspacing
------, ``Helikite spectrum measurements,'' May. 2022, accessed: 2023-01-04.
  [Online]. Available:
  \url{https://sites.google.com/ncsu.edu/aerpaw-wiki/aerpaw-user-manual/4-sample-experiments-repository/4-4-data-repository/aerpaw-nrdz-research/may-2022-helikite-spectrum-measurements}
\BIBentrySTDinterwordspacing

\bibitem{maeng2023COMSNET}
S.~J. Maeng, O.~Ozdemir, H.~Nandakumar, I.~Guvenc, M.~Sichitiu, R.~Dutta, and
  M.~Mushi, ``{Spectrum Activity Monitoring and Analysis for Sub-6 GHz Bands
  Using a Helikite},'' in \emph{Proc Int. Conf. Commun. Syst. Netw.
  (COMSNETS)}, Bengaluru, India, Jan. 2023.

\bibitem{maeng2022national}
S.~J. Maeng, I.~G{\"u}ven{\c{c}}, M.~Sichitiu, B.~A. Floyd, R.~Dutta,
  T.~Zajkowski, {\"O}.~{\"O}zdemir, and M.~J. Mushi, ``National radio dynamic
  zone concept with autonomous aerial and ground spectrum sensors,'' in
  \emph{IEEE Int. Conf. Communications Workshops (ICC Workshops)}, 2022, pp.
  687--692.

\end{thebibliography}

\end{document}